\documentclass[conference]{IEEEtran}

\IEEEoverridecommandlockouts

\usepackage{amsmath,amssymb,amsfonts,amsthm}
\usepackage{algorithmic}
\usepackage{graphicx}
\usepackage{textcomp}
\usepackage{xcolor}
\usepackage{orcidlink}
\usepackage{xcolor}
\usepackage{siunitx}
\usepackage{acronym}
\usepackage{placeins}
\usepackage{subfigure}
\usepackage{multirow}
\usepackage{float}
\usepackage{todonotes}
\usepackage{booktabs, tabularx}
\newcolumntype{Y}{>{\raggedright\arraybackslash}X}
\newcolumntype{Z}{>{\centering\arraybackslash}X}  
\usepackage{lipsum}
\usepackage{adjustbox}
\usepackage[capitalise]{cleveref}
\Crefname{equation}{Eq.}{Eqs.}
\Crefname{figure}{Fig.}{Figs.}
\Crefname{tabular}{Tab.}{Tabs.}
\usepackage[normalem]{ulem}

\usepackage[style=ieee, sorting=nty, dashed=false, maxcitenames=1, mincitenames=1,isbn=false,url=true,doi=false,date=year]{biblatex}

\AtEveryBibitem{
    \clearlist{language}
}
\usepackage{todonotes}

\usepackage{pgfplots, pgfplotstable} 
\usepackage{tikz}
\pgfplotsset{compat=1.15}
\usetikzlibrary{shapes.geometric, arrows, patterns}
\usetikzlibrary{pgfplots.groupplots, matrix}
\usetikzlibrary{arrows,backgrounds}
\tikzset{every picture/.style={/utils/exec={\footnotesize}}}

\definecolor{rwthBlue}{cmyk}{1.0,0.5,0.0,0.0}
\definecolor{rwthGreen}{cmyk}{0.7,0.0,1.0,0.0}
\definecolor{rwthOrange}{cmyk}{0.0,0.4,1.0,0.0}
\definecolor{rwthRed}{cmyk}{0.15,1.0,1.0,0.0}

\newcolumntype{L}[1]{>{\raggedright\arraybackslash}p{#1}}

\def\BibTeX{{\rm B\kern-.05em{\sc i\kern-.025em b}\kern-.08em
    T\kern-.1667em\lower.7ex\hbox{E}\kern-.125emX}}

\def\soamid{\textit{architecture platform}}
\def\soamids{\textit{architecture platforms}}
\def\commid{\textit{communication middleware}}
\def\commids{\textit{communication middlewares}}
\def\COMMIDS{\textit{Communication Middlewares}}
\def\COMMID{\textit{Communication Middleware}}
\def\Soamid{\textit{Architecture platform}}
\def\Soamids{\textit{Architecture platforms}}

\def\Commids{\textit{Communication middlewares}}
\def\shsoamid{\textit{platform}}
\def\shsoamids{\textit{platforms}}
\def\pattern{software design pattern}

\def\patterns{software design patterns}
\def\PATTERN{Software Design Pattern}
\def\system{automotive software system}
\def\systems{automotive software systems}
\def\commpat{communication pattern}
\def\commpats{communication patterns}

\def\Commpats{Communication patterns}
\def\COMMPAT{Communication Pattern}

\def\ImageSize{0.5\textwidth}

\usepackage{tcolorbox}
\tcbuselibrary{theorems}

\theoremstyle{definition}
\newtheorem{definition}{Definition}[section]

\makeatletter
\def\ps@IEEEtitlepagestyle{%
  \def\@oddfoot{\mycopyrightnotice}%
  \def\@evenfoot{}%
}
\def\mycopyrightnotice{%
	\begin{minipage}{\textwidth}
		\centering \scriptsize
		© 2022 IEEE. Personal use of this material is permitted. Permission from IEEE must be obtained for all other uses, in any current or future media, including reprinting/republishing this material for advertising or promotional purposes, creating new collective works, for resale or redistribution to servers or lists, or reuse of any copyrighted component of this work in other works.\hfill
	\end{minipage}
	\gdef\mycopyrightnotice{}
}
\makeatother

\begin{document}

\title{Modern Middlewares for Automated Vehicles: A Tutorial
\thanks{$^{1}$ The authors are with the Chair of Embedded Software, RWTH Aachen University, \texttt{kluener@embedded.rwth-aachen.de, molz@embedded.rwth-aachen.de, kampmann@embedded.rwth\-aachen.de and kowalewski@embedded.rwth-aachen.de} }
\thanks{$^{2}$ The author is with the Department of Aerospace Engineering, University of the Bundeswehr Munich, Germany, {\texttt{bassam.alrifaee@unibw.de}}}
}

\author{
	\IEEEauthorblockN{
         David Philipp Klüner$^{1}$\,\orcidlink{0009-0006-3451-9435},
         Marius Molz$^{1}$\,\orcidlink{0009-0005-8406-8727},~\IEEEmembership{Student~Members,~IEEE},
         Alexandru Kampmann$^{1}$\orcidlink{0009-0008-8340-1913},\\
         Stefan Kowalewski$^{1}$\orcidlink{0000-0001-9397-2009}, and Bassam Alrifaee$^{2}$\,\orcidlink{0000-0002-5982-021X},~\IEEEmembership{Senior Member,~IEEE}
    }
}

\maketitle

\begin{abstract}

This paper offers a tutorial on current middlewares in automated vehicles. 
Our aim is to provide the reader with an overview of current middlewares and to identify open challenges in this field.
We start by explaining the fundamentals of software architectures in distributed systems and the distinguishing requirements of \aclp{AV}.
We then distinguish between \commids\ and \soamids\ and highlight their key principles and differences.
Next, we present five state-of-the-art middlewares as well as their capabilities and functions.
We explore how these middlewares could be applied in the design of future vehicle software and their role in the automotive domain.
Finally, we compare the five middlewares presented and discuss open research challenges.

\end{abstract}

\section{Introduction}
\label{sec:intro}

\subsection{Motivation}
New challenges in vehicular computing arise as \acfp{AV} provide new automated driving functions, requiring advanced perception and complex decision making \cite{bello_recent_2019,liu_impact_2022}. 
These functions require resource-intensive processing of sensor data and the implementation of advanced decision-making algorithms.
Changed perspectives lead customers of \acp{AV} to increasingly expect their vehicles to behave like other modern computing products with ongoing updates and enhancements to their capabilities \cite{aaron_aboagye_facing_2017}.
These new computing challenges have consequences on the \ac{E/E} domain and the software domain in \acp{AV} \cite{burkacky_rethinking_nodate}.
In the \ac{E/E} domain, these challenges affect the transition from domain-based architectures to centralized zone controller architectures \cite{zhu_requirements-driven_2021,wang_review_2024}.
In current \ac{E/E} distributed architectures, which are still prevalent today, embedded microcontrollers are used to perform some selected vehicle functions \cite{zhu_requirements-driven_2021}.
The hardware offers very limited excess compute capacity, as the hardware is cost-optimized to the software requirements.
Updating a \ac{ECU} in a current architecture requires flashing the entire \ac{ECU}, and due to resource constraints, offers limited space for new functionalities \cite{benckendorff_comparing_2019}.
In contrast, zone controller \ac{E/E} architectures use more powerful multi-function \acp{ZCU} for compute intensive tasks.
These platforms provide significantly more compute resources by incorporating many-core processors, different \acp{OS}, and accelerators such as \acp{FPGA} and \acp{GPU}.
In these architectures, traditional \acp{ECU} at the edge of the system interact with hardware sensors and actuators, while zone controllers or central vehicle computers perform high-level functions \cite{benckendorff_comparing_2019,zhu_requirements-driven_2021}.

In the software domain, the automotive industry is moving from current signal-oriented software architectures to new software architectures built on modern automotive middlewares \cite{wang_review_2024}.
In signal-oriented software architectures, the software on each \ac{ECU} is designed to implement a limited number of functions and, due to the strong dependence on the microcontroller, is often not reusable \cite{vetter_development_2020}.
To alleviate these challenges and support new \ac{E/E} architectures, new middlewares such as AUTOSAR \ac{AP} seek to address them \cite{autosar_ap_explanation_sw_arch}.
Modern middlewares have not yet reached market dominance in the automotive domain and are still under active development \cite{autosar_ap_explanation_sw_arch}.
However, in other domains, such as robotics, middlewares have achieved the status of ubiquity and default tooling.
In the robotics domain, the \ac{ROS 2} middleware has emerged as the de facto standard middleware that attempts to solve similar challenges \cite{malavolta_how_2020}.

Middlewares should serve as foundational software for the development of automotive software \cite{burkacky_rethinking_nodate}. 
They define elements of the vehicle software architecture and provide crucial functions such as communication, update, and security for automotive applications. 
Consequently, choosing the correct middleware for automotive application development is important for manufacturers, developers, and researchers. 
Current state-of-the-art middlewares vary in supported features, software architecture, and license requirements.
This paper provides an introduction to middlewares in \acp{AV} and an overview of the current state-of-the-art. 

\subsection{Main Contributions}
\label{sec:intro:contrib}
The main contributions of this article include:
\begin{enumerate} 
    \item Introduction to \ac{E/E} architectures, middlewares, and underlying concepts.
    \item Overview of five state-of-the-art middlewares from the robotics and automotive domain and comparison of their features and performance.
    \item Examples showing how middlewares affect the software architecture of a system.
    \item Discussion of current challenges facing modern middlewares.
\end{enumerate}

\subsection{Outline}
\label{sec:intro:outline}
This article is structured as follows:
\Cref{sec:background} provides the background information necessary to understand middlewares in the automotive domain.
\Cref{sec:middleware} presents the fundamental concepts of middlewares and their function in a modern automotive software stack. It also provides an overview of the state-of-the-art middlewares FastDDS, \ac{SOME/IP}, Zenoh, ROS 2, and AUTOSAR \ac{AP}.
In \cref{sec:middleware_comp}, we compare these middlewares and discuss their features and \cref{sec:challenges} presents current research challenges for middlewares in the \ac{AV} domain. 

\section{Fundamentals}
\label{sec:background}
Before addressing middlewares and system software architectures in \acp{AV}, this section outlines the fundamental aspects of the software and computing hardware utilized in modern vehicles. 
The computing hardware is defined by the vehicle's \ac{E/E} architecture, which we explore through the following questions:
What defines an \ac{E/E} architecture and what is the current state of \ac{E/E} architectures?
Then we consider what challenges are associated with \ac{E/E} architectures and what future direction attempts to address these challenges.
Once the hardware and \ac{E/E} architecture are understood, we focus on software architecture, addressed through the following questions:
What defines a software architecture and why is a well-defined software architecture essential?
To remain in the automotive domain, we consider how the software architecture influences an automotive software systems.

\begin{figure}[tb]
    \centering
    \includegraphics[width=0.48\textwidth]{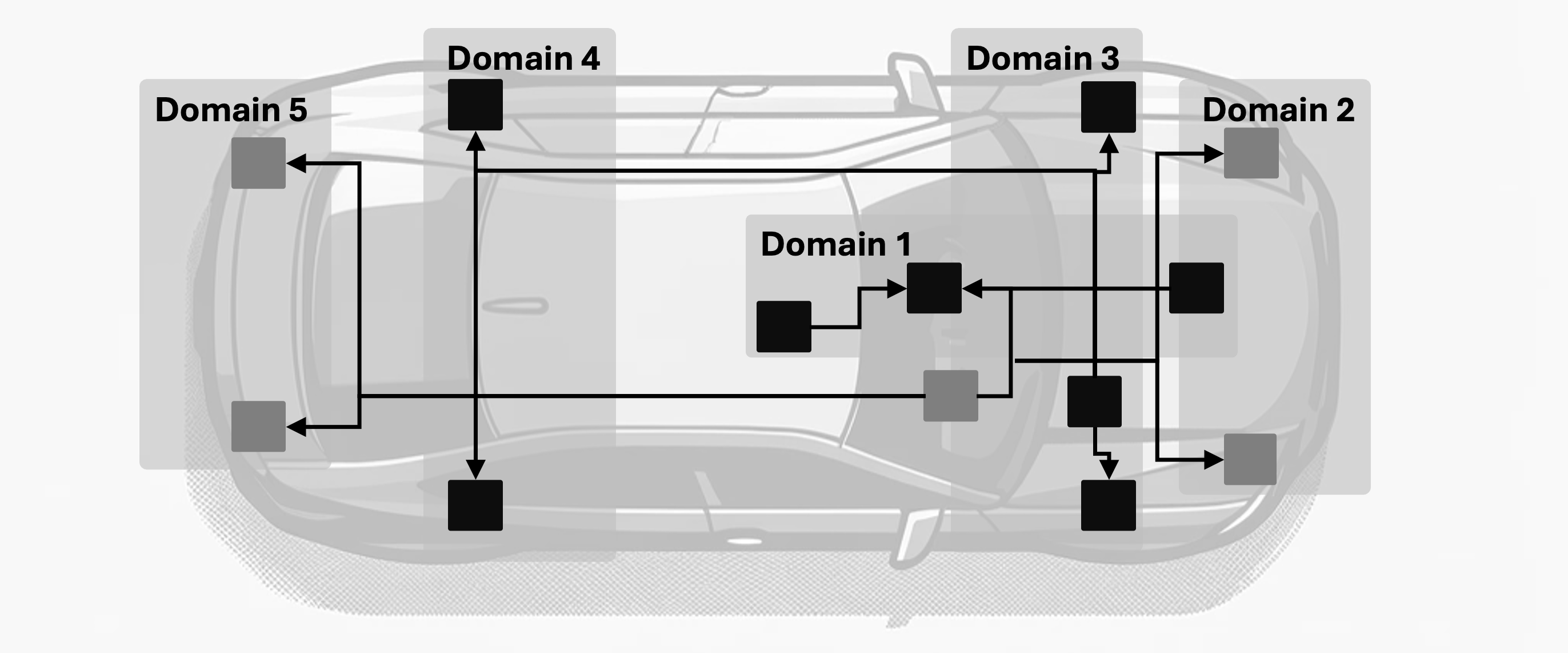}
    \caption{Illustration of a domain-based \ac{E/E} architecture. Actuator controllers are depicted in black with gray representing a sensor. The components are connected within their own domain. The domains are determined by the function that the \acp{ECU} contribute to, for example drive train control while cooperation between domains is limited \cite{wang_review_2024}.}
    
    \label{fig:domain_based}
\end{figure}

\subsection{What defines an \ac{E/E} architecture?}
\label{sec:background:arch}

\begin{definition}[\textit{\ac{E/E} Architecture}]
According to \citeauthor{zhu_requirements-driven_2021} \cite{zhu_requirements-driven_2021}, the E/E architecture can be defined as the organization of the electric and electronic components of the vehicle and their interactions. Key components of the architecture include \acp{ECU}, sensors, actuators, power systems, and  \ac{IVN}. The \ac{IVN} allows for interactions between the components and enables the components to cooperate to perform their expected functions.
\end{definition}

\noindent The core elements of the E/E architecture are various control units known as \acp{ECU} \cite{vetter_development_2020}.
\acp{ECU} are traditionally responsible for executing control functions within the vehicle, including management of the engine, drive train, and body electronics \cite{wang_review_2024}. 
These units vary in computational power, ranging from traditional embedded \acp{ECU} to high-performance zone controllers.

\begin{definition}[\ac{ECU}]
Electronic control units are the fundamental compute platforms in the vehicle. Classically based on microcontrollers and constrained in compute resources and interconnected via fieldbus systems, they operate fundamental components in the vehicle \cite{autosar_explanationPlatformDesign_2024}. 
\end{definition}

\noindent Furthermore, the \ac{E/E} architecture specifies the interconnection of \acp{ECU}. 
Communication is achieved through a fieldbus system, such as \ac{CAN}, FlexRay, or an Ethernet-based \ac{IVN}, which provides higher bandwidth, or a combination of these systems \cite{kampmann_dynamic_2019}. The interconnection and placement of \acp{ECU} is also referred to as the \ac{ECU} topology.
\begin{definition}[Automotive Software System]
The automotive software system refers to all software components in a vehicle. These components range from the operating system, individual functions, to algorithms in the vehicle.
\end{definition}

The choice between these communication systems and \acp{ECU} is driven by the requirements of the \system\ \cite{wang_review_2024}.
Initially, \cref{sec:background:sig_oriented} discusses the current domain-based \ac{E/E} architectures and the challenges introduced by \acp{AV} and increased software demands in vehicles.
Subsequently, \cref{sec:background:ee_challenges} discusses these challenges extensively.
Finally, \cref{sec:background:zone_based} introduces the zone-based \ac{E/E} architecture, which in cooperation with middlewares, addresses these challenges.

\subsection{What is the current state of \ac{E/E} architectures?}
\label{sec:background:sig_oriented}

In current vehicles, the domain-based architecture is dominant \cite{benckendorff_comparing_2019,zhu_requirements-driven_2021}.
The core component of this current architecture is the dedicated single-application \ac{ECU}, where the processor and hardware have been narrowly selected, with limited additional computing resources, to support a single or few functions \cite{van_dijk_future_2017, zhu_requirements-driven_2021,kampmann_asoa_2022}.
An example of such a single-application \ac{ECU}, is the engine control unit, which is specifically tasked only with monitoring and controlling the internal combustion engine in a vehicle.
These \acp{ECU} are commonly embedded microcontroller-based devices that run real-time operating systems or software directly on the hardware.
Changes in these embedded devices often require a reflash of the microcontrollers, making changes to the software difficult.

Communication in this architecture is characterized by static fieldbus systems with fixed mappings between senders and receivers \cite{vetter_development_2020}.
This architecture pattern is also referred to as a signal-oriented architecture and is characterized by its statically defined \ac{IVN} between \acp{ECU}.
These signals follow predetermined routes from senders to receivers, organized in a communication matrix shared among all \acp{ECU}. 
The statically defined communication matrix allows for real-time constraints and predictable behavior \cite{vetter_development_2020,wang_review_2024}.

The \acp{ECU} and their communication are structured by domains such as engine control or body control. Domain-based architectures require many component interconnections, as components in a single domain may not be physically close in location.

Different domains are interconnected by gateway \acp{ECU}, which bridge messages between domains. Generally, domains are isolated from each other with gateways that provide limited interconnection between networks.

\begin{definition}[ Domain-based \ac{E/E} Architecture]
A domain-based \ac{E/E} architecture typically consists of multiple single-application \acp{ECU} interconnected to each other based on their domain. Communication is characterized by static fieldbus systems with fixed communication schedules saved in communication matrices. 
\end{definition}

\Cref{fig:domain_based} illustrates an exemplary domain-based \ac{E/E} architecture, where multiple \acp{ECU} are interconnected using multiple \ac{CAN} fieldbus systems according to the domain. 
The cooperation between applications in separate domains is generally limited due to bandwidth limitations. 
Data from connected sensors is primarily processed locally within each ECU and is not commonly shared across the \ac{IVN}.

\subsection{What challenges are associated with \ac{E/E} architectures?}
\label{sec:background:ee_challenges}
New challenges for vehicles in general also present challenges for \ac{E/E} architectures.
Providing increasingly automated functions requires additional compute resources, and software updates require new types of software architecture.
\citeauthor{zhu_requirements-driven_2021} \cite{zhu_requirements-driven_2021} define these challenges as follows:

\begin{enumerate}
    \item The bandwidth bottleneck in modern vehicles is a significant challenge, as current \acp{IVN} have limited capacity to handle the data demands of advanced perception systems in \acp{AV}. This limitation requires more efficient data transmission methods to support increased bandwidth requirements.
    
    \item With the addition of numerous \acp{ECU} to support new vehicle functions, the complexity of wiring within vehicles has increased. This increase in complexity not only makes maintenance more challenging, but also contributes to more complicated vehicle architectures, negatively affecting performance and reliability.
    
    \item Ensuring low deterministic latency in \acp{IVN} is critical, especially in the context of highly concurrent computations, and large traffic volumes, such as sensor data. The requirement for deterministic latencies poses a challenge, necessitating improved \ac{IVN} management and data prioritization strategies.
    
    \item The design of future automotive architectures must prioritize flexibility and scalability, facilitating online updates, maintenance, and dynamic reconfiguration.
\end{enumerate}

\subsection{What is the near-future of \ac{E/E} architectures?}
\label{sec:background:zone_based}

The currently emerging \ac{E/E} architecture, seeking to address current challenges, is the zone-based \ac{E/E} architecture \cite{zhu_requirements-driven_2021}. 
The components in this \ac{E/E} architecture are structured in vehicle zones by proximity and \acp{ECU} interconnected within these zones.
In contrast to the current domain-based architecture, components are no longer integrated based on function, but rather location.

In the zone-based \ac{E/E} architecture, traditional \acp{ECU} at the edge of the vehicle are supported within a zone by \acp{ZCU} and central vehicle computers. 
\acp{ZCU} represent higher performance \acp{ECU} supporting multiple vehicle functions rather than a single one. As required for co-hosting several functions at once, these architectures also support more modern software architectures. The \acp{ZCU} may employ a PC \ac{OS}, such as Linux or QNX. \cite{wang_review_2024,zhu_requirements-driven_2021}

Communication in zone-based \ac{E/E} architectures often uses automotive ethernet for its \ac{IVN} or hybrid communication architectures.
Hybrid topologies that integrate different fieldbus and ethernet networks, such as an ethernet \ac{IVN} between \acp{ZCU} and fieldbus \acp{IVN} to the edge \acp{ECU}, are also employed \cite{wang_review_2024}.
Automotive ethernet is an extension to classical ethernet, allowing higher data rates and increasing the possible number of interconnected components compared to fieldbus \acp{IVN} \cite{hu_gatekeeper_2022,noauthor_ieee_2022}.

\acp{ECU} within a zone are connected to their zones \ac{ZCU}, while the connection between zones is established by connecting \acp{ZCU} either directly or using a central vehicle computer. 
Due to this simplified topology, the wiring complexity is greatly reduced and software updates and new communication paths can be more easily implemented using the \acp{ZCU}.
However, more communication between \acp{ECU} and \acp{ZCU} is required, as cooperating \acp{ECU} are likely to be in separate zones \cite{wang_review_2024,zhu_requirements-driven_2021}.
For example, in hybrid topologies, the in-zone communication between edge \acp{ECU} to the \ac{ZCU} could be implemented using fieldbus-based networks, while the \acp{ZCU} use automotive ethernet to address greater bandwidth and latency requirements.

\begin{figure}[tb]
    \centering
    \includegraphics[width=0.48\textwidth]{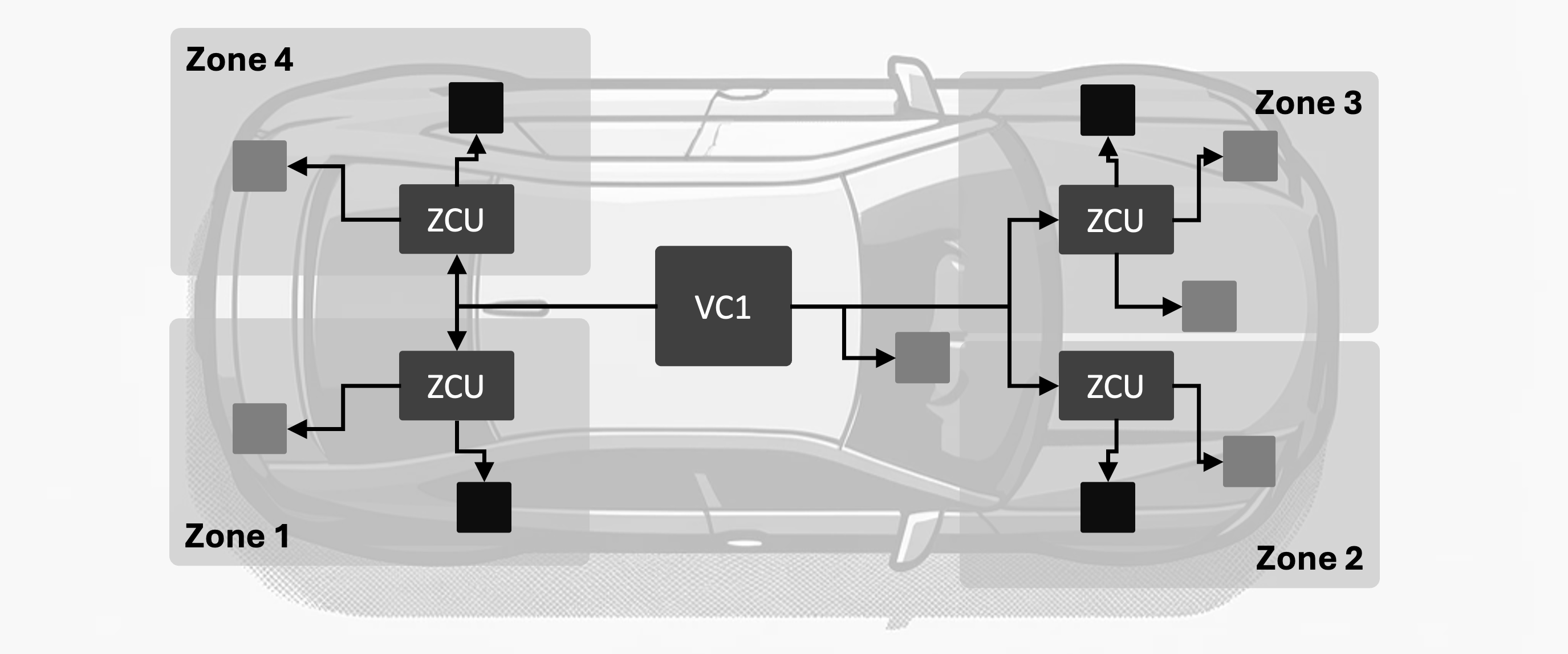}
    \caption{Illustration of a zone-based \ac{E/E} architecture. Actuator controllers are depicted in black with gray representing a sensor. Each component is connected to a zone control unit, which in turn is connected to a central vehicle computer VC1. This architecture divides the vehicle into four separate zones connected according to proximity with a high degree of cross-zone cooperation \cite{wang_review_2024}.}
    \label{fig:zone_based}
\end{figure}

\Cref{fig:zone_based} illustrates an example zone-based \ac{E/E} architecture. 
Based on the division of the vehicle into zones, the example depicts multiple \acp{ZCU} performing multiple functions and interacting with \acp{ECU} at the edge of the vehicle.

Following the zone-based \ac{E/E} architecture, the centralized architecture has emerged as the target architecture for multiple \acp{OEM} \cite{zhu_requirements-driven_2021}. However, the exact implementation of this architecture remains an area of active research and development.

\subsection{ What defines a software architecture?}
\label{sec:background:softarch}

\begin{definition}[\textit{Software Architecture}]
A \textit{software architecture} is defined by \citeauthor{bass_software_2003} \cite{bass_software_2003} as the structure or structures of the system. The structure is defined by the software elements, their properties, and relationships.
\end{definition}

Similarly to the \ac{E/E} architecture in \cref{sec:background:arch}, design decisions include the core components of the system and the key properties of each component.
Components can take different forms, such as classes in an application, services for a web application, or individual applications composing a distributed system.
An architecture also specifies the interactions between the components of the software system.
For classes as components, interactions could be made via function calls or, in the case of web applications, HTTPS API calls.
The software architecture ensures a clear division of the responsibilities and paths of interaction in the software system.

\begin{figure}[tb]
    \centering
    \includegraphics[width=0.48\textwidth]{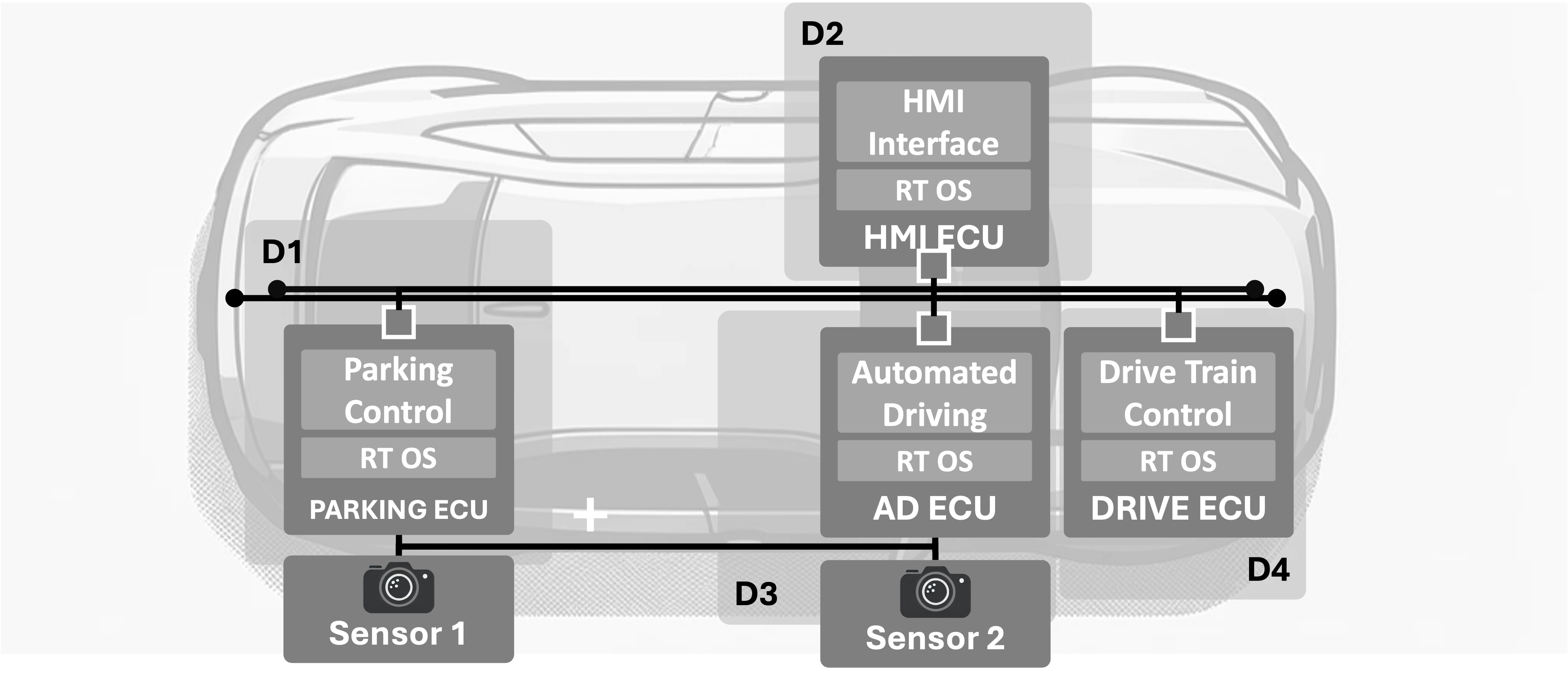}
    \caption{Illustration of the interaction of software and hardware in a domain-based \ac{E/E} architecture. The \system\ depicted is based on the requirements outlined in \cref{sec:background:signal_oriented_example}.}
    \label{fig:domain_based_with_sw}
\end{figure}

\subsection{Why is a well-defined software architecture essential?}
\label{sec:background:arch_adv}
An automotive software architecture is based on design decisions which often have to be made quite early in the development process and have far-reaching effects on the quality of the resulting software.
It is well known that architectural design decisions determine most non-functional attributes of a \system, and therefore can either support or impair properties like maintainability, updateability or reuseability \cite{garlan_software_2014}.
According to \citeauthor{garlan_software_2014} \cite{garlan_software_2014}, the advantages of a well-thought-out and well-defined software architecture include:

\begin{itemize}
    \item With the appropriate top-level design decision, the software architecture can ensure the \textit{updateability} of the software system. 
    The architecture can constrain the components and specify that the components must be designed with update mechanisms and modularity.
    Taking into account that continued development during the vehicle lifecycle is one of the drivers of automotive software, especially for \ac{SDV}, an architecture can contribute significantly to this goal.
    \item Architecting a software system with well-defined paths of interaction offers additional advantages for \textit{reuse}. A software component with clear interfaces explicitly defines what the software system needs to provide and what the component offers the software system. For automotive software development, for example, multiple software components that share the same interfaces and interface definitions can easily be exchanged. 
    \item The division of responsibilities aids in the \textit{construction} of software. Each component has well-defined tasks and responsibilities in the software system, allowing a developer to know what functions must be implemented and which are the responsibility of another developer.
\end{itemize}

\begin{figure*}[t]
    \centering
    \includegraphics[width=0.9\textwidth]{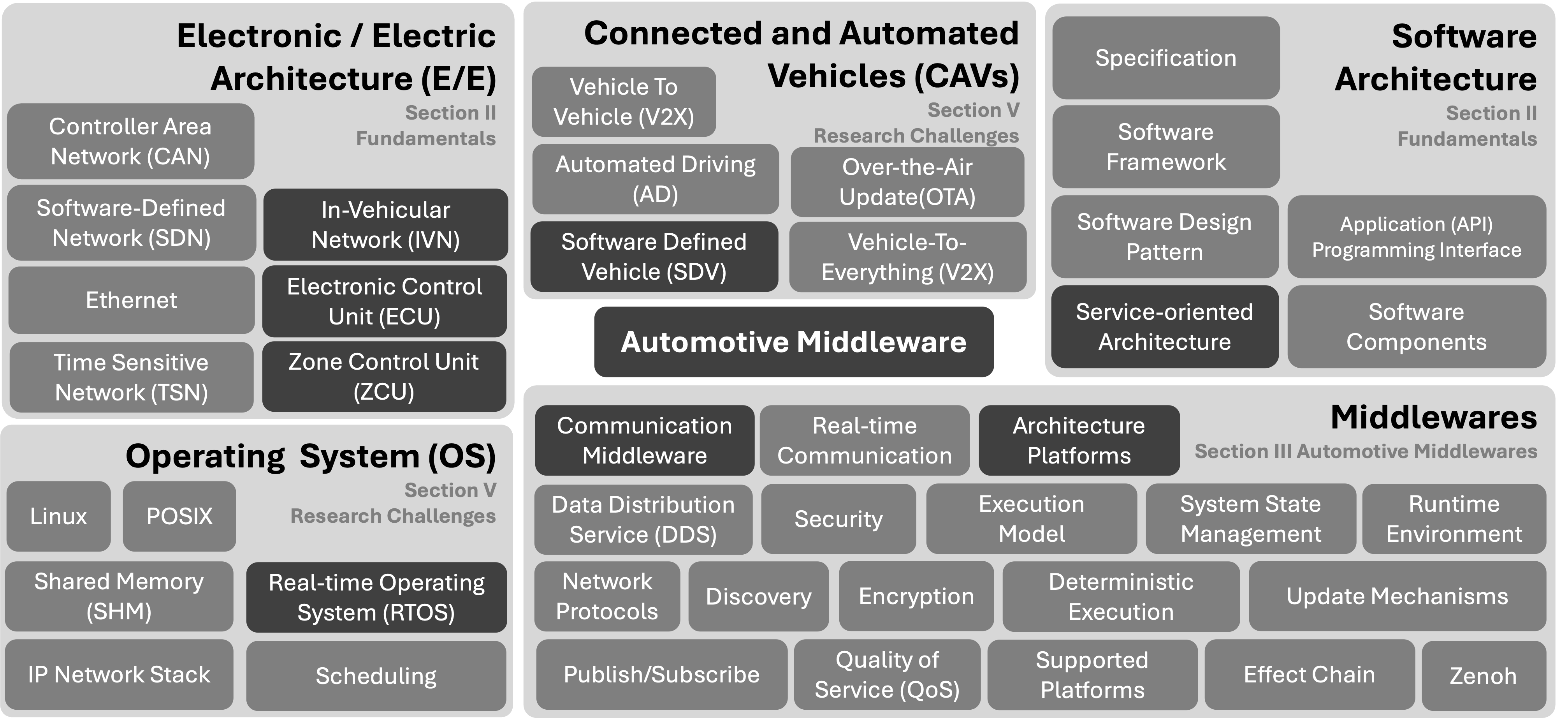}
    \caption{Illustration of the different domains interacting with middlewares. Domains are positioned according to their topical closeness. Black signifies increased importance from the perspective of the authors.}
    \label{fig:topics_mindmap}
\end{figure*}

\subsection{How does the software architecture influence automotive software systems?}
\label{sec:background:signal_oriented_example}
To show how a software architecture influences the capabilities and maintainability of a software system, we consider an example \system.
The example consists of an \ac{AD} system, and we explore how the software affects an expansion of its functionality. 
As the expansion and improvement of software functions is a major consideration in automotive software architectures, this example highlights the advantages and disadvantages of the three presented software architectures.

In this example the \system\ is responsible for the automated driving functions of the vehicle and can operate safely in the operational domain of a highway. 
This example has been derived from the UNICARagil project, where automated vehicles were developed \cite{woopen_unicaragil_2018}.
For this purpose, the \ac{AV} has long-range narrow field-of-view cameras.
Additionally, the \ac{AV} can park using short-range, near-field cameras and sensors, to achieve automated parking functions.
For safe operation, multiple requirements must be met and cooperation of multiple vehicle \acp{ECU} is required.
We define the basic requirements for the original, unmodified \system\ as follows.

\begin{enumerate}
    \item \textit{The \system\ must detect lane boundaries, road users, and traffic signs reliably on highways.}
    \item \textit{The \system\ must plan a safe and legal trajectory through the detected highway environment.}
    \item \textit{The \system\ must execute this trajectory safely.}
\end{enumerate}

To analyze the expansion of an architecture, we introduce one new requirement: \textit{The vehicle must drive automatically in city environments.}
We assume that this new requirement must be satisfied by updating the \system.

First, we consider a \system\ built on a domain-based \ac{E/E} and signal-oriented software architecture.

In this architecture, each function is executed on a single \ac{ECU} as depicted in \cref{fig:domain_based_with_sw}.
The \acp{ECU} for each function are connected via a \ac{CAN} fieldbus with a fixed communication schedule.
The \system\ architecture is illustrated in \cref{fig:domain_based_with_sw} involving the following \acp{ECU}:

\begin{itemize}
    \item The automation \ac{ECU} performs the automated driving functions in the vehicle on the highway. In this example, these include the perception and planning functions. It transmits trajectory commands using the fieldbus to the drivetrain module. Directly connected to this system are long range cameras for the highway domain.
    \item The \ac{HMI} \ac{ECU} provides the drive interface functions, such as information on the vehicles' states, and other interfaces, such as the infotainment system.
    \item The drive train \ac{ECU} is responsible for controlling the drive train, including the vehicles engine, and communicating the state of the drive train back to other components of the \system. In this example, it is also responsible for executing the planned trajectory of the vehicle.
    \item The parking \ac{ECU} implements automated parking functions. For this purpose it has access to short range, near field cameras, and ultra sound distance sensors which enable the vehicle to park itself automatically.
\end{itemize}

To satisfy the new requirement of city operations, the functions of the automated driving module must be expanded. 
To accomplish this, the core functionality of the automated driving module could be updated. 
The functionality of the perception component must be expanded for additional requirements of city environments, and the AD \ac{ECU} requires access to the near-field sensors and cameras of the parking module.
Motion planning capabilities must be expanded to accommodate the additional complexity of city roads and crosswalks.

As \acp{ECU} in domain-based \ac{E/E} architectures are commonly sized for their specific task, this addition of requirements could require switching to another \ac{ECU} or even adding more \acp{ECU}.
Both cases would then have to be integrated into the existing vehicle \ac{CAN} fieldbus network, requiring additional gateways or consuming more bandwidth than is available. 
The benefits of this architecture are the avoidance of resource contention between applications, as each functionality is assigned to a dedicated \ac{ECU} that is appropriately sized. 
Additionally, each \ac{ECU} can be cost-optimized to its specific function.
Changes to the communication matrix and the vehicle gateways also avoid possible \ac{IVN} contention.
However, the costs of switching hardware platforms, adding a new \ac{ECU}, additional wiring or gateways required to connect sensors or cameras, and integration with other \acp{ECU} are high.
Consequently, implementing such changes post-manufacture is highly impractical and cost prohibitive in such an architecture.

To solve the issues highlighted by this example, new approaches to software are equally required, as well as new \ac{E/E} architectures. 
The following \cref{sec:middleware} will present such a new approach in the form of modern automotive middlewares in conjunction with modern \ac{E/E} architectures.
First, we introduce automotive middlewares at large, then discuss how we classify middlewares and compare five state-of-the-art middlewares in detail.

\section{Automotive Middlewares}
\label{sec:middleware}
Middlewares are used in various computing domains, including \ac{IoT}, web services, and \ac{CPS} with similar objectives \cite{perera_context_2014,zeng_qos-aware_2004,kang_rdds_2012}.

\begin{definition}[\textit{Middlewares}]
\citeauthor{neely_adaptive_2006} broadly define middlewares as the layer between the application and system software, abstracting the underlying system to enable developers to focus on application-specific tasks \cite{neely_adaptive_2006}.
\end{definition}

\begin{figure}[t]
    \centering
    \includegraphics[width=0.48\textwidth]{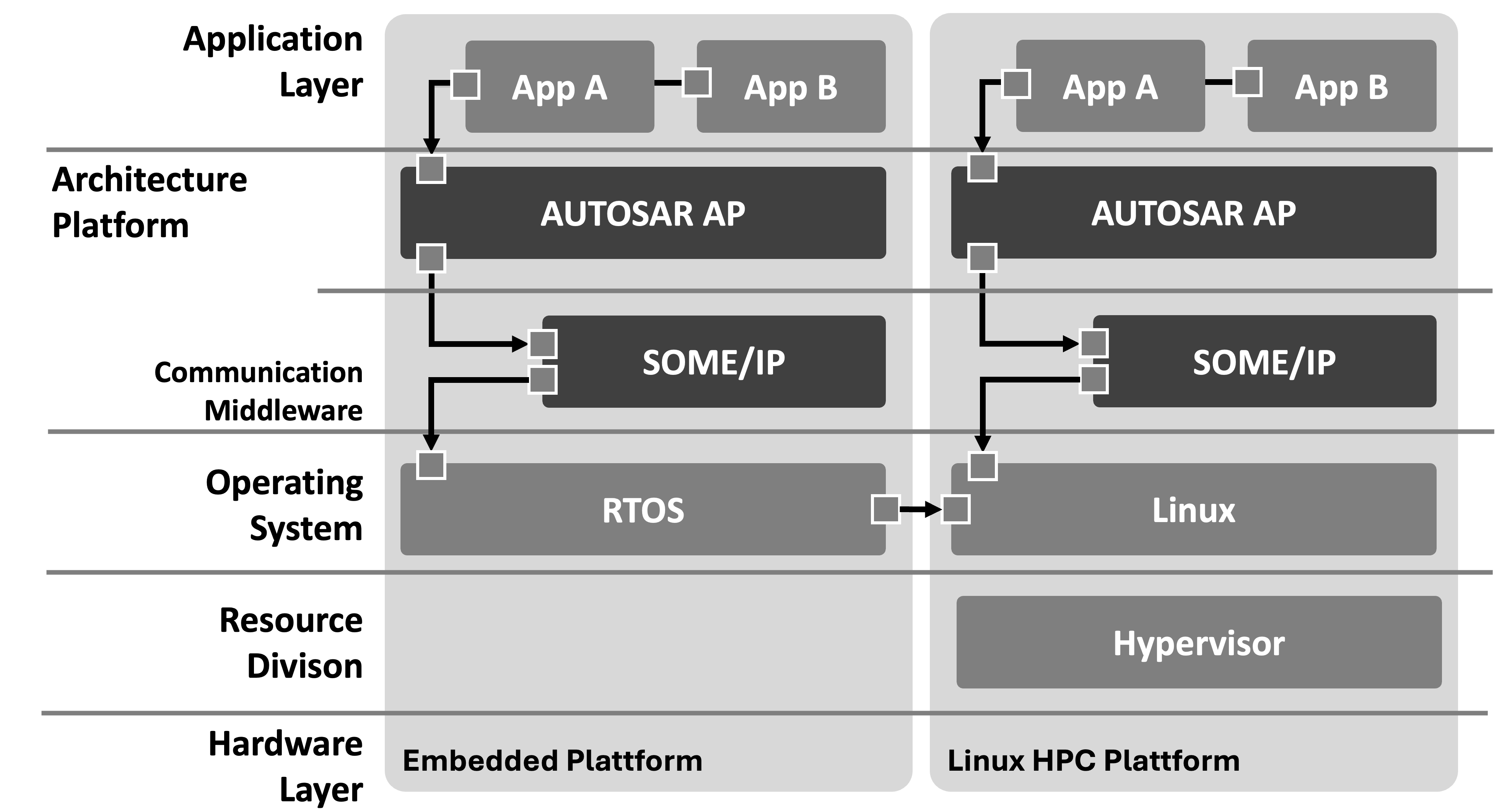}
    \caption{Illustration depicting the role of a middleware in the development of distributed software. The middleware functions as an intermediary between the application and the operating system. In this case AUTOSAR \ac{AP} makes use of DDS and the operating systems network stack to communicate across the two devices.}
    \label{fig:middleware_idea}
\end{figure}

Built on modern \ac{E/E} architectures, a new generation of automotive middlewares has emerged as the foundation of new software architectures in the automotive domain \cite{henle_architecture_2022}. 
In contrast to other domains, the automotive domain is differentiated by distributed E/E architectures, heterogeneous computing platforms, and high safety requirements.
\Cref{fig:middleware_idea} presents the software stack underlying such software architectures using the example of two communicating applications in a distributed system.

\begin{definition}[\textit{Automotive Software Stack}]
We define the automotive software stack as the layers of software used to support applications. Starting with either the \ac{OS} or hypervisor, additional software frameworks and middlewares are added to provide functions to the application.
\end{definition}

The application uses an \ac{API} offered by the middleware layer to send messages from one \ac{ECU} to another application on a second \ac{ECU}.
AUTOSAR \ac{AP} in this case forwards the message to \ac{SOME/IP}, which uses the operating systems network stack to relay the message from the microcontroller to the \ac{HPC}.

This example represents a typical automotive software stack. 
In practice, as shown in \Cref{fig:middleware_idea}, the automotive middleware is separated into two layers. 
The lower layer is a software framework concerned with communication in distributed systems, in this example \ac{SOME/IP}. 
The upper layer, in this case AUTOSAR \ac{AP} is built on the communication framework, to offer features beyond communication, such as \ac{OTA} updates, resource control, and security.

In the literature and in the automotive domain at large, the nomenclature for these two complementary layers is not well established, and terms are used loosely.
While some authors refer to the combination of both layers as middlewares, others describe only the lower layer as a middleware, while referring to the upper layer using new, different nomenclature. 
This inconsistency extends to developers, as the developers of \ac{ROS 2} mostly refer to it as a \textit{set of libraries and tools}, while AUTOSAR \ac{AP} uses \textit{platform} \cite{autosar_ap_explanation_sw_arch, open_robotics_ros_2024}.
In academia, researchers refer to \ac{ROS 2} as a middleware \cite{wu_oops_2021}, while others use different nomenclature, such as architecture platform \cite{henle_architecture_2022}.

In an effort to establish uniform terminology, we propose the term middleware as a general descriptor of the class, in keeping with its original definition \cite{neely_adaptive_2006}. 
We also propose a new term for the lower communication layer, making use of the \textit{\soamid} nomenclature defined by \citeauthor{henle_architecture_2022} \cite{henle_architecture_2022} to refer to the upper layer.
Consequently, we define automotive middleware by the following characteristics:

\begin{figure}[t]
    \centering
    \includegraphics[width=0.48\textwidth]{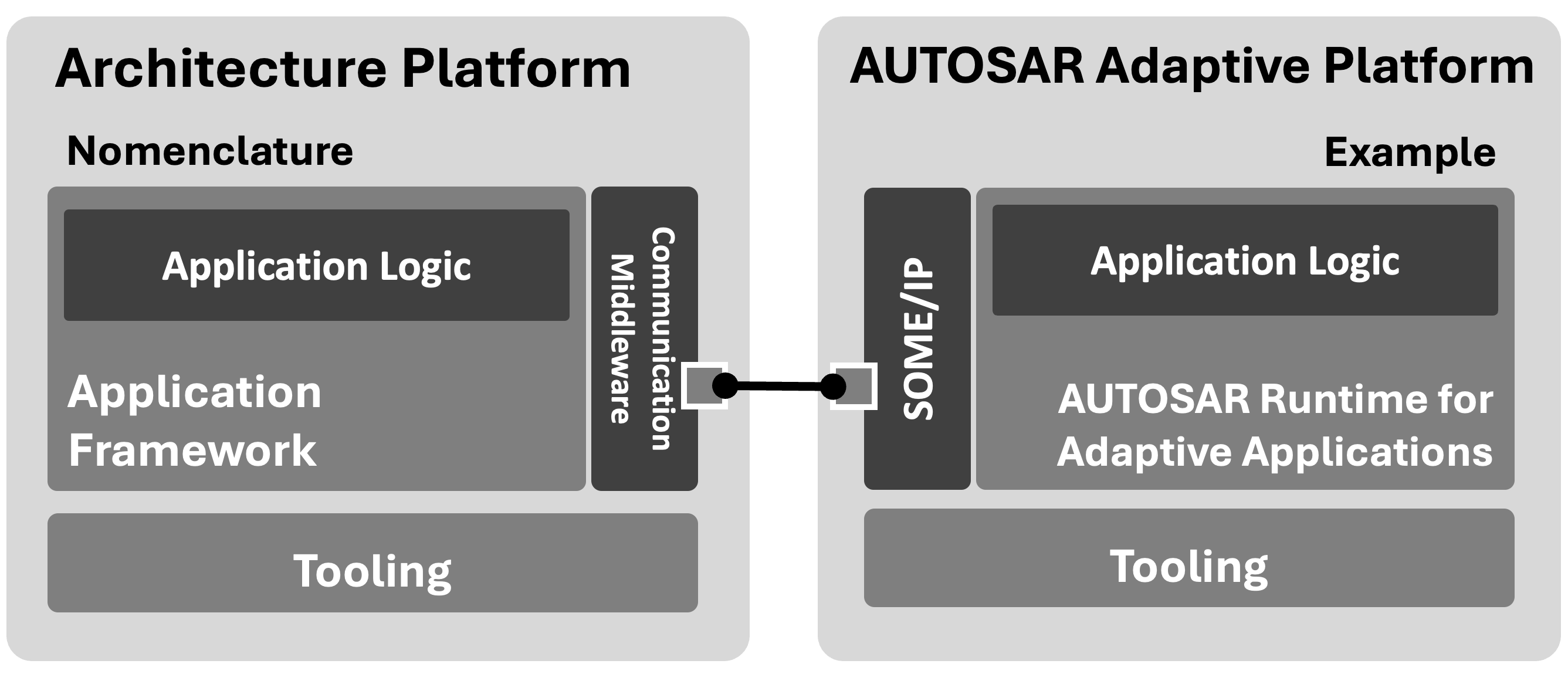}
    \caption{Illustration depicting elements of the \system with our proposed nomenclature. We distinguish between \commids, \soamids\ and application frameworks, to refer to different elements of the automotive software stack. In the left half of the illustration, the nomenclature is shown. In the right half, AUTOSAR \ac{AP} is used as an example. }
    \label{fig:middleware_nome}
\end{figure}

\begin{definition}[\textit{Automotive Middlewares}]
We define automotive middlewares as as the software layer between the application and system software. These middlewares operate in distributed, heterogeneous automotive \ac{E/E} architectures and due to the distributed system architecture, communication and structured software execution are their core functions.
\end{definition}

As a consequence of this position, middlewares interact with a wide variety of domains. 
\cref{fig:topics_mindmap} illustrates these domains and highlights the section in which they are discussed.
Middlewares accomplish communication in distributed system by using new \commpat\ on \acp{IVN} to interconnect applications on different \acp{ECU}. 
Some middlewares are comprehensive platforms that address many software domains in automated vehicles, such as communication, security, updates, and resource control. 

To highlight the communication function of the lower layer framework, we define them as \commids\ and discuss them in detail in \cref{sec:middleware:communication}.
We use the definition of \citeauthor{henle_architecture_2022} to refer to the higher layer, built on \commids, as \soamids \cite{henle_architecture_2022}.
Beyond communication, \shsoamids\ provide additional features such as resource control, deterministic execution, and security functions.
These \shsoamid\ also commonly rely on other \patterns .

\begin{definition}[\textit{\PATTERN}]
Software design patterns are generalized solutions for common design problems.
They are defined by \citeauthor{bass_software_2012} as descriptions of elements, for example, clients and services, and relations, such as requests between these elements, with constraints on how they can be used. They commonly describe only a subset or repeating pattern in a software architecture and classic patterns examples include the client-server pattern, the n-tier pattern, or service-orientation \cite{bass_software_2012}.
\end{definition}

Employing such \patterns, \soamids\ determine high-level architecture decisions, such as the components and communication methods of the software system.
As such, the \soamid\ is an integral part of the \system.
This class of middleware is discussed in \cref{sec:middleware:soa}.

\begin{figure}[bp]
    \centering
    \includegraphics[width=0.48\textwidth]{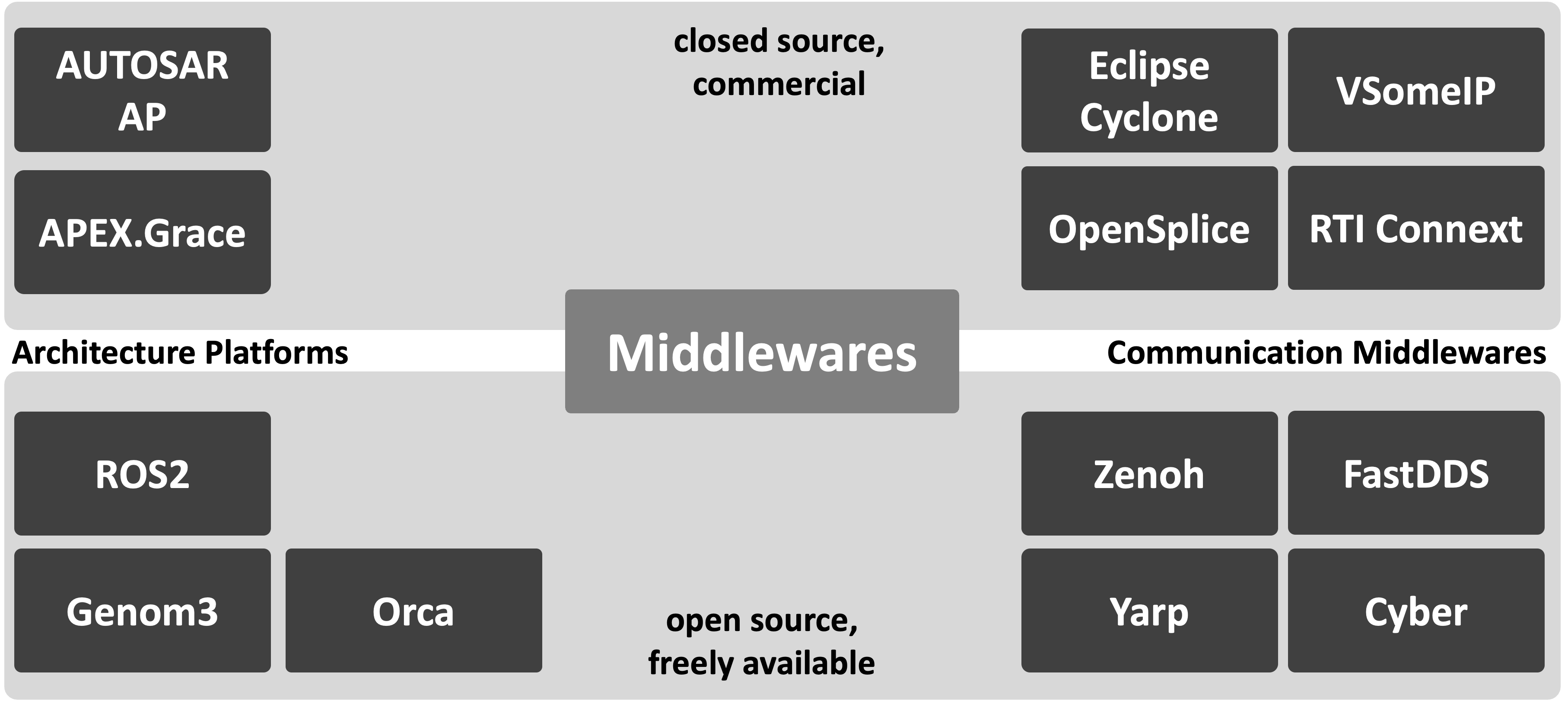}
    \caption{Overview over common \commids \ and \soamids\ in the automotive and CPS domain. The illustration divides middlewares according to the \commid \ and \soamid\ distinction as well as according to their access limitations.}
    \label{fig:middleware_overview}
\end{figure}

\begin{figure*}[tb]
    \centering
    \includegraphics[width=0.98\textwidth]{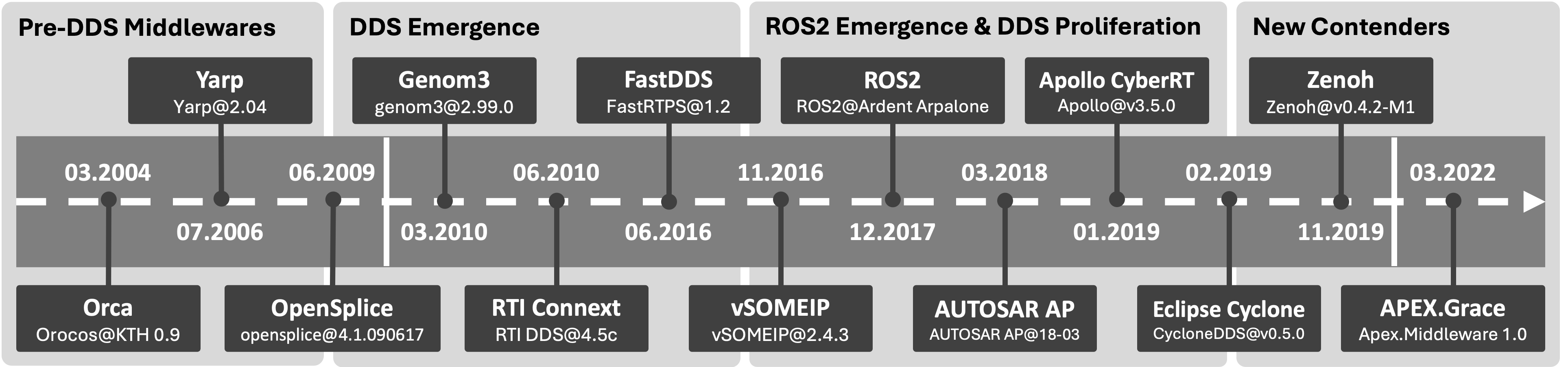}
    \caption{Illustration depicting middleware releases and developments over time. The figure lists the current name and name and version at time of the first publicly available release.}
    
    \label{fig:timeline_mws}
\end{figure*}

In \cref{fig:middleware_overview}, we present an illustration of common middlewares. The figure distinguishes between closed-source and open-source middlewares as well as our classification into \soamids\ and \commids. \cref{fig:timeline_mws} provides a chronological overview and provides context on the recency of the middlewares.

\subsection{What are \COMMIDS?}
\label{sec:middleware:communication}

\noindent \Commids\ are core to automotive software architectures, facilitating data exchange between various \acp{ECU} \cite{autosar_consortium_ip_2023}.
They accomplish this by separating the application from the underlying \ac{IVN} topology. 
This is commonly implemented by discovery, where communication paths are not predetermined but are dynamically discovered at runtime. 
Alternatively, centralized definitions of network topology can be used to separate the \ac{IVN} from the application.
This conceptually simple \pattern\ allows the application built on the middleware to be independent of the specific ECU and \ac{IVN} topology, since all other applications in the network are dynamically discovered.
\begin{definition}[\COMMIDS]
\Commids\ are middlewares designed to implement communication within distributed systems. 
They provide \commpats\ such as publish-subscribe or request-response communication and configurable \ac{QoS} \cite{eprosima_dds_2024}.
\end{definition}

\begin{definition}[\textit{Discovery}]
Discovery in a modern middleware allows an application to automatically discover and connect to all peer applications, commonly in the same network, without manual configuration of connections \cite{eprosima_5_2024}.

\end{definition}
\noindent As a result, no changes to communication parameters are necessary if an application is updated or moved to another \ac{ECU}.
For simplified communication, these middlewares implement \commpats\ that define how communication is structured.
\begin{definition}[\textit{\COMMPAT}]
We define communication patterns, derived from the software pattern definition, as the structure and pattern of exchanged messages between applications, as well as if it is one-sided communication or an exchange of messages, and how recipients of these messages are determined. 
\end{definition}
\noindent Simple \commpats\ allow developers to focus on the core functions of their applications instead of \ac{IVN} configuration.
The two most common \commpats\ in modern middlewares are publish-subscribe and request-response communication. These \commpats\ are present in most middlewares presented in this tutorial.

\begin{figure}[tb]
    \centering
    \includegraphics[width=0.48\textwidth]{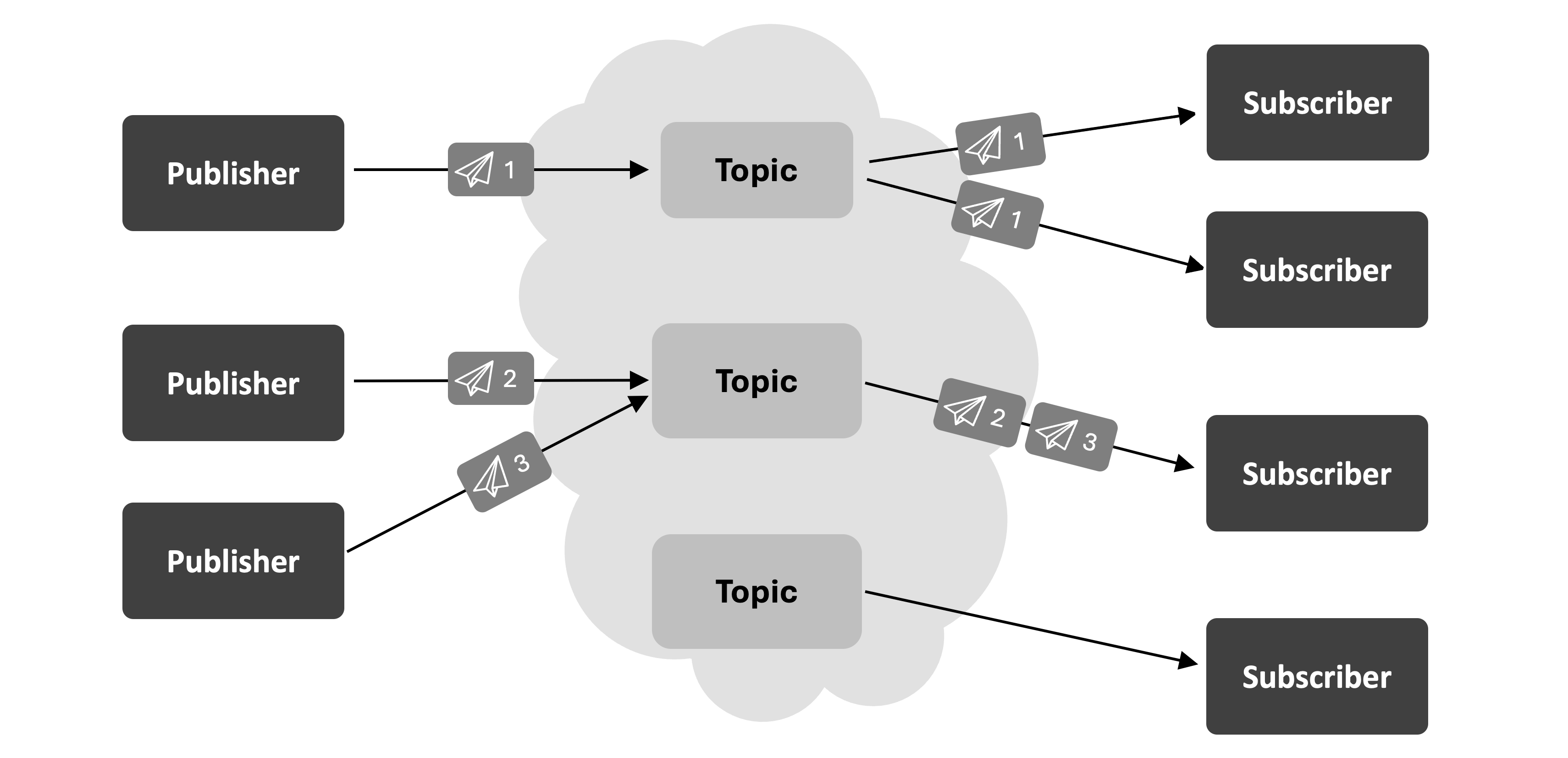}
    \caption{Illustration depicting topic-based communication between publishers and subscribers components in a DDS domain. }
    
    \label{fig:example_dds}
\end{figure}

In \textit{Publish-Subscribe} communication, data is separated into messages, which are structured using topics \cite{eprosima_dds_2024}. 
Topics serve as separate channels of communication that allow access to all messages on them. 
\Cref{fig:example_dds} depicts an example of a topic-based communication.
A participant can publish messages on a topic, and a participant that wants to receive the message can subscribe to it to be notified.
As a result, messages are not directly associated with a recipient, but only with their topic. 
Thus, allowing new participants to easily join existing networks, as they can just interact with already existing topics.
In \textit{Request-Response} communication, a participant can send a request to another remote participant, which will return a response to the request \cite{autosar_consortium_ip_2023}.
This \commpat\ is commonly used to call remote methods on other machines, similar to regular function calls.

\Commids\ may also address reliability and security considerations in \acp{IVN}. 
To facilitate functional safety in communication, some middlewares offer configurable \ac{QoS} which can ensure the reliability of message transmission. 
For security considerations, the middlewares may directly support encryption of exchanged messages, while others may offer optional extensions to supplement security features.

The implementation of \commids\ is commonly built on the IP network stack in the host operating system to access the ethernet-based \ac{IVN}.
\ac{TCP}, \ac{UDP}, and shared memory are commonly used as transport layers. 
Typically, middlewares automatically determine which transport to use based on the \ac{IVN} and \ac{E/E} architecture.

The following sections present three \commids. 
FastDDS, an open source implementation of the \ac{DDS} standard, \ac{SOME/IP} the automotive industries approach, and Zenoh an emerging middleware.

\subsubsection{SOME/IP}
\label{sec:middleware:someip}
A standard that defines a \commid\ is the \acfp{SOME/IP}. 
It was defined as an automotive \commid\ standard for the AUTOSAR project and maintained in connection with the project. 
It is being developed by the AUTOSAR consortium consisting of major \acp{OEM} and automotive tier one suppliers. 
It represents an approach to middlewares by the automotive industry.
It shares some concepts with other middlewares, such as \ac{DDS} discussed in \cref{sec:middleware:fastdds}. 
These similarities include \commpats\ and methods, such as the underlying protocols and the use of the \ac{OS} network stack \cite{autosar_consortium_ip_2023}. 

In its core design principles, \ac{SOME/IP} was specifically developed to work with stringent resource budgets of embedded devices and aims to ensure compatibility across a wide range of use cases and communication partners.
Due to its origin in the AUTOSAR project, it is specifically and narrowly designed for the automotive domain.
Additional design objectives due to the origin of the middleware are compatibility with its predecessor AUTOSAR classic and the ability to scale to full \system\ with hundreds of applications.

To transport \ac{SOME/IP} messages, different transport protocols can be used, currently supporting \ac{UDP} and \ac{TCP}. 
\ac{SOME/IP} can be configured to use both. However, developers recommend configurations depending on the network characteristics.

\ac{SOME/IP} supports both publish-subscribe and request-response communication but uses proprietary names for these \commpats, detailed in \cref{sec:middleware:communication}. In addition to these common \commpats, it also supports \patterns\ to distribute values globally throughout a \system.
These values, called fields, are accessible by every application in a \ac{SOME/IP} software system. These \commpats\ are implemented in three features:

\underline{Methods} are a \textit{request-response} communication paradigm and allow remote methods to be called. 
A request sent to a local or remote application is processed and an optional response is returned to the calling service.
This \textit{request-response} communication also supports \textit{fire-and-forget}, where only the request is transmitted without a response.
The standard specifies this with an initial request message and a corresponding response message.

\underline{Events} are transmitted by notification to inform other applications of an event in the \system. 
Notifications follow the \textit{publish-subscribe} concept, where applications can decide to subscribe to the notification.
For example, these events could be used to notify other components of a brake event when an automated system detects a possible collision.
Notifications can be sent according to different strategies, such as cyclical, on-change, and epsilon-change, when the value considered changes beyond an epsilon threshold.

\underline{Fields} are shared values in an automotive system based on \ac{SOME/IP}.
They are accessible via the \textit{getter and setter functions} and use event notification to distribute values throughout the software system. 
They allow applications to share values and receive updates on changes to this value.
The \textit{getter and setter} functionality builds on the methods functions, and updates are sent via events.

An exhaustive overview of the capabilities of \ac{SOME/IP} can be found in \cite{autosar_consortium_ip_2023}.
\ac{SOME/IP} also supports with an extension the automatic discovery of applications in a network \cite{autosar_consortium_ip_2023}. 
The SOME/IP-SD (SOME/IP-Service Discovery) extension allows the application to locate service instances inside the \ac{IVN} on other \acp{ECU}. 
It also enables \ac{SOME/IP} to determine whether a service instance is running and subsequently implement the publish-subscribe handling.

For its implementation, \ac{SOME/IP} uses the network stack of the underlying POSIX operating system.
It supports \ac{UDP} and \ac{TCP} communication and the standard documentation recommends the use of both under specific conditions.
The service discovery functionality is implemented using \ac{UDP} multi-cast messages.

\ac{SOME/IP} does not directly provide communication security features. In the AUTOSAR \ac{AP} stack, security is provided by the \ac{CM} with the addition of another layer of security protocols layered on \ac{SOME/IP} \cite{autosar_communicationManagement_2024}.

To discuss the performance of \ac{SOME/IP}, we focus on vSOME/IP as the representative implementation \ac{SOME/IP}.
Literature providing comparative evaluations against other middleware protocols is relatively scarce. 
Existing publications on the \commid\ predominantly propose improvements to specific components of vSOME/IP, while only few studies assess its performance in relation to other middlewares. 
One of the few comparative studies that provides results across middlewares is our own work \cite{kluner_automotive_2025}.
In this study, vSOME/IP showed substantially higher transmission latency, slower discovery, and lower throughput than both Zenoh and FastDDS. 
Resource usage was in line with FastDDS and better than Zenohs.
These findings are influenced by vSOME/IPs internal routing manager and its design focus on smaller message sizes than those used in our evaluation \cite{kluner_automotive_2025}.

\subsubsection{FastDDS}
\label{sec:middleware:fastdds}
An open-source \commid\ standard in competition with \ac{SOME/IP} is the \ac{DDS} standard defined by the \ac{OMG}.
FastDDS is an open-source implementation of this standard. 
The \ac{DDS} standard defines \commpats\ in distributed systems and a wire protocol \cite{eprosima_dds_2024}.
Due to the open-source nature, large community, and inclusion in the \ac{ROS 2} project, it represents a good choice for communication middleware.

Communication in \ac{DDS} is data centric. 
Data in \ac{DDS} is structured using the publish-subscribe \commpat. 
It is associated with a topic and is accessible to all participants who are subscribed to the topic. 
This type of communication can also be classified as anonymous communication because the sender is not required to determine a specific receiver \cite{eprosima_11_2024}.

The core component for communication on the DDS layer is the domain participant which an application can create to begin FastDDS communication.
To send data to a topic, a participant can create a publisher for this topic \cite{eprosima_3_2024}. 
To receive this data, a participant that requires this data can create a subscriber to be informed when new data is available.
This mechanism continues to work even for newly joined FastDDS applications, which can subscribe and/or publish to existing topics and seamlessly interact with the existing network at runtime.

Integration of new participants is enabled by dynamic service discovery, where FastDDS applications advertise their presence using multicast \ac{UDP} messages to other applications on the same network \cite{eprosima_5_2024}.
Discovery takes place in two phases.
In the first phase, other applications in the same network using FastDDS are discovered. 
Once these are known, information about endpoints such as subscribers and publishers is exchanged and communication channels are established. 
FastDDS also offers other mechanisms, such as manual, static, or predefined discovery, to reduce discovery times and to have more control over the discovery process.
This allows for the configuration of a fixed communication topology, eliminating the need for the discovery process, or a dynamic discovery process, where new participants are connected at runtime into the communication at the cost of overhead discovery traffic.
DDS specifies the RTPS protocol on top of \ac{TCP} or \ac{UDP} for communication, but additionally FastDDS also supports shared memory \cite{eprosima_6_2024}. 
It seamlessly switches between these depending on the network topology, similar to \ac{SOME/IP}.

To address mixed-criticality traffic with different requirements, FastDDS offers control over \ac{QoS} to configure transmission reliability, message history storage behavior, among other settings for each topic, endpoint, and participants in the network \cite{eprosima_12_2024}.
Using these \ac{QoS} settings, the application can configure whether a history of past messages sent should be kept, whether the communication must be reliable, or if only the best transmission effort should be made. 
These configurations present a trade off, as reliability may incur overhead while best-effort may result in messages lost in case of contention.
FastDDS offers extensive \ac{QoS} parameters, defined in \cite{eprosima_12_2024}.

To address security challenges in \acp{IVN}, the \ac{DDS} standard offers multiple extensions. 
From these extensions, FastDDS implements tools for the authentication of participants when joining a \ac{DDS} network \cite{eprosima_8_2024}.
In addition, it implements access control methods to prevent participants from performing protected operations, cryptographic extensions that enable the encryption of traffic between endpoints, and additional security features. 

In terms of FastDDS performance, a substantial body of research has examined the performance of \ac{DDS} implementations, particularly the FastDDS stack that serves as the default \ac{ROS 2} \ac{RMW}. 
In a recent comparative study, \citeauthor{kluner_automotive_2025} \cite{kluner_automotive_2025} compared FastDDS against Zenoh and vSOME/IP, finding that FastDDS delivers comparable throughput in wired environments while exhibiting slightly reduced reliability at elevated transmission rates.  
Related work by \citeauthor{chisalita_stepping_2025} \cite{chisalita_stepping_2025} compared Zenoh with FastDDS and reported lower latency and jitter for Zenoh, suggesting a more favorable real‑time performance profile.  
A broader throughput investigation by \citeauthor{liang_performance_2023} \cite{liang_performance_2023} confirmed that Zenoh achieves higher bandwidth on wired links, while \citeauthor{zhang_comparison_2023} \cite{zhang_comparison_2023} observed analogous wired results but highlighted a improved wireless performance for Zenoh. 
In conclusion, these studies indicate that FastDDS and Zenoh provide comparable performance in wired networks, with Zenoh offering a distinct advantage in wireless deployments and lower latency characteristics.

\subsubsection{Zenoh}
\label{sec:middleware:Zenoh}

\begin{figure}[tbp]
    \centering
    \includegraphics[width=0.48\textwidth]{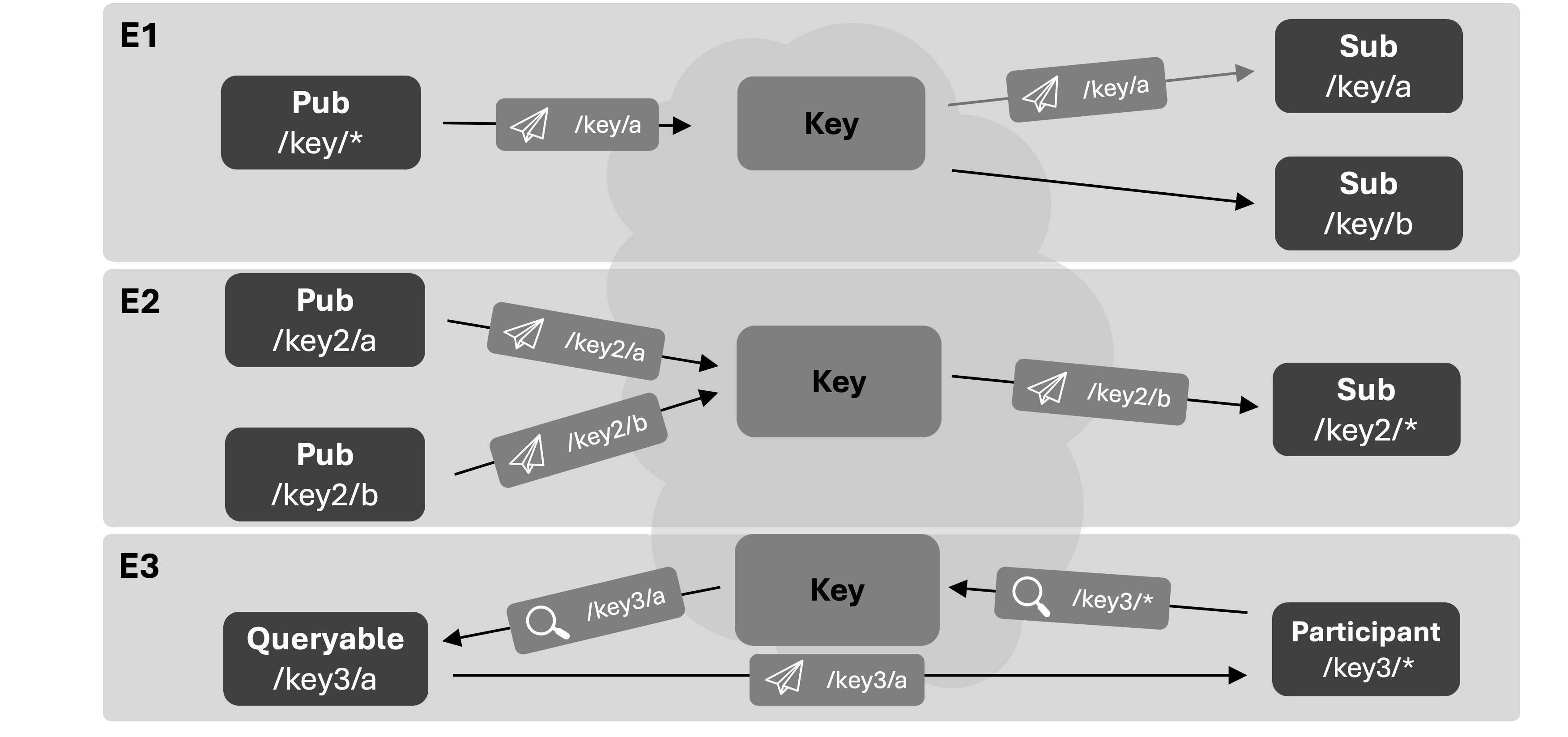}
    \caption{Non-comprehensive illustration of Zenohs resource-structured \commpats. Publishers and subscribers communicating using keys and selectors and Queryables operating as sources of Zenohs resources.}
    \label{fig:zenoh_paradigm}
\end{figure}

Zenoh is a recent addition to the domain of \commids. 
According to the developers \citeauthor{corsaro_zenoh_2023} \cite{corsaro_zenoh_2023}, it was designed with the lessons learned from more established middlewares, such as \ac{DDS} and \ac{MQTT}.
It aims to improve performance by decreasing protocol transmission overhead and reducing latency in contrast to existing protocols.
In addition, Zenoh seeks to address a multitude of challenges, such as scale, network topology, resource constrained devices, and data in motion and at rest.

In Zenoh, data is structured in \textit{resources} $(\texttt{key},\texttt{value})$ and associated with keys \cite{corsaro_zenoh_2023}.
Keys are used to identify data, such as \textit{/key/a} and accessed using either the key or a selector \cite{corsaro_zenoh_2023}. 
A selector allows for the specification of matching operators that enable access to multiple keys at once.
For instance, to select multiple sub resources at once, \textit{/key/*} could be used. 
\cref{fig:zenoh_paradigm} illustrates the communication paradigms used by Zenoh, such as resource-structured publish-subscribe and queryables.

Supporting this data model, the core components of Zenoh are publishers, subscribers, and queryables.
Publishers can be understood as origins of Zenoh resources, creating resources for a single key or for a key expression. 
This is shown in \cref{fig:zenoh_paradigm} in example one.
Similarly, subscribers can be understood as the sink of resources, again delivering all resources that match either a single key or an expression. This is shown in \cref{fig:zenoh_paradigm} in example two by a subscriber for \textit{/key2/*}, as this subscriber receives both messages from \textit{/key2/a} and \textit{/key2/b} publishers.
Queryables deviate from the known publish-subscribe \commpat\ by delivering a resource if a key to which their expression matches is queried \cite{corsaro_zenoh_2023}. \cref{fig:zenoh_paradigm} again illustrates this in example three, where a participant queries for \textit{/key3/*}, to which the queryable that generates \textit{/key3/a} resources responds with a \textit{/key3/a} resource.

The components in Zenoh are built from a limited set of primitives. 
These allow for the declaration of the previously discussed components, such as subscribers, publishers, resources, and queryables.
However, they can also be used directly as atomic primitives. 
These primitives are \textit{put}, \textit{delete}, and \textit{get}.
The \textit{put} primitive allows for the creation of new resources, while the \textit{delete} primitive allows for destruction. 
The query operation \textit{get} issues queries to the Zenoh system. Parameters allow for the specification of the matching policy.

The Zenoh protocol can be used on the data link, network, or transport layer, enabling it to operate with embedded systems in resource-constrained environments.
To secure communication, Zenoh offers support for authentication and secure channel plugins and can operate on the TLS security layer.

Zenoh’s performance has been evaluated across automotive, robotics, and \ac{IoT} domains, reflecting its use in heterogeneous and resource-constrained environments. 
In automotive systems, studies by \citeauthor{kluner_automotive_2025} \cite{kluner_automotive_2025} and \citeauthor{chisalita_stepping_2025} \cite{chisalita_stepping_2025} examined wired in-vehicle communication and inter-vehicle communication, showing that Zenoh achieves latency and bandwidth comparable to \ac{DDS} while exhibiting more robust behavior at higher data rates with reduced message loss. 
In robotics and \ac{IoT} scenarios \cite{chovet_performance_2025, baron_performance_2025, zhang_comparison_2023}, evaluations focused on congestion control, reliability, and wireless performance. 
\citeauthor{baron_performance_2025} \cite{baron_performance_2025} found that Zenoh can outperform MQTT in both routed and brokered deployments, delivering lower latency under specific connectivity conditions. 
\citeauthor{chovet_performance_2025} \cite{chovet_performance_2025} demonstrated in real-world non-line-of-sight mesh networks that Zenoh as a \ac{ROS 2} \ac{RMW} provides faster communication, greater message reachability, and lower CPU usage, however, with increased RAM use compared to FastDDS and CycloneDDS. 

Overall, existing studies indicate that Zenoh at least matches DDS performance in wired environments while offering advantages in robustness, and outperforms traditional protocols such as MQTT and DDS in wireless or lossy networks.

\subsubsection{\COMMID\ example}
In a \system\ built on a \commid\ and a modern \ac{E/E} architecture, a single \ac{ECU} can support multiple applications and is likely to have additional compute capacity. 
The middleware allows applications to communicate transparently between different \acp{ECU} and new applications can dynamically extend the \system.

\begin{figure}[tb]
    \centering
    \includegraphics[width=0.47\textwidth]{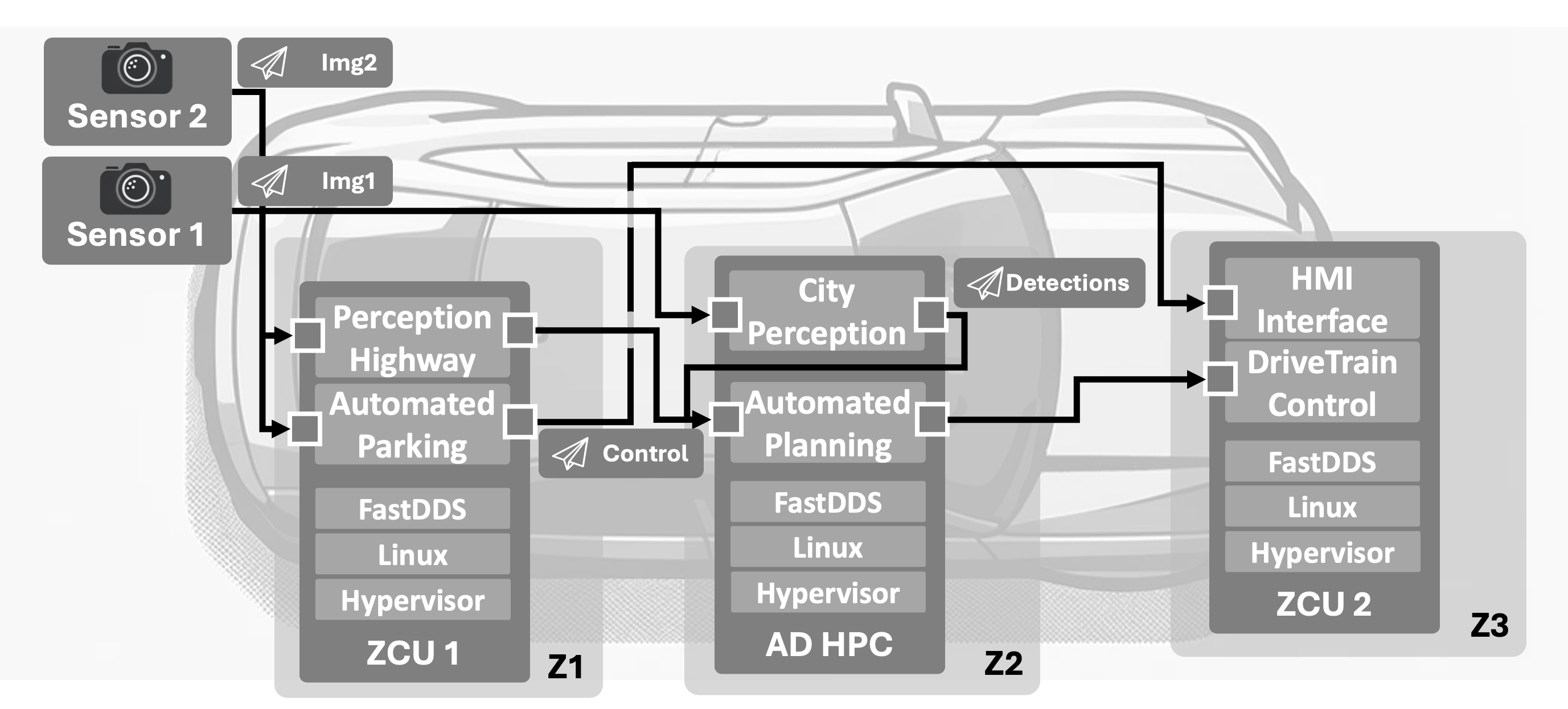}
    \caption{Example software system architecture using a \commid. Multiple applications share \acp{ECU} and communicate using DDS. In DDS, topics serve as the communication keys. In signal-oriented architectures fieldbus systems structure communication, illustrated in \cref{sec:background:signal_oriented_example}.}
    \label{fig:example_commid}
\end{figure}

To implement the expansion of the operational domain to the city, presented in \cref{sec:background:signal_oriented_example}, applications in an architecture based on a \commid\ could be updated.
If sufficient compute resources are available, the updated applications could be deployed to the existing \ac{ECU}.
The deployment of such an application is illustrated in \cref{fig:example_commid}, in contrast \cref{sec:background:signal_oriented_example} illustrates the system in a domain-based architecture.
No changes to the communication infrastructure or other applications are required, as the new application can integrate into the \ac{DDS} domain. 
To accomplish this, the application can subscribe to existing topics and access the data required for its function.
The new application can then easily collect images and sensor data from all sensors in the vehicle, including the near-field sensors, originally included in the vehicle for automated parking.
In addition, it can also send new trajectory commands using the existing \textit{Control} topic to send commands to the drivetrain \ac{ECU}.
This can be accomplished by publishing new messages using \ac{DDS} and can occur without changes to the drivetrain \ac{ECU}.

The core advantage of this approach is that no new dedicated \ac{ECU} is required to add new functions.  
However, \commids\ do not provide automated tools for this process and offer few other assurances, as their focus is on communication.
No changes are required to the remaining applications as a result of the extension. 
However, this approach presents the risk of computing resource contention on the automation module and possible network contention. 
Additionally, \acp{ECU} or vehicle computers running a full operating system and communicating over ethernet introduce new cyber security attack vectors over a signal-oriented architecture.

\subsection{What are \soamids?}
\label{sec:middleware:soa}

\begin{definition}[\Soamids]
\Soamids\ are comprehensive frameworks for the development of automotive software systems.
Beyond communication, they enforce \patterns, offer comprehensive tooling for development and deployment, and often provide execution models, security features, and system state management concepts. 
\end{definition}

\noindent While communication is the core challenge of any middleware, automotive software systems present additional challenges.
Among these challenges are wide-ranging questions such as real-time requirements and deterministic compute in distributed systems, requiring mechanisms to ensure communication and compute deadlines are met, and computations are unaffected by varying propagation delays through the software system.
In the same vein, in a distributed software system, allocation of resources becomes a non-trivial task, especially in a changing software system supporting \ac{OTA} updates.
\Commids\ commonly offer no solutions to these questions, while \soamids\ seek to address them.

\Soamids\ are broad software frameworks that address multiple challenges from different domains, such as software updates, maintenance, resource allocation, and determinism.  
To provide communication, they generally make use of a \commid\ discussed in \cref{sec:middleware:communication}.
This relationship was already illustrated in \cref{fig:middleware_idea}, where AUTOSAR \ac{AP} uses \ac{SOME/IP} for its communication.
Due to this structure, inheriting the advantages these \commids\ offered to applications and building on them.
In addition \soamids\, commonly use \patterns.
An example of these \patterns\ is service-orientation, which proposes a division of software into individual components according to their functionality \cite{kampmann_dynamic_2019}.
This offers advantages for reuse, development, management of responsibilities, and updateability of the software system.

Examples of these \soamids\ include \ac{ROS 2} and AUTOSAR \ac{AP}, which will be discussed in the following sections \cref{sec:middleware:ap,sec:middleware:ROS2}.

\subsubsection{ROS 2}
\label{sec:middleware:ROS2}
The dominant \soamid\ in the robotics domain is \ac{ROS 2} \cite{malavolta_how_2020}. 
We classify it as a \soamid, as it is a feature-rich framework for robotic software development \cite{macenski_robot_2022}.
\ac{ROS 2} is a publicly available open-source project. 
Researchers and organizations have suggested adopting \ac{ROS 2} in the automotive sector, given the success of \ac{ROS 2} in the robotic domain and similarities to the automotive domain \cite{henle_architecture_2022}. 
A private company seeking to bring \ac{ROS 2} to the automotive domain is Apex.AI.
The company provides a proprietary spin-off variant of \ac{ROS 2}, known as Apex.OS, which is tailored for the automotive industry.
Apex.OS is certified to ISO 26262, ASIL D.  
Consequently, this framework is suitable for use in safety-critical automotive software applications \cite{becker_safety-certified_2021,ekberg_autosar_2022}.

The software architecture in a \ac{ROS 2}-based software system follows \ac{SOA} \patterns, as the software is divided into distributed nodes in a compute graph \cite{open_robotics_nodes_2024}.
These nodes perform functions similar to services and are in general independent of their underlying hardware.

\ac{ROS 2} employs a DDS-based \commid\ for its communication. 
This \commid\ provides features as discussed in \cref{sec:middleware:fastdds}, such as discovery and publish-subscribe \commpat.
However, \ac{ROS 2} also builds on \ac{DDS} to implement additional functions and offers the option to exchange \ac{DDS} implementations \cite{open_robotics_interfaces_2024}.
For basic communication between nodes, two \commpats\ are available in the framework. 
\begin{itemize}
    \item Publish-Subscribe communication is used by \ac{ROS 2} for connections between nodes. 
    Consequently, the DDS topic \pattern\ is retained, but \ac{ROS 2} provides its own serialization solution. 
    With this solution, \ac{ROS 2} enables communication between multiple programming languages and allows for cooperation between nodes using different languages.
    Additionally, \ac{ROS 2} provides an execution model that controls communication callbacks to user code.
    In this model, it imposes an execution order on these callbacks and timer activation's using the executor execution model.
    \item Request-Response is built on \ac{DDS} publish-subscribe communication.
    This \pattern\ is implemented in services and actions. 
    Services expose a remote interface possibly on another \ac{ECU}, which can be called and returns a response upon completion.
    Actions behave in a similar way. However, in addition to responding, they provide periodic feedback to the calling node on the progress of the request.
\end{itemize} 

In \ac{ROS 2}, the execution model takes the form of the executor model with a callback-driven interface \cite{macenski_robot_2022,open_robotics_executors_2024}.
User-specified timers, events, and message handlers can be enrolled in the executor instance. The executor itself then processes the callbacks of the enrolled function, such as the callbacks of the message handler, in a partially predictable manner \cite{macenski_robot_2022}. 
Execution models, such as executors, enable assumptions about the behavior of a node and, by conclusion, the \system\ at large.

However, the standard execution model of \ac{ROS 2} does not support a strict execution order or deterministic execution times \cite{henle_architecture_2022}. 
When events or messages must be queued for processing, which could be the case when the frequency and processing time of callbacks exceed the capacity of a single thread, the executor handles them using a round-robin approach. This can have adverse consequences, such as priority inversion and task starvation. 
Consequently, critical task timings may be missed, making it difficult to reliably estimate worst-case execution times for task execution \cite{open_robotics_executors_2024,belsare_micro-ros_2023}.

\begin{figure}[t!]
    \centering
    \includegraphics[width=0.48\textwidth]{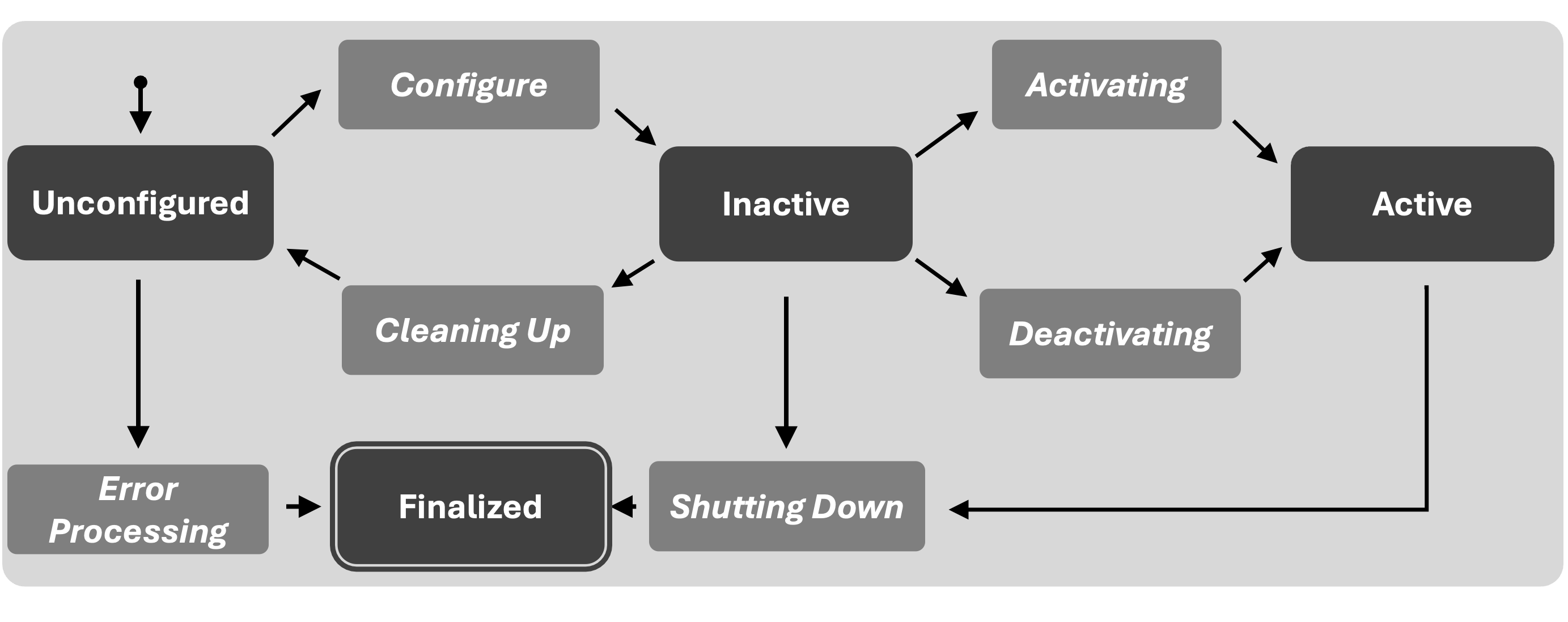}
    \caption{Overview of the ROS 2 lifecycle state machine. Transitory states are shown in grey and primary states in black\cite{macenski_impact_2023}.}
    \label{fig:ROS2Lifecycle}
\end{figure}

An improved version of the \ac{ROS 2} execution model is implemented in the \textit{micro-ROS} framework. Designed for embedded systems, it offers an improved executor compared to the default in \ac{ROS 2} \cite{belsare_micro-ros_2023,macenski_impact_2023}. 
This executor, called \textit{rclc}, enables a fixed execution order and deterministic scheduling \cite{staschulat_rclc_2020}. In addition, it can directly manage the operating system scheduler, enabling custom task prioritization \cite{belsare_micro-ros_2023}.

\ac{ROS 2} also supports configuration management.
It provides a parameter \ac{API} that allows the user to set default parameters and change the parameterization of nodes at runtime \cite{macenski_robot_2022,open_robotics_parameters_nodate}.
To start a \ac{ROS 2} software system of multiple nodes, it provides launch files to define the required nodes and start them automatically \cite{open_robotics_launch_2024}. 

To start a \ac{ROS 2} software system and transition it to an active state, multiple steps may be required, such as initialization of software components or hardware. 
To formalize this initialization process and the life cycle of nodes, \ac{ROS 2} supports a lifecycle model. 
This model is implemented as a state machine in every \textit{lifecycle} node. 
\cref{fig:ROS2Lifecycle} illustrates the transitions between \textit{primary} states. 
These state transitions allow for the representation of tasks at the start and termination of nodes. 
These may include the start sequence for a sensor system or initialization of external software components.
This mechanism can also be used to free system resources if a node is temporarily inactivated or shutdown.
The transitions of the state machine within the node can be controlled both externally by a central controller or internally by the node itself \cite{macenski_robot_2022,macenski_impact_2023}. 

\ac{ROS 2} supports secure communications between nodes using the DDS security standard \cite{open_robotics_ros_2024}.
The standard supports extensions for access control to limit the operations that each entity can perform. Authentication ensures that only permitted entities can join the ROS 2 network. The cryptographic service in the standard allows message encryption to prevent unauthorized access.
The security services in ROS 2 are, consequently, provided by the underlying \commid. 
To enable security in \ac{ROS 2}, each participant in the \ac{ROS 2} software system needs configuration files, which require additional software for setup. However, to simplify the configuration process, \ac{ROS 2} offers the \textit{sros2} package, which includes tools that facilitate the setup of the underlying DDS-Security layer \cite{vilches_sros2_2022,open_robotics_setting_nodate}.

\ac{ROS 2} provides as an expansive open-source project additional tools, such as logging with a common interface. 
To visualize data in an ROS 2 network, the RQT visualization supports many ROS 2 standard messages and can visualize them graphically. 
ROS 2 rosbag supports recording data in a \ac{ROS 2} network and replaying the data to easily replicate and test the nodes. 

\ac{ROS 2} performance has been extensively studied across versions, middleware implementations, and system configurations. 
Early work by \citeauthor{maruyama_exploring_2016} \cite{maruyama_exploring_2016} compared ROS 1 and ROS 2, while \citeauthor{teper_end--end_2022} \cite{teper_end--end_2022} analyzed end-to-end timing chains and showed that performance varies strongly with system utilization. 
\citeauthor{wu_oops_2021} \cite{wu_oops_2021} evaluated multiple \acp{RMW} and identified data copying and serialization as the main communication overheads, with QoS settings having limited impact on local communication. 
\citeauthor{kronauer_latency_2021} \cite{kronauer_latency_2021} further characterized delays across the ROS 2 stack, finding significant latency increases once payloads exceed the UDP fragmentation threshold and showing that DDS processing and ROS 2 message retrieval contribute the largest latency share.
\citeauthor{aartsen_analyzing_2022} \cite{aartsen_analyzing_2022} confirmed that publish–subscribe without security works across RMW combinations and that performance differences remain small and consistent with previous studies \cite{aartsen_analyzing_2022}. 
In contrast, direct comparison to AUTOSAR Adaptive remains difficult due to the limited availability of performance data on the AUTOSAR side.

\subsubsection{AUTOSAR Adaptive Platform}
\label{sec:middleware:ap}
AUTOSAR \ac{AP} is a standard proposed by the automotive industry for an automotive \soamid\ \cite{reichart_progress_2021}.
This standard originated in industry to realize the shift from signal-oriented architectures, standardized in AUTOSAR Classic, to middleware-based architectures. 
It is developed by a large consortium of \acp{OEM} and automotive suppliers specifically for the automotive domain.
Due to its commercial origin, implementations of the standard are commercially available by a number of companies.
AUTOSAR \ac{AP} form a comprehensive standard, providing solutions to many challenges in modern centralized or domain controller architectures \cite{furst_autosar_2016}.
 
\begin{figure}[tbp]
    \centering
    \includegraphics[width=0.5\textwidth]{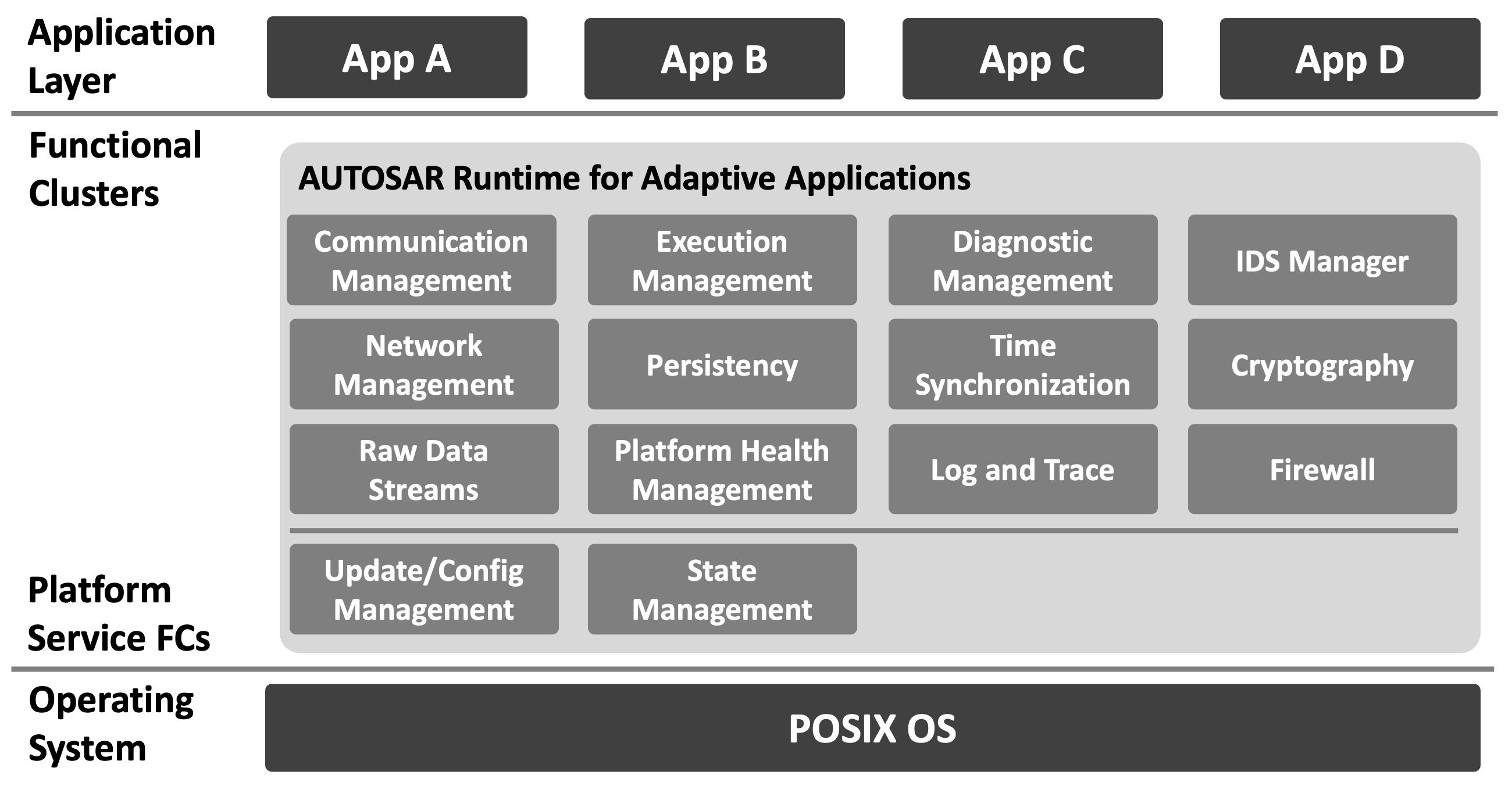}
    \caption{Overview of the AUTOSAR Adaptive Platform Functional Clusters, derived from \cite{autosar_consortium_explanation_2023}.}
    \label{fig:adaptivePlatform}
\end{figure}
    
The explanation of the standards outlines several core \patterns\ \cite{autosar_consortium_explanation_2023}, such as planned dynamics, service-oriented architecture, and parallel processing, among others.
Under the idea of planned dynamics, dynamic operations should be pre-planned in the vehicle such as to ensure their safety and move the computation from the moment of change to before.
In the \pattern\ of service-oriented software architectures, software is divided into separate services to harness simplified updateability and understanding of the software system. 

\acp{AA} are built on AUTOSAR \ac{AP} and use the functions provided by it.
To access functions provided by \ac{AP} \acp{FC}, applications use the \ac{ARA} API \cite{autosar_explanationPlatformDesign_2024}.
In AUTOSAR \ac{AP} several clusters exist that serve purposes such as communication, execution, and state management, with an exhaustive list enumerated in \cref{fig:adaptivePlatform}.
Adaptive applications and functional clusters are implemented on a POSIX \ac{OS} and provide their functions through requests or directly using a library interface.
Both of these components run on the underlying machine as processes and generally communicate via \ac{IPC}. 
This communication method uses operating system functions, such as pipes or shared memory, to efficiently communicate between processes.

As a core functionality, the communication management functional cluster exposes several methods for communication between services \cite{autosar_communicationManagement_2024}.
Based on the \ac{SOME/IP} \commid, it enables inter- and intra-machine communication.
It supports the same \commpats\ in fields, methods, events, and static or dynamic application discovery as \ac{SOME/IP}.
In addition, it also ensures the type and allowable range of each field in a message. 
To support communication with AUTOSAR Classic \acp{ECU}, \ac{S2S} components can also be used.
These enable transparent mapping between \ac{SOME/IP} ethernet-based communication and fieldbus systems.

To manage all adaptive applications and functional clusters in a vehicle and ensure their correct startup procedure, the state management and execution management functional clusters have been developed \cite{autosar_executionManagement_2024}.
The \ac{EM} cluster controls execution in adaptive applications, starting and pausing applications as \ac{OS} processes, while the \ac{SM} decides which function group to start.
In AUTOSAR \ac{AP} the state management is modeled as a state machine, which controls the initial start and the configuration of the function group at run-time. 
As in \ac{ROS 2}, the lifecycle can be modeled by transitioning the state machine between multiple discrete states, such as \textit{Startup}, \textit{Shutdown}, or \textit{Restart} \cite{henle_architecture_2022,autosar_executionManagement_2024}.
Additional responsibilities of the state management include the enforcement of resource constraints and ensuring that only trusted applications can be executed.

AUTOSAR \ac{AP} also supports methods for deterministic execution in the form of data and time determinism \cite{autosar_communicationManagement_2024}.
In the case of execution determinism, the same internal state and input data deterministically produce a predictable result, whereas time determinism ensures that the computation completes before the deadline.
To this end, the execution management supports event-based and cyclically triggered execution using the deterministic client interface. 
To allow for parallelism, the execution management also supports worker pools that a main thread can defer execution to.
This approach enables deterministic execution, which is currently not achievable in \ac{ROS 2} \cite{macenski_impact_2023}.
However, \citeauthor{menard_achieving_2020} claim that this approach is insufficient to yield a deterministic system and present an approach to address this shortcoming \cite{menard_achieving_2020}.

To ensure security, AUTOSAR \ac{AP} contains functional clusters for cryptography, firewall, and intrusion detection. 
These clusters enable encryption of \ac{IVN} traffic, ensure that all machines only expose necessary ports to the \ac{IVN} and attempt to detect intrusion attempts \cite{autosar_ap_explanation_sw_arch}. 

The \ac{UCM} functional cluster in AUTOSAR \ac{AP} performs package management functions. 
It supports operations such as updating, installing, removing, and keeping records, similar to traditional package managers. 
\ac{UCM} maintains a local registry of packages, which serve as the installation units within the AUTOSAR \ac{AP}. 
Using this mechanism, \ac{OTA} updates can be applied \cite{henle_architecture_2022,autosar_consortium_explanation_2023}.

AUTOSAR Adaptive also offers clusters for persistence in vehicles, automatic configuration of \ac{IVN}, time syncing, diagnostic services, logging, and efficient data transmission using raw data streams. \cref{fig:adaptivePlatform} presents a comprehensive overview of all functional clusters specified in AUTOSAR \ac{AP}.

Due to the closed-source nature of AUTOSAR \ac{AP}, publicly available performance evaluations remain rare, with existing literature consisting largely of industry publications and offering few direct comparisons to other middlewares. 
One such rare comparative effort is presented by \citeauthor{schulik_efficient_2025} in a non-peer-reviewed publication, who propose a function interface for \ac{ROS 2} and AUTOSAR \ac{AP} to integrate a single application into both frameworks and evaluate the behavior of both systems \cite{schulik_efficient_2025}. 
Their results indicate that \ac{ROS 2} exhibits higher memory consumption, increased CPU load, and lower application-level runtime performance within this integrated setup.

\subsubsection{\Soamid\ example}
\label{sec:middleware:soamid_example}

\begin{figure}[tbp]
    \centering
    \includegraphics[width=\ImageSize]{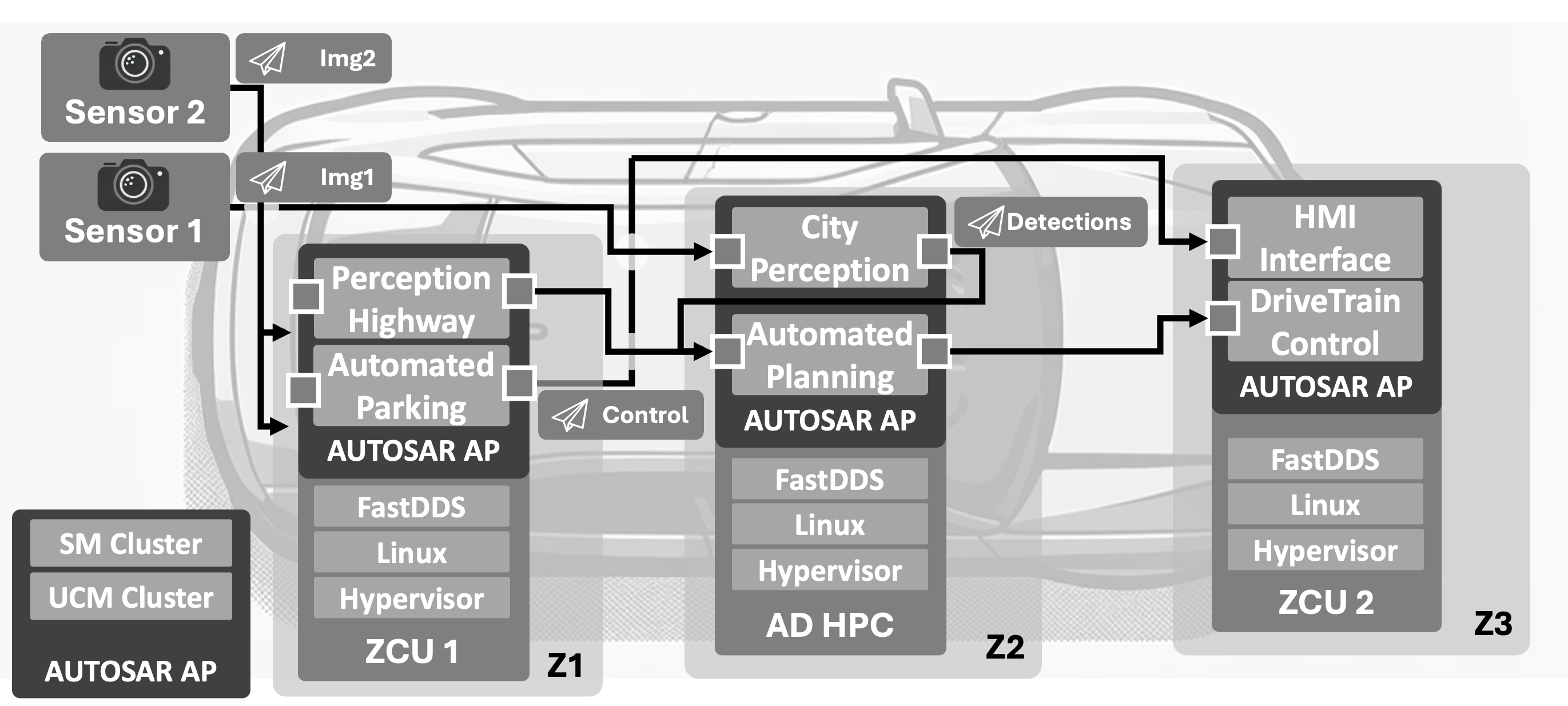}
    \caption{Example software system architecture using a \soamid. All components are finely divided into individual service-oriented components and have an individual runtime environment provided by the \shsoamid.}
    \label{fig:example_soamid}
\end{figure}

In a software system built on a \soamid, the \shsoamid\ commonly enforces a service-oriented architecture of applications.
Furthermore, \soamids\ often have a run-time environment in which the application runs. 
This environment supports additional functions to help in the deployment, maintenance, and development of the \system\ and its applications.

\noindent In our example, using a \soamid\ would make it possible to implement the expansion of the operational design domain by automatically deploying a new service to the automated driving \ac{ECU}.
This deployment can be performed using \ac{OTA} updates using the AUTOSAR \acp{AP} \ac{UCM} cluster, and the service is provisioned with the required resources by the execution management functional cluster.
This approach inherits the benefits of the underlying \commid, such as modular and flexible communication, as depicted in \cref{fig:example_soamid}.
Applications can still use the topic-based publish-subscribe paradigm for message exchange, and the service could access all sensors in the vehicle.
It would also be able to issue commands seamlessly, similar to the \commid\ example.

The core advantage of this approach is that \soamids\ require the division of software components into services and provides support for these services.
The service-oriented architecture ensures a clear division of responsibilities and therefore simplifies updates, maintenance, and even development.
The \soamid\ introduces additional capabilities for \ac{OTA} updates, software system parameterization, and resource management.
The runtime environment aids in the deployment of the new service, employing virtualization techniques to efficiently allocate and manage computational resources.
However, the increased cyber-security risk is also retained, due to the increased complexity of components involved in supporting an application.

\section{Comparison of Modern Middlwares}
\label{sec:middleware_comp}

\begin{table*}[!htb] 
\renewcommand{\arraystretch}{1.10}
\caption{Middlewares in the Automotive Domain}
\label{tab:middleware-comparison}

\begin{footnotesize}
\begin{tabularx}{\textwidth}{Y *{5}{Z}}
\toprule
Criteria & Robot Operating System 2 & AUTOSAR Adaptive Platform &  DDS (FastDDS) & SOME/IP & Zenoh \\  
\midrule
Non-Technical Criteria & & & & & \\
\midrule
Type of Middleware & \textit{Platform} & \textit{Platform} & \textit{Communication} & \textit{Communication} & \textit{Communication} \\
Operation Domain  & robotics \cite{macenski_robot_2022} & automotive \cite{autosar_consortium_explanation_2023}  & general \cite{eprosima_11_2024}  & automotive \cite{autosar_consortium_explanation_2023}  &  general \cite{corsaro_zenoh_2023} \\

License and Accessibility & open-source, free \cite{macenski_robot_2022} & closed-source, commercial \cite{autosar_consortium_explanation_2023} & open-source, free \cite{eprosima_11_2024} & open-source, free  & open-source, free \cite{eclipse_foundation_abstractions_2024} \\

Documentation & open, comprehensive documentation \cite{macenski_robot_2022} & standards, commercial partner\textsuperscript{2} \cite{autosar_consortium_explanation_2023} & open, comprehensive documentation \cite{eprosima_11_2024} & standards, commercial partner\textsuperscript{2} & open documentation \cite{eclipse_foundation_abstractions_2024} \\

Software Tooling & extensive \cite{open_robotics_ros_2024} & commercial partner\textsuperscript{2} & extensive & commercial partner\textsuperscript{2}  & limited \\
Industry Adoption & widely, robotics \cite{malavolta_how_2020} & emerging, automotive industry \cite{zhu_requirements-driven_2021} & widely, various industries & emerging, automotive industry  & emerging, various industry \cite{eclipse_foundation_zenoh_2024} \\

ISO 26262 & APEX.AI fork: ASIL-D \cite{becker_safety-certified_2021,ekberg_autosar_2022} & supported\textsuperscript{2} \cite{autosar_consortium_overview_2022} & Safe DDS: ASIL-D certified \cite{tuv_sud_certificate_2024,eprosima_eprosima_nodate} & unspecified & certification in progress 
\cite{tttech_auto_zetta_2024,eclipse-zenoh_iso_nodate}  \\
\midrule
Technical Criteria & & & &  & \\
\midrule

\Commpats & publish-subscribe, request-response \cite{macenski_robot_2022} & methods, fields, events \cite{autosar_communicationManagement_2024} & publish-subscribe \cite{eprosima_11_2024} & methods, fields, events \cite{autosar_someip_2024}  & publish-subscribe, put-get, queryables \cite{corsaro_zenoh_2023, eclipse_foundation_abstractions_2024} \\

Real-Time Communication & soft real-time \cite{macenski_robot_2022} & soft real-time \cite{autosar_communicationManagement_2024} & soft real-time \cite{eprosima_11_2024} & soft real-time \cite{autosar_someip_2024} & unsupported \\

Quality of Service & derived from DDS \cite{open_robotics_quality_2024} & derived from DDS \cite{autosar_communicationManagement_2024} & extensive \cite{eprosima_12_2024} & limited / unsupported \cite{autosar_someip_2024}  & limited - Reliable / BestEffort \cite{eclipse_foundation_zenoh_2024} \\

Communication Protocols & DDS - RTPS \cite{macenski_robot_2022} & \ac{DDS} - RTPS, SOME/IP & RTPS - TCP / UDP, shared memory \cite{eprosima_6_2024} & proprietary - TCP / UDP \cite{autosar_someip_2024}  & proprietary \cite{corsaro_zenoh_2023} \\


\midrule

Execution Model & executors concept \cite{open_robotics_executors_2024} & EM deterministic client \cite{autosar_executionManagement_2024} & unsupported & unsupported  & Zenoh-Flow \cite{eclipse_foundation_zenoh-flow_2024} \\

Real-Time Capability & microROS: Callback-group-level executor \cite{yang_exploring_2020} & supported, not enforced \cite{autosar_consortium_specification_2024} & supported \cite{eprosima_157_2025} & unspecified\textsuperscript{2} & supported \cite{baron_performance_2025} \\

System State Management & lifecycle nodes \cite{macenski_robot_2022} & State Management \cite{autosar_ap_explanation_sw_arch} & unsupported & unsupported  & unsupported  \\

Application Configuration \\ Management & node parametrization \cite{macenski_robot_2022} & UCM \cite{autosar_ap_explanation_sw_arch} & unsupported & unsupported & unsupported \\

\midrule

Application Discovery & automatic \cite{macenski_robot_2022,eprosima_5_2024} & automatic \cite{autosar_consortium_ip_2023} & manual / automatic \cite{eprosima_5_2024} & manual / automatic - SOME/IP-SD \cite{autosar_consortium_ip_2023} & automatic \cite{eclipse_foundation_deployment_2024} \\

Application Update Mechanism & unsupported & UCM \cite{autosar_ap_explanation_sw_arch} & unsupported & unsupported & unsupported \\

Feature Extension & supported\textsuperscript{3} \cite{macenski_robot_2022} & unsupported\textsuperscript{1} \cite{autosar_ap_explanation_sw_arch} & supported\textsuperscript{3} \cite{eprosima_11_2024} & unsupported\textsuperscript{1}   \cite{autosar_ap_explanation_sw_arch} & supported \cite{eclipse_foundation_zenoh_2024} \\

Application Resource \\ Management & unsupported  & supported, EM \cite{autosar_executionManagement_2024} & unsupported & unsupported & unsupported \\

\midrule

Communication Encryption & supported, SROS 2 \cite{open_robotics_ros_2024,macenski_robot_2022} & supported, \ac{CM} \cite{autosar_communicationManagement_2024} & supported \cite{eprosima_8_2024} & unsupported\textsuperscript{4} & supported, TLS \cite{corsaro_zenoh_2023} \\

Application Authentication & supported, SROS 2 \cite{open_robotics_ros_2024,macenski_robot_2022} & supported, \ac{CM} \cite{autosar_communicationManagement_2024} & supported, DDS Sec \cite{eprosima_8_2024} & unsupported\textsuperscript{4} & supported, TLS \cite{corsaro_zenoh_2023} \\

Additional Security Features &  DDS Sec, SROS 2 \cite{open_robotics_ros_2024,macenski_robot_2022} &  IDS, Firewall, Network & access control \cite{eprosima_8_2024} & unsupported\textsuperscript{4} & unsupported \\

\midrule

Operating System Support & Linux, MacOS, Windows, RTOS \cite{belsare_micro-ros_2023,open_robotics_installation_2024,microros_supported_2025}
& POSIX OS \cite{autosar_ap_explanation_sw_arch} & Linux, MacOS, Windows, Android, QNX \cite{eprosima_dependencies_2025} & POSIX OS & Desktop OS, embedded devices \cite{corsaro_zenoh_2023} \\

ISA Support & amd64, arm64, arm32 
\cite{open_robotics_rep_2023,microros_supported_2025} & unspecified, derived from POSIX OS & amd64, amd32, arm64 \cite{eprosima_dependencies_2025}, arm32 \cite{belsare_micro-ros_2023,microros_supported_2025} & unspecified\textsuperscript{2} & amd64, amd32, arm64, arm32 \cite{noauthor_nokiaeclipse-zenoh-zenoh-c_2024,noauthor_eclipse-zenohzenoh-java_2025}\\
\bottomrule 
\end{tabularx}
\end{footnotesize}

\textsuperscript{1} Outside of the AUTOSAR Consortium and the cooperation processes.
\textsuperscript{2} As a standard, the implementation and support provided depends on the individual company that completed the implementation and the tools they provide.
\textsuperscript{3} As open source software, additions are possible without membership in a committee.
\textsuperscript{4} Handled in the AUTOSAR AP by the Communication Management functional cluster.
\textsuperscript{5} An open-source implementation of \ac{SOME/IP} exists, in the form of VSOMEIP by BMW.

\end{table*}

\begin{table*}[!htb] 
\renewcommand{\arraystretch}{1.10}
\caption{Middlewares~Communication~Performance}
\label{tab:middleware-performance}

\begin{footnotesize}
\begin{tabularx}{\textwidth}{Y *{5}{Z}}
\toprule
Criteria & Robot Operating System 2 & AUTOSAR Adaptive Platform & (DDS) FastDDS & (SOME/IP) vSOME/IP & Zenoh \\
\midrule
Communication & & & & & \\
\midrule
Discovery Latency
& n/r\textsuperscript{1} 
& n/r\textsuperscript{1}
& $+$/$\bigcirc$ \cite{kluner_automotive_2025} 
& $-$ \cite{kluner_automotive_2025, seyler_insights_2015}
& $+$/$\bigcirc$ \cite{kluner_automotive_2025} \\
Transmission Latency
& n/i\textsuperscript{2} \cite{kronauer_latency_2021, kouril_performance_2024, maruyama_exploring_2016, zhang_comparison_2023, wu_oops_2021}
& n/i\textsuperscript{2} \cite{castillo-sanchez_swarm_2024, lee_optimizing_2025}
& $+$/$\bigcirc$ \cite{kluner_automotive_2025, zhang_comparison_2023, chisalita_stepping_2025, liang_performance_2023}
& $-$ \cite{kluner_automotive_2025, yang_xvsomeip_2024}
& $+$/$\bigcirc$ \cite{kluner_automotive_2025, baron_performance_2025, zhang_comparison_2023, chisalita_stepping_2025, liang_performance_2023} \\
Throughput
& n/i\textsuperscript{2} \cite{maruyama_exploring_2016}
& n/i\textsuperscript{2} \cite{lee_optimizing_2025}
& $\bigcirc$ \cite{zhang_comparison_2023}
& $-$ \cite{kluner_automotive_2025, yang_xvsomeip_2024}
& $+$ \cite{kluner_automotive_2025, zhang_comparison_2023, chisalita_stepping_2025, liang_performance_2023} \\
Message Loss
& n/i\textsuperscript{2} \cite{wu_oops_2021}
& n/r\textsuperscript{1}
&  $\bigcirc$ \cite{kluner_automotive_2025, chisalita_stepping_2025}
& $-$ \cite{kluner_automotive_2025}
& $+$ \cite{kluner_automotive_2025, baron_performance_2025, chisalita_stepping_2025} \\
\midrule
Resource Use & & & & & \\
\midrule
CPU Utilization
& $\bigcirc$ \cite{wu_oops_2021, schulik_efficient_2025}
& $+$ \cite{schulik_efficient_2025,merakanapalli_transitioning_2025}
& $+$ \cite{kluner_automotive_2025,chovet_performance_2025}
& $\bigcirc$ \cite{kluner_automotive_2025}
& $-$ \cite{kluner_automotive_2025, chovet_performance_2025} \\
Memory Utilization
& $\bigcirc$ \cite{wu_oops_2021, schulik_efficient_2025}
& $+$ \cite{schulik_efficient_2025, merakanapalli_transitioning_2025}
& $+$ \cite{kluner_automotive_2025,chovet_performance_2025}
& $\bigcirc$ \cite{kluner_automotive_2025}
& $-$ \cite{kluner_automotive_2025, chovet_performance_2025} \\
\bottomrule
\end{tabularx}
\end{footnotesize}

$\bigcirc$: similar performance, $+$: better than other solutions, $-$: worse than other solutions. Comparisons were done between \commids\ and \soamids\ respectively.  
\textsuperscript{1}: Not reported: Our survey yielded no studies that evaluate these performance criteria.  
\textsuperscript{2}: Not interpretable: Due to a lack of comparable studies a conclusive result could not be reached. 
\end{table*}

We compare the five introduced middlewares, \ac{ROS 2}, AUTOSAR \ac{AP}, FastDDS, \ac{SOME/IP} and Zenoh in \cref{tab:middleware-comparison} and \cref{tab:middleware-performance} from an automotive perspective.
In \cref{tab:middleware-comparison} each column contains information about a single middleware and is sorted according to our \commid\ and \soamid\ classification.
The middlewares were assessed based on non-technical and technical criteria, reflecting their design purposes, adoption in the community, \commpats, real-time capabilities, and other significant features.
\cref{tab:middleware-performance} presents an overview over middleware communication performance. 
Due to high study-to-study variance between experimental setups, hardware compute performance, network performance, measurement accuracy, and overall methodology, only conclusions from the literature are presented. Performance measurements are strongly dependent on these factors and cannot be directly compared across studies \cite{beyer_reliable_2019}.

In conclusion of the presented \commids, DDS and its implementations are the most fully featured and proven middleware. 
Its large set of discovery features, highly configurable quality of service settings, and large number of security features are among its advantages.
In addition, its use in \ac{ROS 2} and now the inclusion of \ac{DDS} in the AUTOSAR \ac{AP} standard attest to the maturity and feature richness of \ac{DDS}.
This maturity is also present in its performance. In the literature, FastDDS demonstrated mostly good performance in the criteria surveyed. It performs worse than competitors only in high data rate applications and some wireless scenarios.
In contrast, \ac{SOME/IP} offers a reduced feature set, especially in the context of security, shared memory transport, and limited quality of service configurability.
Additionally, for a time no open-source \ac{SOME/IP} implementation was available, while multiple DDS implementations are freely accessible, well supported, and tested. 
vSOME/IPs performance in literature was also worse than its competitors. In all measured criteria, it performs slower and with reduced robustness relative to the other two \commids.
However, due to the support of the AUTOSAR Consortium and its focus specifically on the automotive domain, it may remain relevant in the domain.
The newest and least proven \commid\ in this comparison is Zenoh. In our opinion, Zenoh does not have fundamentally different paradigms compared to existing \commids, but may present a performance improvement over \ac{DDS} implementations or \ac{SOME/IP}.
In the literature, these claims of increased performance could only be partially substantiated. 
In wired networks, Zenoh shows minimal performance gains in transmission latency, discovery, and throughput. 
However, it performs better in wireless networks and at very high data rates than both FastDDS and vSOME/IP, with a slightly higher memory usage.

When comparing the two presented \soamids, the main difference is the part of the development cycle at which these \shsoamids\ are aimed. 
While AUTOSAR \ac{AP} provides a large number of features for the deployment of applications to production vehicles, \ac{ROS 2} is focused on development.
This is reflected in the license model of these \shsoamids. \ac{ROS 2} targets research and development, is free and open-source, while AUTOSAR \ac{AP} is only accessible through the purchase from commercial partners, as it targets \acp{OEM} and automotive suppliers. 
Consistent with this focus, AUTOSAR \ac{AP} specifies many features for resource control, configuration management, and state management, to ensure that applications can be correctly deployed in real vehicles. 
\ac{ROS 2} on the other hand, offers many features such as visualizations, standard messages, already implemented packages, and a large ecosystem to support the easy and fast development of new applications. 
Although provisions for starting applications are implemented, for example, no system health monitoring or complete software system configuration tools, as can be found in AUTOSAR \ac{AP}, is build into \ac{ROS 2}. 
This reduces the suitability of \ac{ROS 2} for deploying applications in production vehicles.
In this context, the developments of APEX.AI should be examined, as their proprietary \ac{ROS 2} fork combines some advantages of both software systems.
When discussing \soamid\ performance, comparing the communication performance of AUTOSAR \ac{AP} and \ac{ROS 2} proves challenging. 
Due to the closed-source nature of AUTOSAR \ac{AP}, only very few studies report its performance characteristics, in contrast to \ac{ROS 2}, for which extensive evaluations exist. 
The comparison is further complicated by the presence of multiple implementations of the AUTOSAR Adaptive standard, each exhibiting potentially different performance properties. 
In our tutorial, we identified only a non-peer-reviewed publication comparing both approaches \cite{schulik_efficient_2025}, but the available data did not allow us to draw conclusive results about their relative performance.

\section{Research Challenges}
\label{sec:challenges}

Middleware-based software architectures in cooperation with zone-based \ac{E/E} architectures address many current challenges for \acp{AV}.
However, despite being the solution to considerable challenges, middlewares also face open research challenges themselves.
These challenges range from real-time communication in ethernet-based \acp{IVN} to resource allocation in a \system\ where services can be arbitrarily assigned to \acp{ECU}.
The following section presents some of these challenges, organized into five categories, and research seeking to address them.

\subsection{Challenges in Real-time Communication}
Despite the disadvantages of legacy \ac{E/E} architectures, an advantage is better predictability and control over message scheduling.
However, in regular ethernet-based \acp{IVN} the same property is not guaranteed, as network congestion, packet loss, and mutual interference in communication between \acp{ECU} are possible \cite{wang_review_2024}.
To address this challenge, researchers propose the use of \ac{TSN}, an ethernet extension that introduces additional controls on communication and flow \cite{kugele_data-centric_2018}.
\citeauthor{brunner_automotive_2017} \cite{brunner_automotive_2017} proposed an initial research on this topic, which highlights the advantages of \ac{TSN} use in the automotive domain. 
\citeauthor{migge_insights_2018} \cite{migge_insights_2018}, present details on the performance and configuration of \ac{TSN} in the ethernet \acp{IVN}. 
However, currently no standard solution has been introduced in ROS 2 and AUTOSAR AP.
Another approach to provide more guarantees for \ac{IVN}s is the use of \ac{SDN} features. 
\citeauthor{hackel2023dynamic} \cite{hackel2023dynamic} investigate the combination of \ac{TSN} and \ac{SDN} combination to assess its suitability for the automotive domain.
\citeauthor{rotermund_requirements_2020} \cite{rotermund_requirements_2020} investigate the performance of \ac{SDN} controllers for vehicles. 
To achieve heterogeneous systems, where middlewares operate on all devices in \systems, support for embedded devices is necessary. \citeauthor{kampmann_portable_2019} \cite{kampmann_portable_2019} implement \ac{DDS} for multiple automotive platforms.

\subsection{Challenges in Real-time Execution}
The introduction of \ac{ZCU} and high performance \acp{ECU} in vehicles \ac{E/E} architecture also requires the use of full-scale operating systems, such as Linux \cite{furst_autosar_2016}. 
This change of platform directly affects the execution of software and associated hard real-time requirements. 
However, the use of Linux in safety-critical systems remains also an area of active research \cite{schlosser_considering_2024}.
By default, Linux does not guarantee real-time execution, and with OTA updates and updates to vehicle software, tight control over thread execution, timing, and predictability are not guaranteed \cite{blass_automatic_2021}.
Furthermore, with possible resource congestion both on the \ac{IVN} and between applications, new approaches are necessary to guarantee reliable and deterministic execution.
To address this challenge and ensure intelligent resource provisioning, \citeauthor{blass_automatic_2021} propose a latency management framework for ROS 2 \cite{blass_automatic_2021}.
However, ensuring resource provisioning and local execution orders do not guarantee the repeatable and deterministic behavior of the entire system. 
Due to varying execution and transmission latencies, distributed systems suffer from an inherent non-determinism.
This challenge is addressed by \citeauthor{menard_achieving_2020} \cite{menard_achieving_2020} in the context of the AUTOSAR Adaptive framework.
Since that work, the pursuit of deterministic systems using middleware has broadened and has been recognized as an important objective \cite{lee_determinism_2021}. 
For instance, \citeauthor{staschulat_rclc_2020} \cite{staschulat_rclc_2020} developed a deterministic executor for microROS as a building block towards a deterministic system. 
\citeauthor{gemlau_system-level_2021} \cite{gemlau_system-level_2021} introduce system‑level logical execution time to enforce determinism in the AUTOSAR Adaptive Platform. 
Outside of middleware-specific solutions, \citeauthor{lohstroh_toward_2021_b} \cite{lohstroh_toward_2021_b} propose Lingua Franca, a reactor‑oriented coordination language that similarly seeks to achieve determinism.
However, in the case of ROS 2 and AUTOSAR Adaptive, no holistic implementation has been integrated into the framework and the standard yet.
It remains an active research question as to how best to achieve a deterministic distributed automotive software system and which approach to select.

\subsection{Challenges in Security and Safety}
Another consequence of the inclusion of Ethernet-based \acp{IVN} and high performance \acp{ECU} is the increase in cyber-security attack vectors \cite{dieber_application-level_2016,vilches_sros2_2022,rumez_overview_2020}.
While in traditional \acp{ECU} the code was deeply embedded in the hardware, now the middleware, the OS, and the network present new attack vectors.
To survey these challenges, \citeauthor{rumez_overview_2020} \cite{rumez_overview_2020} provides an overview of the security implications of the application of service-orientation to automotive software architectures.
Approaches, such as \ac{IDS}, firewalls and encryption can already be found in some middleware, for instance AUTOSAR \ac{AP} offers functional clusters for these challenges \cite{autosar_ap_explanation_sw_arch} \cite{pullen_security_2023}.
\citeauthor{vilches_sros2_2022} \cite{vilches_sros2_2022} propose a security approach for \ac{ROS 2}, while \ac{ROS 2} also offers some features based on the \ac{DDS} security extension.
However, more research is required to identify whether these solutions are comprehensive and what additional approaches might be required.

\subsection{Challenges in Orchestration and Resource Management}
Applications built on middlewares are, unless they require interaction with specific sensors, actuators, or compute accelerators, such as GPUs, independent of the platform that executes them \cite{burkacky_rethinking_nodate}.
This independence also presents a resource allocation challenge for applications in the automotive software system and the specific \ac{ECU}.
An approach to managing resource use and separation of applications is containerization and virtualization \cite{berger_containerized_2017}.
Enabled by containerized applications, multiple research questions emerge such as where to deploy individual applications and how to orchestrate them.
To address this question, \citeauthor{nayak_automotive_2023} in \cite{nayak_automotive_2023} presents an overview of the requirements, challenges, and directions of containerization for automotive applications. 
Similarly to the ability to control containerized applications, ROS 2 lifecycle nodes and AUTOSAR \ac{AP} applications offer lifecycle controls.
These controls enable a controller, such as an Orchestrator, Kubernetes, or AUTOSAR AP \ac{SM} to free parts of the resources used by a service by deactivating it \cite{autosar_ap_explanation_sw_arch, kampmann_optimization-based_2022}. An approach of \citeauthor{kampmann_dynamic_2019} using an orchestrator to manage an \system, is the \ac{ASOA} \cite{kampmann_dynamic_2019}.
More research is required to achieve an optimal assignment of resources and to conform the software system to the requirements of the specific situation \cite{jatzkowski_integration_2021}.

\subsection{Challenges in Machine Learning Applications}
Modern software systems and software in automated vehicles are increasingly based on artificial intelligence and \ac{ML} capabilities.
Applications built on these technologies present unique challenges to software architecture in general \cite{amershi_software_2019}, and specifically to the automotive domain, as recognized by \citeauthor{kugele_service-orientation_2017} \cite{kugele_service-orientation_2017}.
\citeauthor{serban_adapting_2022} \cite{serban_adapting_2022} and \citeauthor{amershi_software_2019} \cite{amershi_software_2019} discuss in detail the unique challenges of artificial intelligence that applications pose.
The challenges include the limited interpretability of \ac{DL} models and their potential non-robustness on unseen data.
In a software system with multiple \ac{DL} components, the authors predict a potential cascade of failures as each component produces adverse results for the next.
From these requirements, they derive increased monitoring requirements and the need to increase robustness using, for instance, n-versioning.
These challenges are interwoven with the application design, but middlewares could offer solutions to these challenges.
Another emerging research direction is the integration of \acp{LLM} into automotive software development processes, a topic that has seen substantial growth in the past year. 
Recent works explore how \acp{LLM} can support or automate different phases of the development lifecycle: \citeauthor{petrovic_llm-based_2025} \cite{petrovic_llm-based_2025} propose their use for process automation to improve maintainability, while \citeauthor{asad_advancing_2025} \cite{asad_advancing_2025} follow a related idea and investigate automated impact analysis of software updates. 
\citeauthor{patil_towards_2025} \cite{patil_towards_2025} focus on software development by proposing \ac{LLM}-assisted code generation for embedded automotive software combined with formal verification.
This idea is further extended by \citeauthor{kirchner_generating_2025} \cite{kirchner_generating_2025}, who introduce a feedback-driven pipeline that integrates \acp{LLM} into testing, simulation, and verification workflows. 
Focusing on the security of automotive software, \citeauthor{scarano_assessing_2025} \cite{scarano_assessing_2025} demonstrate that \acp{LLM} have substantial domain knowledge of automotive cybersecurity risks and propose their use in security-related processes.
In conclusion, machine learning is not just relevant for function development, but with \acp{LLM} also relevant for the entire process of developing, testing, and verifying automotive software.

\section{Conclusions}
\label{sec:conclusion}

This article presented a survey on middlewares for automated vehicles. 
The survey introduced the fundamentals of automotive computing, gave an overview of five modern middleware, and compared them.
We classified middlewares into \commids\ and \soamids, to highlight the former's communication purpose and the multi-domain, framework nature of the latter and discussed their relative performance.
Among the \commids\ we presented, \ac{DDS} remains the most fully-featured solution, while the Zenoh \commid\ represents a new challenger to \ac{DDS} delivering some performance gains.
Among \soamids\ we concluded that the place in the development cycle determines the \shsoamid\ to choose. 
\ac{ROS 2} focuses on features for development and testing, while AUTOSAR Adaptive offers comprehensive features for deployment.
Although middleware addresses many challenges in vehicles, research questions remain.
Finally, the article presented open research questions for middlewares.
The most prominent questions were related to real-time or deterministic communication and execution in automotive software systems. The increasing complexity and capability of automotive software systems also require new approaches to safety and security. Middleware-based architectures also present resource allocation and orchestration challenges. Machine learning applications and \acp{LLM} also pose new challenges and possibilities.

\section{ACKNOWLEDGMENT}
This research is accomplished within the project ”autotech.agil” (FKZ 01IS22088x). We acknowledge the financial support for the project by the Federal Ministry of Research, Technology and Space of Germany (BMFTR).

\section{ACRONYMS}

\begin{acronym}[asdfasdvs]
    \setlength{\itemsep}{0pt} 
    \acro{AA}{Adaptive Application}
    \acro{AD}{Automated Driving}
    \acro{AP}{Adaptive Platform}
    \acro{API}{Application Programming Interface}
    \acro{ARA}{AUTOSAR Runtime for Adaptive Applications}
    \acro{ASOA}{Automotive Service Oriented Architecture}
    \acro{AV}{Automated Vehicle}
    \acro{CAN}{Controller Area Network}
    \acro{CM}{Communication Management}
    \acro{CPS}{Cyber Physical Systems}
    \acro{DDS}{Data Distribution Service}
    \acro{DL}{Deep Learning}
    \acro{E/E}{Electrical/Electronic}
    \acro{ECU}{Electronic Control Unit}
    \acro{EM}{Execution Management}
    \acro{FC}{Function Cluster}
    \acro{FPGA}{Field Programmable Gate Array}
    \acro{GPU}{Graphics Processing Unit}
    \acro{HMI}{Human Machine Interface}
    \acro{HPC}{High-Performance Computer}
    \acro{IDS}{Intrusion Detection System}
    \acro{IoT}{Internet of Things}
    \acro{IPC}{Inter-Process Communication}
    \acro{IVN}{In-Vehicular Network}
    \acro{LLM}{Large Language Model}
    \acro{ML}{Machine Learning}
    \acro{MQTT}{Message Queuing Telemetry Transport}
    \acro{OEM}{Original Equipment Manufacturer}
    \acro{OMG}{Object Management Group}
    \acro{OS}{Operating System}
    \acro{OTA}{Over-The-Air}
    \acro{QoS}{Quality of Service}
    \acro{RMW}{ROS Middleware Interface}
    \acro{ROS 2}{Robot Operating System 2}
    \acro{S2S}{Service-to-Signal}
    \acro{SDN}{Software-Defined Network}
    \acro{SDV}{Software Defined Vehicles}
    \acro{SM}{State Management}
    \acro{SOA}{Service-oriented Architecture}
    \acro{SOME/IP}{Scalable service-Oriented MiddlewarE over IP}
    \acro{TCP}{Transmission Control Protocol}
    \acro{TSN}{Time-sensitive Networking}
    \acro{UCM}{Update and Configuration Management}
    \acro{UDP}{User Datagram Protocol}
    \acro{ZCU}{Zone Control Unit}
\end{acronym}

\printbibliography

@article{zeng_qos-aware_2004,
	title = {{QoS}-aware middleware for {Web} services composition},
	volume = {30},
	number = {5},
	journal = {IEEE Transactions on Software Engineering},
	author = {Zeng, Liangzhao and Benatallah, B. and Ngu, A.H.H. and Dumas, M. and Kalagnanam, J. and Chang, H.},
	month = may,
	year = {2004},
	note = {Conference Name: IEEE Transactions on Software Engineering},
	pages = {311--327},
}

@inproceedings{schlosser_considering_2024,
	address = {Wiesbaden},
	title = {Considering {Linux} for functional safety relevant system architecture: {Pitfalls} and {Potential}},
	isbn = {978-3-658-45196-7},
	shorttitle = {Considering {Linux} for functional safety relevant system architecture},
	doi = {10.1007/978-3-658-45196-7_12},
	language = {de},
	booktitle = {Automatisiertes {Fahren} 2024},
	publisher = {Springer Fachmedien},
	author = {Schlosser, Joachim and Petersohn, Jens},
	editor = {Heintzel, Alexander},
	year = {2024},
	pages = {149--160},
}

@misc{liang_performance_2023,
	title = {A {Performance} {Study} on the {Throughput} and {Latency} of {Zenoh}, {MQTT}, {Kafka}, and {DDS}},
	url = {http://arxiv.org/abs/2303.09419},
	doi = {10.48550/arXiv.2303.09419},
	urldate = {2024-11-12},
	publisher = {arXiv},
	author = {Liang, Wen-Yew and Yuan, Yuyuan and Lin, Hsiang-Jui},
	month = mar,
	year = {2023},
	note = {arXiv:2303.09419},
}

@article{perera_context_2014,
	title = {Context {Aware} {Computing} for {The} {Internet} of {Things}: {A} {Survey}},
	volume = {16},
	issn = {1553-877X},
	shorttitle = {Context {Aware} {Computing} for {The} {Internet} of {Things}},
	doi = {10.1109/SURV.2013.042313.00197},
	number = {1},
	journal = {IEEE Communications Surveys \& Tutorials},
	author = {Perera, Charith and Zaslavsky, Arkady and Christen, Peter and Georgakopoulos, Dimitrios},
	year = {2014},
	note = {Conference Name: IEEE Communications Surveys \& Tutorials},
	pages = {414--454},
}

@article{liu_impact_2022,
	title = {Impact, {Challenges} and {Prospect} of {Software}-{Defined} {Vehicles}},
	volume = {5},
	issn = {2522-8765},
	doi = {10.1007/s42154-022-00179-z},
	language = {en},
	number = {2},
	journal = {Automotive Innovation},
	author = {Liu, Zongwei and Zhang, Wang and Zhao, Fuquan},
	month = apr,
	year = {2022},
	pages = {180--194},
}

@article{kang_rdds_2012,
	title = {{RDDS}: {A} {Real}-{Time} {Data} {Distribution} {Service} for {Cyber}-{Physical} {Systems}},
	volume = {8},
	issn = {1941-0050},
	shorttitle = {{RDDS}},
	doi = {10.1109/TII.2012.2183878},
	number = {2},
	journal = {IEEE Transactions on Industrial Informatics},
	author = {Kang, Woochul and Kapitanova, Krasimira and Son, Sang Hyuk},
	month = may,
	year = {2012},
	note = {Conference Name: IEEE Transactions on Industrial Informatics},
	pages = {393--405},
}

@article{neely_adaptive_2006,
	title = {Adaptive middleware for autonomic systems},
	volume = {61},
	issn = {1958-9395},
	doi = {10.1007/BF03219883},
	language = {en},
	number = {9},
	journal = {Annales Des Télécommunications},
	author = {Neely, Steve and Dobson, Simon and Nixon, Paddy},
	month = oct,
	year = {2006},
	pages = {1099--1118},
}

@inproceedings{kampmann_portable_2019,
	address = {Auckland, New Zealand},
	title = {A {Portable} {Implementation} of the {Real}-{Time} {Publish}-{Subscribe} {Protocol} for {Microcontrollers} in {Distributed} {Robotic} {Applications}},
	isbn = {978-1-5386-7024-8},
	doi = {10.1109/ITSC.2019.8916835},
	booktitle = {2019 {IEEE} {Intelligent} {Transportation} {Systems} {Conference} ({ITSC})},
	publisher = {IEEE},
	author = {Kampmann, Alexandru and Wustenberg, Andreas and Alrifaee, Bassam and Kowalewski, Stefan},
	month = oct,
	year = {2019},
	pages = {443--448},
}

@book{bass_software_2003,
	title = {Software {Architecture} in {Practice}, 2nd {Edition}},
	author = {Bass, Len and Clements, Paul and Kazman, Rick},
	month = apr,
	year = {2003},
}

@misc{zhang_comparison_2023,
	title = {Comparison of {DDS}, {MQTT}, and {Zenoh} in {Edge}-to-{Edge} and {Edge}-to-{Cloud} {Communication} for {Distributed} {ROS} 2 {Systems}},
	url = {http://arxiv.org/abs/2309.07496},
	doi = {10.48550/arXiv.2309.07496},
	urldate = {2024-11-08},
	publisher = {arXiv},
	author = {Zhang, Jiaqiang and Yu, Xianjia and Ha, Sier and Queralta, Jorge Pena and Westerlund, Tomi},
	month = sep,
	year = {2023},
	note = {arXiv:2309.07496},
}

@article{bello_recent_2019,
	title = {Recent {Advances} and {Trends} in {On}-{Board} {Embedded} and {Networked} {Automotive} {Systems}},
	volume = {15},
	issn = {1941-0050},
	doi = {10.1109/TII.2018.2879544},
	number = {2},
	journal = {IEEE Transactions on Industrial Informatics},
	author = {Bello, Lucia Lo and Mariani, Riccardo and Mubeen, Saad and Saponara, Sergio},
	month = feb,
	year = {2019},
	note = {Conference Name: IEEE Transactions on Industrial Informatics},
	pages = {1038--1051},
}

@inproceedings{becker_safety-certified_2021,
	address = {Wiesbaden},
	title = {A {Safety}-{Certified} {Vehicle} {OS} to {Enable} {Software}-{Defined} {Vehicles}},
	isbn = {978-3-658-34754-3},
	doi = {10.1007/978-3-658-34754-3_5},
	language = {de},
	booktitle = {Automatisiertes {Fahren} 2021},
	publisher = {Springer Fachmedien},
	author = {Becker, Jan and Sagar, Mehul and Pangercic, Dejan},
	editor = {Bertram, Torsten},
	year = {2021},
	pages = {51--67},
}

@misc{ekberg_autosar_2022,
	title = {{AUTOSAR} and {ROS} 2 for {Software}-{Defined} {Vehicle}},
	url = {https://www.apex.ai/post/autosar-and-ros-2-for-software-defined-vehicle},
	language = {en},
	urldate = {2024-11-04},
	journal = {Apex.AI},
	author = {Ekberg, Johan},
	month = may,
	year = {2022},
}

@misc{open_robotics_setting_nodate,
	title = {Setting up security — {ROS} 2 {Documentation}: {Rolling} documentation},
	url = {https://docs.ros.org/en/rolling/Tutorials/Advanced/Security/Introducing-ros2-security.html},
	urldate = {2024-11-04},
	author = {{Open Robotics}},
}

@inproceedings{kampmann_optimization-based_2022,
	title = {Optimization-based {Resource} {Allocation} for an {Automotive} {Service}-oriented {Software} {Architecture}},
	copyright = {https://doi.org/10.15223/policy-029},
	isbn = {978-1-66548-821-1},
	doi = {10.1109/IV51971.2022.9827429},
	language = {en},
	booktitle = {2022 {IEEE} {Intelligent} {Vehicles} {Symposium} ({IV})},
	author = {Kampmann, Alexandru and Luer, Maximilian and Kowalewski, Stefan and Alrifaee, Bassam},
	month = jun,
	year = {2022},
	pages = {678--687},
}

@inproceedings{kampmann_dynamic_2019,
	title = {A {Dynamic} {Service}-{Oriented} {Software} {Architecture} for {Highly} {Automated} {Vehicles}},
	isbn = {978-1-5386-7024-8},
	doi = {10.1109/ITSC.2019.8916841},
	urldate = {2024-01-11},
	booktitle = {2019 {IEEE} {Intelligent} {Transportation} {Systems} {Conference} ({ITSC})},
	author = {Kampmann, Alexandru and Alrifaee, Bassam and Kohout, Markus and Wustenberg, Andreas and Woopen, Timo and Nolte, Marcus and Eckstein, Lutz and Kowalewski, Stefan},
	month = oct,
	year = {2019},
	pages = {2101--2108},
}

@inproceedings{henle_architecture_2022,
	title = {Architecture platforms for future vehicles: a comparison of {ROS2} and {Adaptive} {AUTOSAR}},
	booktitle = {2022 {IEEE} 25th {International} {Conference} on {Intelligent} {Transportation} {Systems} ({ITSC})},
	author = {Henle, Jacqueline and Stoffel, Martin and Schindewolf, Marc and Nagele, Ann-Therese and Sax, Eric},
	month = oct,
	year = {2022},
	pages = {3095--3102},
}

@inproceedings{amershi_software_2019,
	title = {Software {Engineering} for {Machine} {Learning}: {A} {Case} {Study}},
	booktitle = {2019 {IEEE}/{ACM} 41st {International} {Conference} on {Software} {Engineering}: {Software} {Engineering} in {Practice} ({ICSE}-{SEIP})},
	author = {Amershi, Saleema and Begel, Andrew and Bird, Christian and DeLine, Robert and Gall, Harald and Kamar, Ece and Nagappan, Nachiappan and Nushi, Besmira and Zimmermann, Thomas},
	month = may,
	year = {2019},
	pages = {291--300},
}

@inproceedings{serban_adapting_2022,
	title = {Adapting {Software} {Architectures} to {Machine} {Learning} {Challenges}},
	booktitle = {2022 {IEEE} {International} {Conference} on {Software} {Analysis}, {Evolution} and {Reengineering} ({SANER})},
	author = {Serban, Alex and Visser, Joost},
	month = mar,
	year = {2022},
	pages = {152--163},
}

@inproceedings{nayak_automotive_2023,
	title = {Automotive {Container} {Orchestration}: {Requirements}, {Challenges} and {Open} {Directions}},
	booktitle = {2023 {IEEE} {Vehicular} {Networking} {Conference} ({VNC})},
	author = {Nayak, Naresh and Grewe, Dennis and Schildt, Sebastian},
	month = apr,
	year = {2023},
	pages = {61--64},
}

@article{pullen_security_2023,
	title = {A {Security} {Process} for the {Automotive} {Service}-{Oriented} {Software} {Architecture}},
	journal = {IEEE Transactions on Vehicular Technology},
	author = {Püllen, Dominik and Frank, Florian and Christl, Marion and Liu, Wuhao and Katzenbeisser, Stefan},
	year = {2023},
	pages = {1--16},
}

@article{rumez_overview_2020,
	title = {An {Overview} of {Automotive} {Service}-{Oriented} {Architectures} and {Implications} for {Security} {Countermeasures}},
	volume = {8},
	journal = {IEEE Access},
	author = {Rumez, Marcel and Grimm, Daniel and Kriesten, Reiner and Sax, Eric},
	year = {2020},
	pages = {221852--221870},
}

@inproceedings{menard_achieving_2020,
	title = {Achieving {Determinism} in {Adaptive} {AUTOSAR}},
	booktitle = {2020 {Design}, {Automation} {Test} in {Europe} {Conference} {Exhibition} ({DATE})},
	author = {Menard, Christian and Goens, Andrés and Lohstroh, Marten and Castrillon, Jeronimo},
	month = mar,
	year = {2020},
	pages = {822--827},
}

@inproceedings{rotermund_requirements_2020,
	title = {Requirements {Analysis} and {Performance} {Evaluation} of {SDN} {Controllers} for {Automotive} {Use} {Cases}},
	booktitle = {2020 {IEEE} {Vehicular} {Networking} {Conference} ({VNC})},
	author = {Rotermund, Randolf and Häckel, Timo and Meyer, Philipp and Korf, Franz and Schmidt, Thomas C.},
	month = dec,
	year = {2020},
	pages = {1--8},
}

@inproceedings{reichart_progress_2021,
	series = {Proceedings},
	title = {Progress on the {AUTOSAR} {Adaptive} {Platform} for {Intelligent} {Vehicles}},
	doi = {10.1007/978-3-658-34752-9_6},
	language = {de},
	booktitle = {Automatisiertes {Fahren} 2020},
	author = {Reichart, Günter and Asmus, Rinat},
	editor = {Bertram, Torsten},
	year = {2021},
	pages = {67--75},
}

@incollection{belsare_micro-ros_2023,
	series = {Studies in {Computational} {Intelligence}},
	title = {Micro-{ROS}},
	booktitle = {Robot {Operating} {System} ({ROS}): {The} {Complete} {Reference} ({Volume} 7)},
	author = {Belsare, Kaiwalya and Rodriguez, Antonio Cuadros and Sánchez, Pablo Garrido and Hierro, Juanjo and Kołcon, Tomasz and Lange, Ralph and Lütkebohle, Ingo and Malki, Alexandre and Losa, Jaime Martin and Melendez, Francisco and Rodriguez, Maria Merlan and Nordmann, Arne and Staschulat, Jan and von Mendel, Julian},
	editor = {Koubaa, Anis},
	year = {2023},
	doi = {10.1007/978-3-031-09062-2_2},
	pages = {3--55},
}

@article{macenski_robot_2022,
	title = {Robot {Operating} {System} 2: {Design}, {Architecture}, and {Uses} {In} {The} {Wild}},
	volume = {7},
	number = {66},
	journal = {Science Robotics},
	author = {Macenski, Steve and Foote, Tully and Gerkey, Brian and Lalancette, Chris and Woodall, William},
	month = may,
	year = {2022},
}

@inproceedings{benckendorff_comparing_2019,
	title = {Comparing current and future {E}/{EArchitecture} trends of commercial vehicles and passenger cars},
	isbn = {978-3-658-25939-6},
	doi = {10.1007/978-3-658-25939-6_95},
	language = {de},
	booktitle = {19. {Internationales} {Stuttgarter} {Symposium}},
	author = {Benckendorff, Tenny and Lapp, Andreas and Oexner, Thomas and Thiel, Thomas},
	editor = {Bargende, Michael and Reuss, Hans-Christian and Wagner, Andreas and Wiedemann, Jochen},
	year = {2019},
	pages = {1190--1200},
}

@inproceedings{dieber_application-level_2016,
	title = {Application-level security for {ROS}-based applications},
	booktitle = {2016 {IEEE}/{RSJ} {International} {Conference} on {Intelligent} {Robots} and {Systems} ({IROS})},
	author = {Dieber, Bernhard and Kacianka, Severin and Rass, Stefan and Schartner, Peter},
	month = oct,
	year = {2016},
	pages = {4477--4482},
}

@inproceedings{corsaro_zenoh_2023,
	title = {Zenoh: {Unifying} {Communication}, {Storage} and {Computation} from the {Cloud} to the {Microcontroller}},
	booktitle = {2023 26th {Euromicro} {Conference} on {Digital} {System} {Design} ({DSD})},
	author = {Corsaro, Angelo and Cominardi, Luca and Hecart, Olivier and Baldoni, Gabriele and Avital, Julien Enoch Pierre and Loudet, Julien and Guimares, Carlos and Ilyin, Michael and Bannov, Dmitrii},
	month = sep,
	year = {2023},
	pages = {422--428},
}

@inproceedings{kampmann_asoa_2022,
	address = {Aachen, Germany},
	title = {{ASOA} - {A} {Dynamic} {Software} {Architecture} for {Software}-defined {Vehicles}},
	copyright = {All rights reserved},
	booktitle = {31st {Aachen} {Colloquium} {Sustainable} {Mobility} 2022},
	author = {Kampmann, Alexandru and Mokhtarian, Armin and Kowalewski, Stefan and Alrifaee, Bassam},
	year = {2022},
}

@inproceedings{vetter_development_2020,
	title = {Development {Processes} in {Automotive} {Service}-oriented {Architectures}},
	booktitle = {2020 9th {Mediterranean} {Conference} on {Embedded} {Computing} ({MECO})},
	author = {Vetter, Andreas and Obergfell, Philipp and Guissouma, Houssem and Grimm, Daniel and Rumez, Marcel and Sax, Eric},
	month = jun,
	year = {2020},
	pages = {1--7},
}

@inproceedings{kugele_service-orientation_2017,
	title = {On {Service}-{Orientation} for {Automotive} {Software}},
	booktitle = {2017 {IEEE} {International} {Conference} on {Software} {Architecture} ({ICSA})},
	author = {Kugele, S. and Obergfell, P. and Broy, M. and Creighton, O. and Traub, M. and Hopfensitz, W.},
	month = apr,
	year = {2017},
	pages = {193--202},
}

@inproceedings{berger_containerized_2017,
	title = {Containerized {Development} and {Microservices} for {Self}-{Driving} {Vehicles}: {Experiences} \& {Best} {Practices}},
	booktitle = {2017 {IEEE} {International} {Conference} on {Software} {Architecture} {Workshops} ({ICSAW})},
	author = {Berger, Christian and Nguyen, Björnborg and Benderius, Ola},
	month = apr,
	year = {2017},
	pages = {7--12},
}

@inproceedings{blass_automatic_2021,
	address = {Nashville, TN, USA},
	title = {Automatic {Latency} {Management} for {ROS} 2: {Benefits}, {Challenges}, and {Open} {Problems}},
	booktitle = {2021 {IEEE} 27th {Real}-{Time} and {Embedded} {Technology} and {Applications} {Symposium} ({RTAS})},
	publisher = {IEEE},
	author = {Blass, Tobias and Hamann, Arne and Lange, Ralph and Ziegenbein, Dirk and Brandenburg, Bjorn B.},
	month = may,
	year = {2021},
	pages = {264--277},
}

@inproceedings{hackel2023dynamic,
  title={Dynamic service-orientation for software-defined in-vehicle networks},
  author={H{\"a}ckel, Timo and Meyer, Philipp and Mueller, Mehmet and Schmitt-Solbrig, Jan and Korf, Franz and Schmidt, Thomas C},
  booktitle={2023 IEEE 97th Vehicular Technology Conference (VTC2023-Spring)},
  pages={1--5},
  year={2023},
  organization={IEEE}
}

@inproceedings{brunner_automotive_2017,
	title = {Automotive {E}/{E}-architecture enhancements by usage of ethernet {TSN}},
	booktitle = {2017 13th {Workshop} on {Intelligent} {Solutions} in {Embedded} {Systems} ({WISES})},
	author = {Brunner, Stefan and Roder, Jurgen and Kucera, Markus and Waas, Thomas},
	month = jun,
	year = {2017},
	pages = {9--13},
}

@inproceedings{kugele_data-centric_2018,
	title = {Data-{Centric} {Communication} and {Containerization} for {Future} {Automotive} {Software} {Architectures}},
	booktitle = {2018 {IEEE} {International} {Conference} on {Software} {Architecture} ({ICSA})},
	author = {Kugele, S. and Hettler, D. and Peter, J.},
	month = apr,
	year = {2018},
	pages = {65--6509},
}

@misc{vilches_sros2_2022,
	title = {{SROS2}: {Usable} {Cyber} {Security} {Tools} for {ROS} 2},
	shorttitle = {{SROS2}},
	publisher = {arXiv},
	author = {Vilches, Victor Mayoral and White, Ruffin and Caiazza, Gianluca and Arguedas, Mikael},
	month = aug,
	year = {2022},
	note = {arXiv:2208.02615 [cs]},
}

@misc{macenski_impact_2023,
	title = {Impact of {ROS} 2 {Node} {Composition} in {Robotic} {Systems}},
	publisher = {arXiv},
	author = {Macenski, Steve and Soragna, Alberto and Carroll, Michael and Ge, Zhenpeng},
	month = may,
	year = {2023},
	note = {arXiv:2305.09933 [cs]},
}

@inproceedings{maruyama_exploring_2016,
	address = {Pittsburgh Pennsylvania},
	title = {Exploring the performance of {ROS2}},
	booktitle = {Proceedings of the 13th {International} {Conference} on {Embedded} {Software}},
	publisher = {ACM},
	author = {Maruyama, Yuya and Kato, Shinpei and Azumi, Takuya},
	month = oct,
	year = {2016},
	pages = {1--10},
}

@inproceedings{garlan_software_2014,
	address = {New York, NY, USA},
	series = {{FOSE} 2014},
	title = {Software architecture: a travelogue},
	booktitle = {Future of {Software} {Engineering} {Proceedings}},
	publisher = {Association for Computing Machinery},
	author = {Garlan, David},
	month = may,
	year = {2014},
	pages = {29--39},
}

@inproceedings{hu_gatekeeper_2022,
	address = {New York, NY, USA},
	series = {{ASIA} {CCS} '22},
	title = {Gatekeeper: {A} {Gateway}-based {Broadcast} {Authentication} {Protocol} for the {In}-{Vehicle} {Ethernet}},
	booktitle = {Proceedings of the 2022 {ACM} on {Asia} {Conference} on {Computer} and {Communications} {Security}},
	publisher = {Association for Computing Machinery},
	author = {Hu, Shengtuo and Zhang, Qingzhao and Weimerskirch, André and Mao, Z. Morley},
	year = {2022},
	pages = {494--507},
}

@inproceedings{furst_autosar_2016,
	address = {Toulouse, France},
	title = {{AUTOSAR} for {Connected} and {Autonomous} {Vehicles}: {The} {AUTOSAR} {Adaptive} {Platform}},
	isbn = {978-1-5090-3688-2},
	shorttitle = {{AUTOSAR} for {Connected} and {Autonomous} {Vehicles}},
	doi = {10.1109/DSN-W.2016.24},
	booktitle = {2016 46th {Annual} {IEEE}/{IFIP} {International} {Conference} on {Dependable} {Systems} and {Networks} {Workshop} ({DSN}-{W})},
	publisher = {IEEE},
	author = {Furst, Simon and Bechter, Markus},
	month = jun,
	year = {2016},
	pages = {215--217},
}

@inproceedings{malavolta_how_2020,
	address = {New York, NY, USA},
	series = {{ICSE}-{SEIP} '20},
	title = {How do you architect your robots? state of the practice and guidelines for {ROS}-based systems},
	booktitle = {Proceedings of the {ACM}/{IEEE} 42nd {International} {Conference} on {Software} {Engineering}: {Software} {Engineering} in {Practice}},
	publisher = {Association for Computing Machinery},
	author = {Malavolta, Ivano and Lewis, Grace and Schmerl, Bradley and Lago, Patricia and Garlan, David},
	month = sep,
	year = {2020},
	pages = {31--40},
}

@article{wang_review_2024,
	title = {Review of {Electrical} and {Electronic} {Architectures} for {Autonomous} {Vehicles}: {Topologies}, {Networking} and {Simulators}},
	volume = {7},
	shorttitle = {Review of {Electrical} and {Electronic} {Architectures} for {Autonomous} {Vehicles}},
	language = {en},
	number = {1},
	journal = {Automotive Innovation},
	author = {Wang, Wenwei and Guo, Kaidi and Cao, Wanke and Zhu, Hailong and Nan, Jinrui and Yu, Lei},
	month = feb,
	year = {2024},
	pages = {82--101},
}

@article{zhu_requirements-driven_2021,
	title = {Requirements-{Driven} {Automotive} {Electrical}/{Electronic} {Architecture}: {A} {Survey} and {Prospective} {Trends}},
	volume = {9},
	issn = {2169-3536},
	shorttitle = {Requirements-{Driven} {Automotive} {Electrical}/{Electronic} {Architecture}},
	language = {en},
	journal = {IEEE Access},
	author = {Zhu, Hailong and Zhou, Wei and Li, Zhiheng and Li, Li and Huang, Tao},
	year = {2021},
}

@misc{aaron_aboagye_facing_2017,
	title = {Facing digital disruption in mobility as a traditional auto player {\textbar} {McKinsey}},
	url = {https://www.mckinsey.com/industries/automotive-and-assembly/our-insights/facing-digital-disruption-in-mobility-as-a-traditional-auto-player},
	urldate = {2024-09-25},
	journal = {Facing digital disruption in mobility as a traditional auto player},
	author = {{Aaron Aboagye} and {Aamer Baig} and {Asutosh Padhi} and {Russel Hensley} and {Richard Kelly} and {Danish Shafi}},
	month = dec,
	year = {2017},
}

@article{noauthor_ieee_2022,
	title = {{IEEE} {Standard} for {Ethernet}},
	url = {https://ieeexplore.ieee.org/document/9844436},
	doi = {10.1109/IEEESTD.2022.9844436},
	urldate = {2024-07-16},
	journal = {IEEE Std 802.3-2022 (Revision of IEEE Std 802.3-2018)},
	month = jul,
	year = {2022},
	pages = {1--7025},
}

@misc{autosar_communicationManagement_2024,
	type = {{AUTOSAR} {Standard} {Adaptive} {Platform}},
	title = {Specification of {Communication} {Management}},
	shorttitle = {Specification of {Communication} {Management} {AUTOSAR} {AP} {R23}-11},
	url = {https://www.autosar.org/fileadmin/standards/R23-11/AP/AUTOSAR_AP_SWS_CommunicationManagement.pdf},
	language = {en},
	urldate = {2024-03-25},
	author = {{AUTOSAR Consortium}},
	month = nov,
	year = {2023},
}

@misc{autosar_executionManagement_2024,
	type = {{AUTOSAR} {Standard} {Adaptive} {Platform}},
	title = {Specification of {Execution} {Management}},
	url = {https://www.autosar.org/fileadmin/standards/R23-11/AP/AUTOSAR_AP_SWS_ExecutionManagement.pdf},
	language = {en},
	urldate = {2024-03-27},
	author = {{AUTOSAR Consortium}},
	month = nov,
	year = {2023},
}

@misc{eclipse_foundation_zenoh-flow_2024,
	type = {Documentation},
	title = {Zenoh-{Flow} 0.6.0-rc: {Getting} {Started} · {Zenoh} - pub/sub, geo distributed storage, query},
	url = {https://zenoh.io/blog/2024-01-31-zenoh-flow-getting-started/},
	language = {en},
	urldate = {2024-07-24},
	journal = {Zenoh Flow Getting Starget},
	author = {{Eclipse Foundation}},
	month = jul,
	year = {2024},
}

@misc{autosar_someip_2024,
	type = {{AUTOSAR} {Standard} {Foundation}},
	title = {{SOME}/{IP} {Protocol} {Specification}},
	shorttitle = {{SOME}/{IP} {Protocol} {Specification} {AUTOSAR} {FO} {R23}-11},
	url = {https://www.autosar.org/fileadmin/standards/R23-11/FO/AUTOSAR_FO_PRS_SOMEIPProtocol.pdf},
	language = {en},
	urldate = {2024-03-19},
	author = {{AUTOSAR Consortium}},
	month = nov,
	year = {2023},
}

@misc{autosar_explanationPlatformDesign_2024,
	type = {Part of {AUTOSAR} {Standard} {Adaptive} {Platform}},
	title = {Explanation of {Adaptive} {Platform} {Design}},
	shorttitle = {Explanation of {Adaptive} {Platform} {Design} {AUTOSAR} {AP} {R23}-11},
	url = {https://www.autosar.org/fileadmin/standards/R23-11/AP/AUTOSAR_AP_EXP_PlatformDesign.pdf},
	language = {en},
	urldate = {2024-03-27},
	author = {{AUTOSAR Consortium}},
	month = nov,
	year = {2023},
}

@misc{eclipse_foundation_deployment_2024,
	type = {Documentation},
	title = {Deployment · {Zenoh} - pub/sub, geo distributed storage, query},
	url = {https://zenoh.io/docs/getting-started/deployment/},
	urldate = {2024-07-22},
	journal = {Zenoh Documentation},
	author = {{Eclipse Foundation}},
	month = jul,
	year = {2024},
}

@misc{eclipse_foundation_zenoh_2024,
	type = {Documentation},
	title = {Zenoh {Adopters} · {Zenoh} - pub/sub, geo distributed storage, query},
	url = {https://zenoh.io/adopters/},
	language = {en},
	urldate = {2024-07-22},
	journal = {Zenoh Documentation},
	author = {{Eclipse Foundation}},
	month = jul,
	year = {2024},
}

@misc{eclipse_foundation_abstractions_2024,
	title = {Abstractions · {Zenoh} - pub/sub, geo distributed storage, query},
	url = {https://zenoh.io/docs/manual/abstractions/},
	language = {en},
	urldate = {2024-07-22},
	journal = {Zenoh Documentation},
	author = {{Eclipse Foundation}},
	month = jul,
	year = {2024},
}

@misc{open_robotics_installation_2024,
	type = {Documentation},
	title = {Installation — {ROS} 2 {Documentation}: {Iron} documentation},
	url = {https://docs.ros.org/en/iron/Installation.html},
	language = {en},
	urldate = {2024-06-04},
	journal = {Installation — ROS 2 Documentation: Iron documentation},
	author = {{Open Robotics}},
	year = {2024},
}

@misc{autosar_consortium_ip_2023,
	type = {{AUTOSAR} {Standard} {Foundation}},
	title = {{SOME}/{IP} {Service} {Discovery} {Protocol} {Specification}},
	shorttitle = {{SOME}/{IP} {Service} {Discovery} {Protocol} {Specification} {AUTOSAR} {FO} {R23}-11},
	url = {https://www.autosar.org/fileadmin/standards/R23-11/FO/AUTOSAR_FO_PRS_SOMEIPServiceDiscoveryProtocol.pdf},
	language = {en},
	urldate = {2024-03-19},
	author = {{AUTOSAR Consortium}},
	month = nov,
	year = {2023},
}

@misc{open_robotics_interfaces_2024,
	title = {Interfaces — {ROS} 2 {Documentation}: {Iron} documentation},
	url = {https://docs.ros.org/en/iron/Concepts/Basic/About-Interfaces.html},
	language = {en},
	urldate = {2024-03-26},
	journal = {Interfaces — ROS 2 Documentation: Iron documentation},
	author = {{Open Robotics}},
	year = {2024},
}

@misc{open_robotics_launch_2024,
	type = {Documentation},
	title = {Launch — {ROS} 2 {Documentation}: {Iron} documentation},
	url = {https://docs.ros.org/en/iron/Concepts/Basic/About-Launch.html},
	language = {en},
	urldate = {2024-03-27},
	journal = {Launch — ROS 2 Documentation: Iron documentation},
	author = {{Open Robotics}},
	year = {2024},
}

@misc{open_robotics_ros_2024,
	type = {Documentation},
	title = {{ROS} 2 {Security} — {ROS} 2 {Documentation}: {Iron} documentation},
	url = {https://docs.ros.org/en/iron/Concepts/Intermediate/About-Security.html},
	language = {en},
	urldate = {2024-03-27},
	journal = {ROS 2 Security — ROS 2 Documentation: Iron documentation},
	author = {{Open Robotics}},
	year = {2024},
}

@misc{open_robotics_quality_2024,
	type = {Documentation},
	title = {Quality of {Service} settings — {ROS} 2 {Documentation}: {Iron} documentation},
	url = {https://docs.ros.org/en/iron/Concepts/Intermediate/About-Quality-of-Service-Settings.html},
	language = {en},
	urldate = {2024-03-27},
	journal = {Quality of Service settings — ROS 2 Documentation: Iron documentation},
	author = {{Open Robotics}},
	year = {2024},
}

@misc{open_robotics_parameters_nodate,
	type = {Documentation},
	title = {Parameters — {ROS} 2 {Documentation}: {Iron} documentation},
	url = {https://docs.ros.org/en/iron/Concepts/Basic/About-Parameters.html},
	language = {en},
	urldate = {2024-03-26},
	journal = {Parameters — ROS 2 Documentation: Iron documentation},
	author = {{Open Robotics}},
}

@misc{open_robotics_nodes_2024,
	type = {Documentation},
	title = {Nodes — {ROS} 2 {Documentation}: {Iron} documentation},
	url = {https://docs.ros.org/en/iron/Concepts/Basic/About-Nodes.html},
	language = {en},
	urldate = {2024-03-26},
	author = {{Open Robotics}},
	year = {2024},
}

@misc{eprosima_dds_2024,
	type = {Documentation},
	title = {{DDS} {API} — {Fast} {DDS} 2.14.0 documentation},
	url = {https://fast-dds.docs.eprosima.com/en/latest/02-formalia/titlepage.html},
	language = {en},
	urldate = {2024-03-26},
	journal = {DDS API — Fast DDS 2.14.0 documentation},
	author = {{eProsima}},
	month = mar,
	year = {2024},
}

@misc{open_robotics_executors_2024,
	type = {Documentation},
	title = {Executors — {ROS} 2 {Documentation}: {Iron} documentation},
	url = {https://docs.ros.org/en/iron/Concepts/Intermediate/About-Executors.html},
	language = {en},
	urldate = {2024-03-27},
	journal = {Executors — ROS 2 Documentation: Iron documentation},
	author = {{Open Robotics}},
	year = {2024},
}

@misc{autosar_ap_explanation_sw_arch,
	type = {Part of {AUTOSAR} {Standard} {Adaptive} {Platform}},
	title = {Explanation of {Adaptive} {Platform} {Software} {Architecture}},
	url = {https://www.autosar.org/fileadmin/standards/R23-11/AP/AUTOSAR_AP_EXP_SWArchitecture.pdf},
	language = {en},
	urldate = {2024-03-14},
	author = {{AUTOSAR Consortium}},
	month = nov,
	year = {2023},
}

@misc{autosar_consortium_explanation_2023,
	type = {Part of {AUTOSAR} {Standard} {Adaptive} {Platform}},
	title = {Explanation of ara::com {API}},
	shorttitle = {Explanation of ara::com {API} {AUTOSAR} {AP} {R23}-11},
	url = {https://www.autosar.org/fileadmin/standards/R23-11/AP/AUTOSAR_AP_EXP_ARAComAPI.pdf},
	language = {en},
	urldate = {2024-03-27},
	author = {{AUTOSAR Consortium}},
	month = nov,
	year = {2023},
}

@misc{eprosima_8_2024,
	type = {Documentation},
	title = {8. {Security} — {Fast} {DDS} 2.14.0 documentation},
	url = {https://fast-dds.docs.eprosima.com/en/latest/fastdds/security/security.html},
	language = {en},
	urldate = {2024-03-26},
	journal = {8. Security — Fast DDS 2.14.0 documentation},
	author = {eProsima},
	month = mar,
	year = {2024},
}

@misc{eprosima_6_2024,
	type = {Documentation},
	title = {6. {Transport} {Layer} — {Fast} {DDS} 2.14.0 documentation},
	url = {https://fast-dds.docs.eprosima.com/en/latest/fastdds/transport/transport.html},
	language = {en},
	urldate = {2024-03-26},
	journal = {6. Transport Layer — Fast DDS 2.14.0 documentation},
	author = {eProsima},
	month = mar,
	year = {2024},
}

@misc{eprosima_12_2024,
	type = {Documentation},
	title = {12. {PropertyPolicyQos} {Options} — {Fast} {DDS} 2.14.0 documentation},
	url = {https://fast-dds.docs.eprosima.com/en/latest/fastdds/property_policies/property_policies.html},
	language = {en},
	urldate = {2024-03-26},
	journal = {12. PropertyPolicyQos Options — Fast DDS 2.14.0 documentation},
	author = {eProsima},
	month = mar,
	year = {2024},
}

@misc{eprosima_5_2024,
	type = {Documentation},
	title = {5. {Discovery} — {Fast} {DDS} 2.14.0 documentation},
	url = {https://fast-dds.docs.eprosima.com/en/latest/fastdds/discovery/discovery.html},
	language = {en},
	urldate = {2024-03-26},
	journal = {5. Discovery — Fast DDS 2.14.0 documentation},
	author = {eProsima},
	month = mar,
	year = {2024},
}

@misc{eprosima_3_2024,
	type = {Documentation},
	title = {3. {DDS} {Layer} — {Fast} {DDS} 2.14.0 documentation},
	url = {https://fast-dds.docs.eprosima.com/en/latest/fastdds/dds_layer/dds_layer.html},
	language = {en},
	urldate = {2024-03-26},
	journal = {3. DDS Layer — Fast DDS 2.14.0 documentation},
	author = {eProsima},
	month = mar,
	year = {2024},
}

@misc{eprosima_11_2024,
	type = {Documentation},
	title = {1.1. {What} is {DDS}? — {Fast} {DDS} 2.14.0 documentation},
	url = {https://fast-dds.docs.eprosima.com/en/latest/fastdds/getting_started/definitions.html#the-dcps-conceptual-model},
	language = {en},
	urldate = {2024-03-26},
	journal = {1. Getting Started » 1.1. What is DDS?},
	author = {eProsima},
	month = mar,
	year = {2024},
}

@inproceedings{migge_insights_2018,
	title = {Insights on the {Performance} and {Configuration} of {AVB} and {TSN} in {Automotive} {Ethernet} {Networks}},
	url = {https://www.semanticscholar.org/paper/Insights-on-the-Performance-and-Configuration-of-in-Migge-Villanueva/ca961a75d95ac2a74aaf67a5c0d53a2348c1813c},
	urldate = {2024-03-19},
	author = {Migge, J. and Villanueva, Josetxo and Navet, N. and Boyer, M.},
	month = feb,
	year = {2018},
}

@misc{van_dijk_future_2017,
	title = {Future {Vehicle} {Networks} and {ECUs}},
	shorttitle = {Future {Vehicle} {Networks} and {ECUs}},
	url = {https://www.nxp.com/docs/en/white-paper/FVNECUA4WP.pdf},
	language = {en},
	urldate = {2024-03-14},
	publisher = {NXP Semiconductors},
	author = {van Dijk, Luc},
	year = {2017},
}

@misc{burkacky_rethinking_nodate,
	title = {Rethinking car software and electronics architecture {\textbar} {McKinsey}},
	url = {https://www.mckinsey.com/industries/automotive-and-assembly/our-insights/rethinking-car-software-and-electronics-architecture},
	language = {en},
	urldate = {2024-03-14},
	journal = {Rethinking car software and electronics architecture},
	author = {Burkacky, Ondrej and Deichmann, Johannes and Doll, Georg and Knochenhauer, Christian},
}

@inproceedings{teper_end--end_2022,
	address = {Houston, TX, USA},
	title = {End-{To}-{End} {Timing} {Analysis} in {ROS2}},
	isbn = {978-1-66545-346-2},
	doi = {10.1109/RTSS55097.2022.00015},
	urldate = {2024-01-11},
	booktitle = {2022 {IEEE} {Real}-{Time} {Systems} {Symposium} ({RTSS})},
	publisher = {IEEE},
	author = {Teper, Harun and Gunzel, Mario and Ueter, Niklas and Von Der Bruggen, Georg and Chen, Jian-Jia},
	month = dec,
	year = {2022},
	pages = {53--65},
}

@book{bass_software_2012,
	edition = {3rd},
	title = {Software {Architecture} in {Practice}},
	isbn = {978-0-321-81573-6},
	publisher = {Addison-Wesley Professional},
	author = {Bass, Len and Clements, Paul and Kazman, Rick},
	month = sep,
	year = {2012},
}

@inproceedings{aartsen_analyzing_2022,
	title = {Analyzing {Interoperability} and {Security} {Overhead} of {ROS2} {DDS} {Middleware}},
	doi = {10.1109/MED54222.2022.9837282},
	urldate = {2023-12-04},
	booktitle = {2022 30th {Mediterranean} {Conference} on {Control} and {Automation} ({MED})},
	author = {Aartsen, Max and Banga, Kanta and Talko, Konrad and Touw, Dustin and Wisman, Bertus and Meïnsma, Daniel and Björkqvist, Mathias},
	month = jun,
	year = {2022},
	note = {ISSN: 2473-3504},
	pages = {976--981},
}

@misc{kronauer_latency_2021,
	title = {Latency {Analysis} of {ROS2} {Multi}-{Node} {Systems}},
	url = {http://arxiv.org/abs/2101.02074},
	doi = {10.48550/arXiv.2101.02074},
	urldate = {2023-12-03},
	publisher = {arXiv},
	author = {Kronauer, Tobias and Pohlmann, Joshwa and Matthe, Maximilian and Smejkal, Till and Fettweis, Gerhard},
	month = jun,
	year = {2021},
	note = {arXiv:2101.02074 [cs]},
}

@article{wu_oops_2021,
	title = {Oops! {It}'s {Too} {Late}. {Your} {Autonomous} {Driving} {System} {Needs} a {Faster} {Middleware}},
	volume = {6},
	issn = {2377-3766},
	doi = {10.1109/LRA.2021.3097439},
	number = {4},
	journal = {IEEE Robotics and Automation Letters},
	author = {Wu, Tianze and Wu, Baofu and Wang, Sa and Liu, Liangkai and Liu, Shaoshan and Bao, Yungang and Shi, Weisong},
	month = oct,
	year = {2021},
	note = {Conference Name: IEEE Robotics and Automation Letters},
	pages = {7301--7308},
}

@inproceedings{woopen_unicaragil_2018,
	address = {Aachen, Germany},
	title = {{UNICARagil} - {Disruptive} {Modular} {Architectures} for {Agile}, {Automated} {Vehicle} {Concepts}},
	copyright = {All rights reserved},
	doi = {10.18154/RWTH-2018-229909},
	booktitle = {Aachen {Colloquium}},
	author = {Woopen, T. and Lampe, B. and Böddeker, T. and Eckstein, L. and Kampmann, A. and Alrifaee, B. and Kowalewski, S. and Moormann, D. and Stolte, T. and Jatzkowski, I. and Maurer, M. and Möstl, M. and Ernst, R. and Ackermann, S. and Amersbach, C. and Leinen, S. and Winner, H. and Püllen, D. and Katzenbeisser, S. and Becker, M. and Stiller, C. and Furmans, K. and Bengler, K. and Diermeyer, F. and Lienkamp, M. and Keilhoff, D. and Reuss, H. C. and Buchholz, M. and Dietmayer, K. and Lategahn, H. and Siepenkötter, N. and Elbs, M. and v. Hinüber, E. and Dupuis, M. and Hecker, C.},
	year = {2018},
}

@article{gemlau_system-level_2021,
	title = {System-level {Logical} {Execution} {Time}: {Augmenting} the {Logical} {Execution} {Time} {Paradigm} for {Distributed} {Real}-time {Automotive} {Software}},
	volume = {5},
	shorttitle = {System-level {Logical} {Execution} {Time}},
	language = {en},
	number = {2},
	urldate = {2023-04-13},
	journal = {ACM Transactions on Cyber-Physical Systems},
	author = {Gemlau, Kai-Björn and Köhler, Leonie and Ernst, Rolf and Quinton, Sophie},
	month = apr,
	year = {2021},
	pages = {1--27},
}

@inproceedings{staschulat_rclc_2020,
	title = {The rclc {Executor}: {Domain}-specific deterministic scheduling mechanisms for {ROS} applications on microcontrollers: work-in-progress},
	shorttitle = {The rclc {Executor}},
	booktitle = {2020 {International} {Conference} on {Embedded} {Software} ({EMSOFT})},
	publisher = {IEEE},
	author = {Staschulat, Jan and Lütkebohle, Ingo and Lange, Ralph},
	year = {2020},
	pages = {18--19},
}

@inproceedings{yang_exploring_2020,
	title = {Exploring {Real}-{Time} {Executor} on {ROS} 2},
	doi = {10.1109/ICESS49830.2020.9301530},
	booktitle = {2020 {IEEE} {International} {Conference} on {Embedded} {Software} and {Systems} ({ICESS})},
	author = {Yang, Yuqing and Azumi, Takuya},
	month = dec,
	year = {2020},
	pages = {1--8},
}

@inproceedings{jatzkowski_integration_2021,
	title = {Integration of a {Vehicle} {Operating} {Mode} {Management} into {UNICARagil}’s {Automotive} {Service}-oriented {Software} {Architecture}},
	booktitle = {30th {Aachen} {Colloquium} {Sustainable} {Mobility} 2021, 4.–6. {Oktober} 2021},
	author = {Jatzkowski, Inga and Stolte, Torben and Graubohm, Robert and Maurer, Markus and Kampmann, Alexandru and Alrifaee, Bassam and Kowalewski, Stefan and Buchholz, Michael and Dietmayer, Klaus},
	year = {2021},
}

@misc{autosar_consortium_overview_2022,
	title = {Overview of {Functional} {Safety} {Measures} in {AUTOSAR}},
	url = {https://www.autosar.org/fileadmin/standards/R22-11/CP/AUTOSAR_EXP_FunctionalSafetyMeasures.pdf},
	language = {en},
	urldate = {2025-11-15},
	author = {{AUTOSAR Consortium}},
	month = nov,
	year = {2022},
}

@misc{eclipse-zenoh_iso_nodate,
	title = {{ISO} 26262 certification for {Zenoh} · eclipse-zenoh/roadmap · {Discussion} \#167},
	url = {https://github.com/eclipse-zenoh/roadmap/discussions/167},
	language = {en},
	urldate = {2025-11-18},
	journal = {GitHub},
	author = {{eclipse-zenoh}},
}

@misc{tttech_auto_zetta_2024,
	title = {Zetta {Auto} {\textbar} {TTTech} {Auto}},
	url = {https://www.tttech-auto.com/software-products/zetta-auto},
	language = {en},
	urldate = {2025-11-18},
	author = {{TTTech Auto}},
	month = nov,
	year = {2024},
}

@misc{autosar_consortium_specification_2024,
	title = {Specification of {Execution} {Management}},
	url = {https://www.autosar.org/fileadmin/standards/R24-11/AP/AUTOSAR_AP_SWS_ExecutionManagement.pdf},
	language = {en},
	urldate = {2025-11-15},
	author = {{AUTOSAR Consortium}},
	month = nov,
	year = {2024},
}

@misc{eprosima_157_2025,
	title = {15.7. {Real}-time behavior - 3.4.0},
	url = {https://fast-dds.docs.eprosima.com/en/stable/fastdds/use_cases/realtime/realtime.html},
	urldate = {2025-11-18},
	author = {{eProsima}},
    year = {2025},
}

@article{baron_performance_2025,
	title = {On the performance of {Zenoh} in {Industrial} {IoT} {Scenarios}},
	volume = {170},
	issn = {1570-8705},
	url = {https://www.sciencedirect.com/science/article/pii/S1570870525000320},
	doi = {10.1016/j.adhoc.2025.103784},
	urldate = {2025-11-18},
	journal = {Ad Hoc Networks},
	author = {Barón, Miguel and Diez, Luis and Zverev, Mihail and Juárez, José R. and Agüero, Ramón},
	month = apr,
	year = {2025},
	pages = {103784},
}

@misc{microros_supported_2025,
	title = {Supported {Hardware}},
	url = {https://micro.ros.org/docs/overview/hardware/},
	language = {en-US},
	urldate = {2025-11-18},
	journal = {micro-ROS},
	month = sep,
	year = {2025},
	author = {{micro-ROS}}
}

@misc{open_robotics_rep_2023,
	title = {{REP} 2000 -- {ROS} 2 {Releases} and {Target} {Platforms} ({ROS}.org)},
	url = {https://www.ros.org/reps/rep-2000.html},
	urldate = {2025-11-18},
	author = {{Open Robotics}},
	month = nov,
	year = {2023},
}

@misc{eprosima_eprosima_nodate,
	title = {{eProsima} {Safe} {DDS}},
	url = {https://www.eprosima.com/middleware/safe-dds},
	urldate = {2025-11-18},
	author = {{eProsima}},
}

@misc{tuv_sud_certificate_2024,
	title = {Certificate {Explorer} - {Z10} 129613 0001 {Rev}. 00 - {eProsima} {Safe} {DDS}},
	url = {https://www.tuvsud.com/en/customer-hub/ps-cert/?q=Z10%20129613%200001%20Rev.%2000},
	language = {en},
	urldate = {2025-11-18},
	author = {{TÜV SÜD}},
	month = dec,
	year = {2024},
}

@misc{eprosima_dependencies_2025,
	type = {Documentation},
	title = {Dependencies and compatibilities — {Fast} {DDS} 3.4.0 documentation},
	url = {https://fast-dds.docs.eprosima.com/en/stable/notes/versions.html#platform-support},
	language = {en},
	urldate = {2025-11-17},
	journal = {Fast DDS 3.4.0 documentation},
	author = {eProsima},
	month = may,
	year = {2025},
}

@misc{noauthor_nokiaeclipse-zenoh-zenoh-c_2024,
	title = {nokia/eclipse-zenoh-zenoh-c},
	url = {https://github.com/nokia/eclipse-zenoh-zenoh-c},
	urldate = {2025-11-18},
	publisher = {Nokia},
	month = feb,
	year = {2024},
	note = {original-date: 2024-02-14T08:08:30Z},
}

@misc{noauthor_eclipse-zenohzenoh-java_2025,
	title = {eclipse-zenoh/zenoh-java},
	url = {https://github.com/eclipse-zenoh/zenoh-java},
	urldate = {2025-11-18},
	publisher = {Eclipse zenoh},
	month = nov,
	year = {2025},
	note = {original-date: 2020-01-21T15:42:31Z},
}

@article{chisalita_stepping_2025,
	title = {Stepping {Toward} {Zenoh} {Protocol} in {Automotive} {Scenarios}},
	volume = {13},
	urldate = {2025-11-07},
	journal = {IEEE Access},
	author = {Chisăliţă, Andreea-Iasmina and Korodi, Adrian},
	year = {2025},
	pages = {166167--166180},
}

@INPROCEEDINGS{kluner_automotive_2025,
title={Automotive Middleware Performance: Comparison of FastDDS, Zenoh and vSomeIP},
booktitle={2025 IEEE International Conference on Vehicular Electronics and Safety (ICVES)},
author={David Philipp Klüner and Lucas Hegerath and Amin Dieter Hatib and Stefan Kowalewski and Bassam Alrifaee and Alexandru Kampmann},
year={2025},
volume={},
number={},
pages={1-8}
}

@article{chovet_performance_2025,
	title = {Performance {Comparison} of {ROS2} {Middlewares} for {Multi}-robot {Mesh} {Networks} in {Planetary} {Exploration}},
	volume = {111},
	language = {en},
	number = {1},
	urldate = {2025-11-07},
	journal = {Journal of Intelligent \& Robotic Systems},
	author = {Chovet, Loïck Pierre and Garcia, Gabriel Manuel and Bera, Abhishek and Richard, Antoine and Yoshida, Kazuya and Olivares-Mendez, Miguel Angel},
	month = jan,
	year = {2025},
	pages = {18},
}

@article{seyler_insights_2015,
	title = {Insights on the {Configuration} and {Performances} of {SOME}/{IP} {Service} {Discovery}},
	volume = {8},
	language = {English},
	number = {1},
	urldate = {2025-11-07},
	journal = {SAE International Journal of Passenger Cars - Electronic and Electrical Systems},
	author = {Seyler, Jan and Navet, Nicolas and Fejoz, Loïc},
	month = apr,
	year = {2015},
	note = {Publisher: SAE International},
	pages = {124--129},
}

@inproceedings{yang_xvsomeip_2024,
	title = {{XvSomeIP}: {A} {High}-{Performance} {In}-{Vehicle} {Communication} {Middleware} {Based} on {XDP}},
	shorttitle = {{XvSomeIP}},
	urldate = {2025-11-07},
	booktitle = {2024 {IEEE} 22nd {International} {Conference} on {Industrial} {Informatics} ({INDIN})},
	author = {Yang, Guoqing and Zhong, Hongming and Zhou, Qiang and Lv, Pan and Li, Hong and Pan, Zhijie},
	month = aug,
	year = {2024},
	note = {ISSN: 2378-363X},
	pages = {1--6},
}

@inproceedings{kouril_performance_2024,
	title = {Performance evaluation of a {ROS2} based {Automated} {Driving} {System}},
	urldate = {2025-11-07},
	booktitle = {Proceedings of the 10th {International} {Conference} on {Vehicle} {Technology} and {Intelligent} {Transport} {Systems}},
	author = {Kouril, Jorin and Schäufele, Bernd and Radusch, Ilja and Schnor, Bettina},
	year = {2024},
	note = {arXiv:2411.11607 [cs]},
	pages = {52--63},
}

@misc{schulik_efficient_2025,
	title = {Efficient {Integration} of cross platform functions onto service-oriented architectures},
	url = {http://arxiv.org/abs/2510.27344},
	urldate = {2025-11-16},
	publisher = {arXiv},
	author = {Schulik, Thomas and Batchu, Viswanatha Reddy and Dharmapuri, Ramesh Kumar and Gundlapalli, Saran and Nadarajan, Parthasarathy and Pelcz, Philipp},
	month = oct,
	year = {2025},
	note = {arXiv:2510.27344 [cs]},
}

@article{merakanapalli_transitioning_2025,
	title = {Transitioning from {AUTOSAR} {Classic} to {Adaptive} for {Service}-{Based} {Architectures}},
	volume = {6},
	language = {en},
	number = {4},
	urldate = {2025-11-16},
	journal = {International Journal of Emerging Research in Engineering and Technology},
	author = {Merakanapalli, Saibabu and Bodapati, Sai Jagadish},
	month = oct,
	year = {2025},
	pages = {7--17},
}

@misc{lee_optimizing_2025,
	title = {Optimizing {ROS} 2 {Communication} for {Wireless} {Robotic} {Systems}},
	url = {http://arxiv.org/abs/2508.11366},
	urldate = {2025-11-16},
	publisher = {arXiv},
	author = {Lee, Sanghoon and Kim, Taehun and Chae, Jiyeong and Park, Kyung-Joon},
	month = aug,
	year = {2025},
	note = {arXiv:2508.11366 [cs]},
}

@article{castillo-sanchez_swarm_2024,
	title = {Swarm {Robot} {Communications} in {ROS} 2: {An} {Experimental} {Study}},
	volume = {12},
	issn = {2169-3536},
	shorttitle = {Swarm {Robot} {Communications} in {ROS} 2},
	urldate = {2025-11-16},
	journal = {IEEE Access},
	author = {Castillo-Sánchez, José-Borja and González-Parada, Eva and Cano-García, José-Manuel},
	year = {2024},
	pages = {142930--142943},
}

@inproceedings{patil_towards_2025,
	address = {Cham},
	title = {Towards {Specification}-{Driven} {LLM}-{Based} {Generation} of {Embedded} {Automotive} {Software}},
	language = {en},
	booktitle = {Bridging the {Gap} {Between} {AI} and {Reality}},
	publisher = {Springer Nature Switzerland},
	author = {Patil, Minal Suresh and Ung, Gustav and Nyberg, Mattias},
	editor = {Steffen, Bernhard},
	year = {2025},
	pages = {125--144},
}

@article{asad_advancing_2025,
	series = {58th {CIRP} {Conference} on {Manufacturing} {Systems} 2025},
	title = {Advancing {Automotive} {Production}: {An} {LLM}-based {Impact} {Analysis} for {Software} {Updates}},
	volume = {134},
	shorttitle = {Advancing {Automotive} {Production}},
	urldate = {2025-11-17},
	journal = {Procedia CIRP},
	author = {Asad, Aiman El and Hahn, Michael and Zhai, Yi and Reuss, Hans-Christian},
	month = jan,
	year = {2025},
	pages = {127--132},
}

@inproceedings{scarano_assessing_2025,
	address = {New York, NY, USA},
	series = {{CSCS} '25},
	title = {Assessing {LLMs} models’ knowledge of automotive cyberthreats benchmarking {autoISAC} framework},
	urldate = {2025-11-17},
	booktitle = {Proceedings of the 2nd {Cyber} {Security} in {CarS} {Workshop}},
	publisher = {Association for Computing Machinery},
	author = {Scarano, Nicola and Mannella, Luca and Savino, Alessandro and Di Carlo, Stefano},
	month = oct,
	year = {2025},
	pages = {1--7},
}

@misc{petrovic_llm-based_2025,
	title = {{LLM}-{Based} {Approach} for {Enhancing} {Maintainability} of {Automotive} {Architectures}},
	url = {http://arxiv.org/abs/2509.12798},
	urldate = {2025-11-17},
	publisher = {arXiv},
	author = {Petrovic, Nenad and Mazur, Lukasz and Knoll, Alois},
	month = sep,
	year = {2025},
	note = {arXiv:2509.12798 [cs]},
}

@misc{kirchner_generating_2025,
	title = {Generating {Automotive} {Code}: {Large} {Language} {Models} for {Software} {Development} and {Verification} in {Safety}-{Critical} {Systems}},
	shorttitle = {Generating {Automotive} {Code}},
	url = {http://arxiv.org/abs/2506.04038},
	urldate = {2025-11-17},
	publisher = {arXiv},
	author = {Kirchner, Sven and Knoll, Alois C.},
	month = jun,
	year = {2025},
	note = {arXiv:2506.04038 [cs]},
}

@article{beyer_reliable_2019,
	title = {Reliable benchmarking: requirements and solutions},
	volume = {21},
	shorttitle = {Reliable benchmarking},
	language = {en},
	number = {1},
	urldate = {2025-11-17},
	journal = {International Journal on Software Tools for Technology Transfer},
	author = {Beyer, Dirk and Löwe, Stefan and Wendler, Philipp},
	month = feb,
	year = {2019},
	pages = {1--29},
}

@article{lohstroh_toward_2021_b,
	title = {Toward a {Lingua} {Franca} for {Deterministic} {Concurrent} {Systems}},
	volume = {20},
	number = {4},
	urldate = {2025-11-19},
	journal = {ACM Trans. Embed. Comput. Syst.},
	author = {Lohstroh, Marten and Menard, Christian and Bateni, Soroush and Lee, Edward A.},
	month = may,
	year = {2021},
	pages = {36:1--36:27},
}

@article{lee_determinism_2021,
	title = {Determinism},
	volume = {20},
	language = {en},
	number = {5},
	urldate = {2025-11-19},
	journal = {ACM Transactions on Embedded Computing Systems},
	author = {Lee, Edward A.},
	month = sep,
	year = {2021},
	pages = {1--34},
}

\begin{IEEEbiography}
    [{\includegraphics[width=1in,height=1.25in,clip,keepaspectratio]{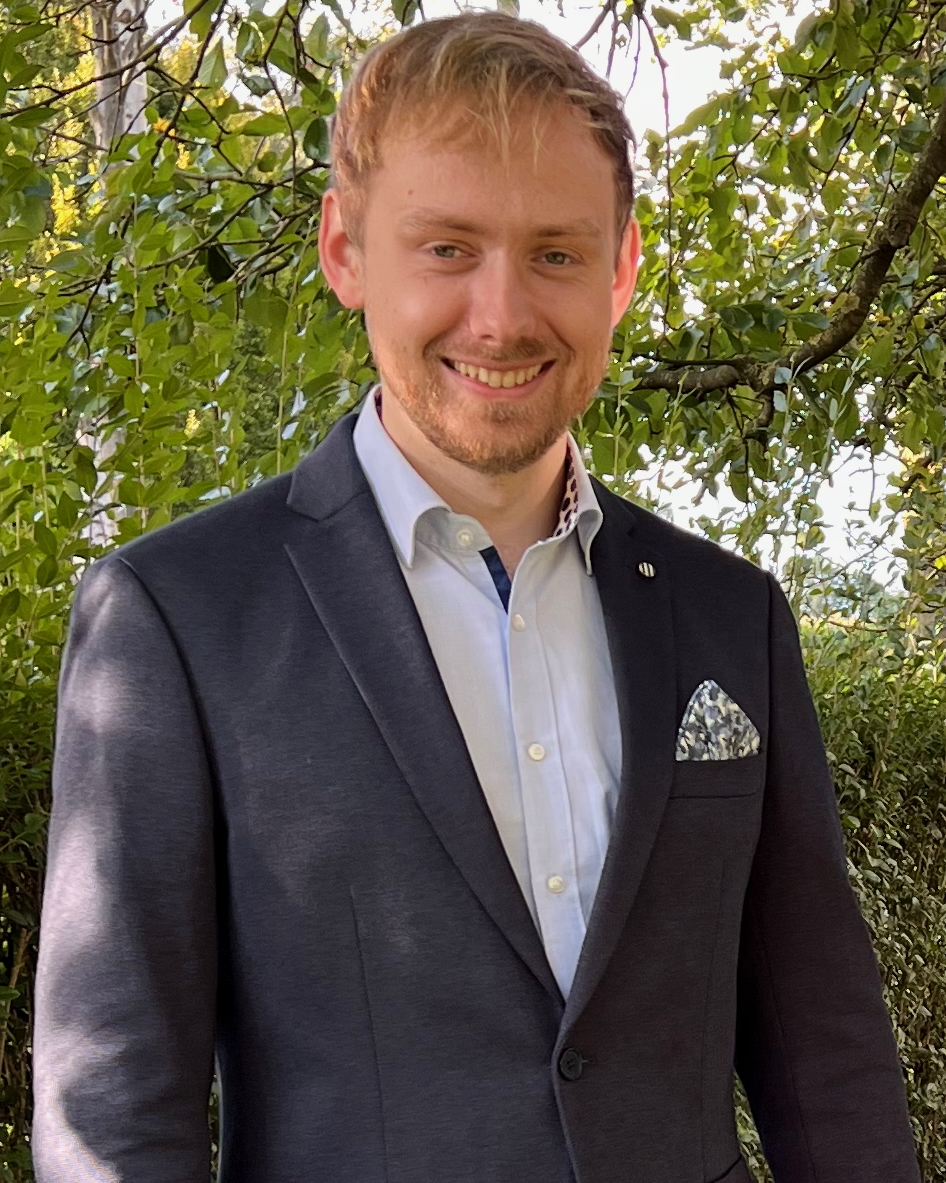}}]{David Klüner} received the B.Sc. and M.Sc. degrees in computer science from RWTH Aachen University, Aachen, Germany. He is currently a Research Associate at Informatik 11–Embedded Software, RWTH Aachen University. His research interest includes automotive middlewares and machine learning in automotive software systems.
\end{IEEEbiography}
\begin{IEEEbiography}
    [{\includegraphics[width=1in,height=1.25in,clip,keepaspectratio]{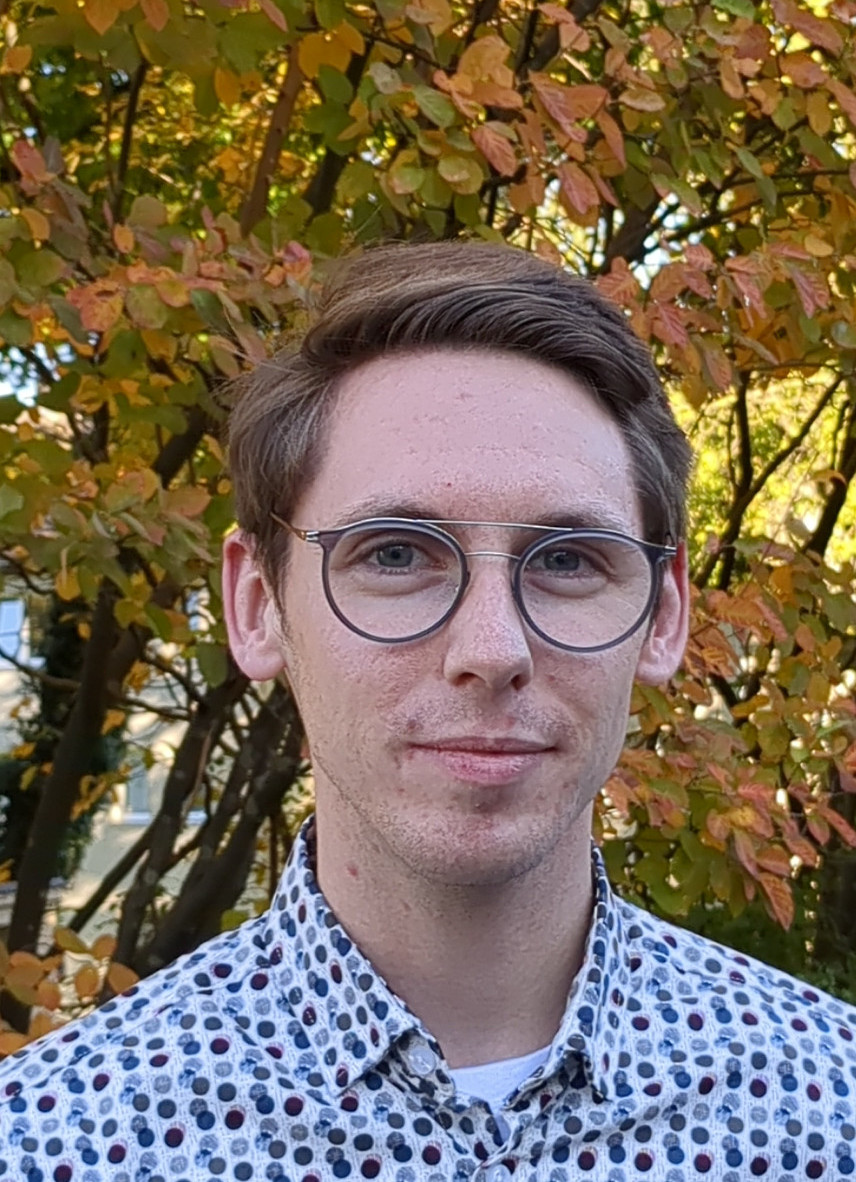}}]{Marius Molz} received the B.Sc. and M.Sc. degrees in computer science from RWTH Aachen University, Aachen, Germany. He is currently a Research Associate at Informatik 11–Embedded Software, RWTH Aachen University. His research interest includes automotive middlewares and safety in automotive software systems.
\end{IEEEbiography}

\begin{IEEEbiography}
    [{\includegraphics[width=1in,height=1.25in,clip,keepaspectratio]{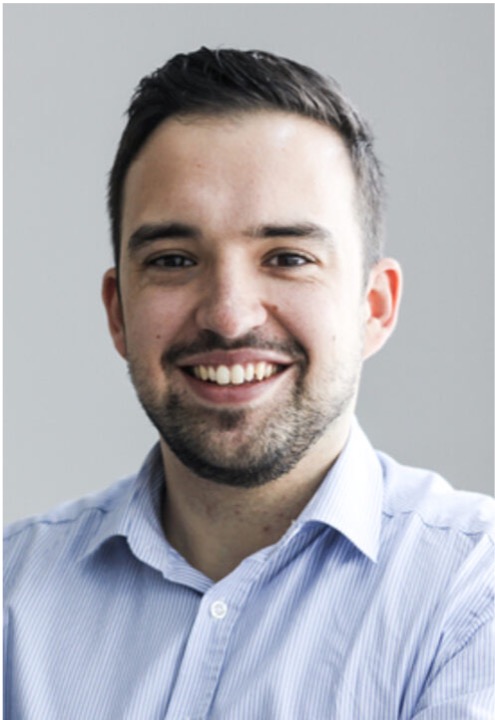}}]{Dr. Alexandru Kampmann} earned his master’s degree in computer science from RWTH Aachen University in 2016, followed by a Ph.D. in 2023. He is currently a research scientist at the Embedded Software Chair at RWTH Aachen University. His primary research interests lies in the area of automotive software engineering, with a focus on middleware and engineering tools.
\end{IEEEbiography}
\begin{IEEEbiography}
    [{\includegraphics[width=1in,height=1.25in,clip,keepaspectratio]{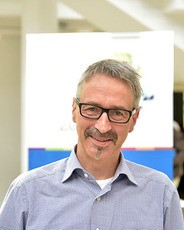}}]{Prof. Stefan Kowalewski} is a full professor of embedded software at RWTH Aachen University. He earned his Dipl.-Ing. degree in Electrical Engineering from the University of Karlsruhe and holds both a Dr.-Ing. and a habilitation degree in Control and Safety Engineering from the University of Dortmund. Before joining RWTH Aachen, he served as a group manager in the Corporate Research and Advanced Engineering division of Robert Bosch GmbH in Frankfurt, Germany, where he specialized in automotive software engineering. His primary research
interests include the design and analysis of software-intensive embedded systems, with a focus on the application of formal methods and safety-critical systems.
\end{IEEEbiography}
\begin{IEEEbiography}
    [{\includegraphics[width=1in,height=1.25in,clip,keepaspectratio]{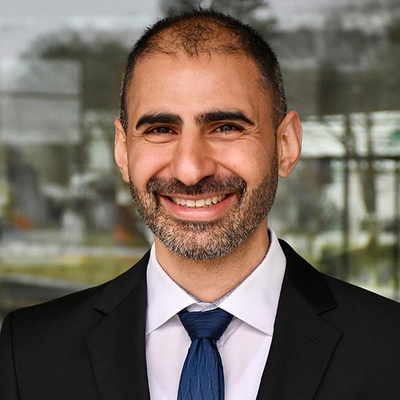}}]{Prof. Bassam Alrifaee} holds the Professorship for Adaptive Behavior of Autonomous Vehicles in the Department of Aerospace Engineering at the University of the Bundeswehr (UniBw) Munich. His research focuses on the intelligent control of autonomous systems, with particular emphasis on distributed control, cooperative localization, software architectures, and experimental validation.
Before joining UniBw Munich in 2024, he served as a Senior Researcher and Lecturer at RWTH Aachen University, where he founded the Cyber-Physical Mobility (CPM) group and the CPM Lab. In 2023, he was a Visiting Scholar at the Information and Decision Science Laboratory at the University of Delaware, USA.
Prof. Alrifaee has secured research grants from various institutions and received awards for his advisory and editorial contributions. He is a Senior Member of the IEEE.
\end{IEEEbiography}
\vfill\null
\end{document}